\documentclass[a4paper,11pt]{article}
\usepackage{amsmath}
\usepackage{graphicx}
\usepackage{amsbsy}
\usepackage{cite}
\usepackage{amsfonts}
\usepackage{amssymb}

\begin{document}
\thispagestyle{empty}  
\setcounter{page}{0}  
\begin{flushright}
CP3-11-12\\March 2011 \\
\end{flushright}

\vskip 2.2 true cm

\begin{center}
{\huge The $s\rightarrow d\gamma$ decay in and beyond the Standard Model}\\[1.9cm]

\textsc{Philippe Mertens}$^{1}$\textsc{ and Christopher Smith}$^{2}$%
\\[12pt]$^{1}$\textsl{Center for Cosmology, Particle Physics and Phenomenology
(CP3), }

\textsl{Universit\'{e} catholique de Louvain, Chemin du Cyclotron 2, 1348
Louvain-la-Neuve, BELGIUM}\\[9pt]
$^{2}$\textsl{Universit\'{e} Lyon 1 \& CNRS/IN2P3, UMR5822 IPNL,}

\textsl{Rue Enrico Fermi 4, 69622 Villeurbanne Cedex, FRANCE}\\[1.9cm]

\textbf{Abstract}
\end{center}

\begin{quote}
\noindent\ The New Physics sensitivity of the $s\rightarrow d\gamma$ transition and its accessibility through hadronic processes are thoroughly investigated. Firstly, the Standard Model predictions for the direct CP-violating observables in radiative $K$ decays are systematically improved. Besides, the magnetic contribution to $\varepsilon^{\prime}$ is estimated and
found subleading, even in the presence of New Physics, and a new strategy to resolve its electroweak versus QCD penguin fraction is identified. Secondly, the signatures of a series of New Physics scenarios, characterized as model-independently as possible in terms of their underlying dynamics, are investigated by combining the information from all the FCNC transitions in the $s\rightarrow d$ sector.\footnotetext[1]{philippe.mertens@uclouvain.be} \footnotetext[2]{c.smith@ipnl.in2p3.fr}\newpage%

\setcounter{tocdepth}{2}
\rule{\linewidth}{0.3mm}
\tableofcontents
\rule{\linewidth}{0.3mm}
\end{quote}

\section{Introduction}

Quantum electrodynamics is among the most successful theories ever designed. At very low energy, up to a few MeV, its predictions have been tested and confirmed to a fantastic level of precision. At higher energies, with the advent of the Standard Model (SM) arises the possibility for the
electromagnetic current to induce flavor transitions. This peculiar phenomenon requires a delicate interplay at the quantum level between the three families of matter particles. So delicate in fact that in the presence of physics beyond the Standard Model, significant deviations are expected. As for the past 150 years, electromagnetism could thus once more guide our quest for unification, and enlighten our understanding of Nature.

For this reason, the $b\rightarrow s\gamma$ and $\mu\rightarrow e\gamma$ transitions have received considerable attention. The former is known to NNLO precision in the SM~\cite{bsg}, and has been measured accurately at the $B$ factories~\cite{bsgEXP}. It is now one of the most constraining observables for New Physics (NP) models. The latter, obviously free of hadronic uncertainties, is so small in the SM that its experimental observation would immediately signal the presence of NP~\cite{meg}. Further, most models do not suppress this transition as effectively as the SM, with rates within reach of the current MEG experiment at PSI~\cite{megEXP}.

The $s\rightarrow d\gamma$ process is complementary to $b\rightarrow s\gamma$ and $\mu\rightarrow e\gamma$, as the relative strengths of these transitions is a powerful tool to investigate the NP dynamics. However, two issues have severely hampered its abilities up to now. First, the $s\rightarrow d\gamma$ decay takes place deep within the QCD non-perturbative regime, and thus
requires control over the low-energy hadronic physics. Second, these hadronic effects strongly enhance the SM contribution, to the point that identifying a possible deviation from NP is very challenging both theoretically and experimentally. To circumvent those difficulties is one of the goals of the present paper.

Indeed, the experimental situation calls for improved theoretical treatments. The recent experimental results~\cite{ExpKppg} for the $K^{+}\rightarrow \pi^{+}\pi^{0}\gamma$ decay, driven by the $s\rightarrow d\gamma$ process, should be exploited. More importantly, several $K$ decay experiments will start in the next few years, NA62 at CERN, K0TO at J-Parc, and KLOE-II at the
LNF. In view of their expected high luminosities, new strategies may open up to constrain, or even signal, the NP in the $s\rightarrow d\gamma$ transition. This requires identifying the most promising observables, both in terms of theoretical control over the SM contributions and in terms of sensitivity to NP effects. These are the two other goals of the paper.

In the next section, the anatomy of the $s\rightarrow d\gamma$ process in the SM is detailed, together with the tools required to deal with the long-distance QCD effects. From these general considerations, the best windows to probe the $s\rightarrow d\gamma$ decays are identified. These observables are then analyzed in details in the following section, where predictions for their SM contributions are obtained. Particular attention is paid to their sensitivity to short-distance effects, and thereby to possible NP contributions. This is put to use in the last (mostly self-contained) section, where the signatures of several NP scenarios are characterized in terms of
correlations among the rare and radiative $K$ decays, as well as $\operatorname{Re}(\varepsilon^{\prime}/\varepsilon)$.

\section{The flavor-changing electromagnetic currents}

\begin{figure}[t]
\centering       \includegraphics[width=9.2cm]{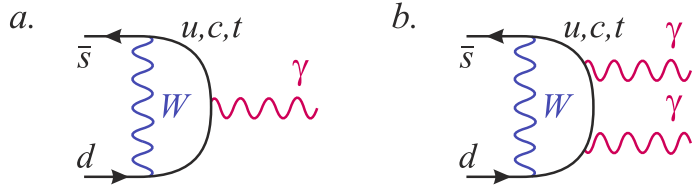}  \caption{The
flavor-changing electromagnetic currents in the Standard Model.}
\label{Fig1}
\end{figure}

In the SM, the flavor changing electromagnetic current arises at the loop level, as depicted in Fig.~\ref{Fig1}. When QCD is turned off, and $m_{s,d}\ll m_{u,c,t}$, the single photon penguin can be embedded into local effective interactions of dimension greater than four:%
\begin{equation}
\mathcal{H}_{eff}^{\gamma}=C_{\gamma}^{\pm}Q_{\gamma}^{\pm}+C_{\gamma^{\ast}}^{\pm}Q_{\gamma^{\ast}}^{\pm}+h.c.\;,\label{HeffG}
\end{equation}
with the magnetic and electric operators defined as%
\begin{equation}
Q_{\gamma}^{\pm}=\frac{Q_{d}e}{16\pi^{2}}\;(\bar{s}_{L}\sigma^{\mu\nu}d_{R}\pm\bar{s}_{R}\sigma^{\mu\nu}d_{L})\,F_{\mu\nu}\;,\;\;Q_{\gamma^{\ast}}^{\pm}=\frac{Q_{d}e}{16\pi^{2}}\;(\bar{s}_{L}\gamma^{\nu}d_{L}\pm\bar{s}_{R}\gamma^{\nu}d_{R})\,\partial^{\mu}F_{\mu\nu}\;,\label{HeffG2}
\end{equation}
and $2\sigma^{\mu\nu}=i[\gamma^{\mu},\gamma^{\nu}]$, $Q_{d}=-1/3$ the down-quark electric charge. For a real photon emission, $\partial^{\mu}F_{\mu\nu}=0$ so only the magnetic operators contribute. The corresponding Wilson coefficients are~\cite{BuchallaBL96}
\begin{equation}
Q_{d}(C_{\gamma}^{+}-C_{\gamma}^{-})=\sqrt{2}G_{F}\lambda_{i}D_{0}^{\prime}\left(  x_{i}\right)  m_{s}\;,\;\;Q_{d}(C_{\gamma}^{+}+C_{\gamma}^{-})=\sqrt{2}G_{F}\lambda_{i}D_{0}^{\prime}\left(  x_{i}\right)  m_{d}\;,\label{SMc}
\end{equation}
and
\begin{equation}
Q_{d}(C_{\gamma^{\ast}}^{+}+C_{\gamma^{\ast}}^{-})=-2\sqrt{2}G_{F}\lambda_{i}D_{0}\left(  x_{i}\right) \;,\;\;Q_{d}(C_{\gamma^{\ast}}^{+}-C_{\gamma^{\ast}}^{-})\approx0\;,\label{SMc2}
\end{equation}
where $i=u,c,t$, $\lambda_{i}=V_{is}^{\ast}V_{id}$ the CKM matrix elements, and $D_{0}^{(\prime)}(x_{i}\equiv m_{i}^{2}/M_{W}^{2})$ the loop functions (see e.g. Ref.~\cite{BuchallaBL96} for their expressions). Summing over the three up-quark flavors, it is their dependences on the quark masses which ensure the necessary GIM breaking, since otherwise CKM unitarity $\lambda_{u}+\lambda_{c}+\lambda_{t}=0$ would force them to vanish. In this respect, $D_{0}^{\prime}(x)$ is suppressed for light quarks, while $D_{0}(x)$ breaks GIM logarithmically both for $x\rightarrow\infty$ and $x\rightarrow0$. However, QCD corrections significantly soften the quadratic GIM breaking of
$D_{0}^{\prime}(x)$ in the $x\rightarrow0$ limit~\cite{D0QCD}, and exacerbate the logarithmic one of $D_{0}(x)$~\cite{DEIP98}, making light-quark contributions significant for both operators.

In the presence of NP, new mechanisms could produce the $s\rightarrow d\gamma$ transition. Since the NP energy scale is presumably above the electroweak scale, these effects would simply enter into the Wilson coefficients of the same effective local operators~(\ref{HeffG}). This is the shift we want to extract phenomenologically. In this respect, the magnetic operators are a priori most sensitive to NP for two reasons. First, the electric transition is essentially left-handed and the magnetic operators are very suppressed in the SM because right-handed external quarks $(s,d)_{R}$ are accompanied by the chiral suppression factor $m_{s,d}$. These strong suppressions may be lifted
in the presence of NP, where larger chirality flip mechanisms can be available. Second, the magnetic operators are formally of dimension five, and thus a priori less suppressed by the NP energy scale than the dimension six electric operators. Sizeable NP effects could thus show up, as will be quantitatively analyzed in Sec.~4.

With the help of the standard QED interactions, the $\mathcal{H}_{eff}^{\gamma}$ operators also contribute to processes with more than one photon, where they compete with the effective operators directly involving several photon fields. For example, for two real photons, the dominant operators are%
\begin{equation}
Q_{\gamma\gamma,||}^{\pm}=(\bar{s}_{L}d_{R}\pm\bar{s}_{R}d_{L})F_{\mu\nu}F^{\mu\nu},\;\;Q_{\gamma\gamma,\perp}^{\pm}=(\bar{s}_{L}d_{R}\pm\bar{s}_{R}d_{L})F_{\mu\nu}\tilde{F}^{\mu\nu}\;,\label{Qgg}
\end{equation}
with $\tilde{F}^{\mu\nu}=\varepsilon^{\mu\nu\rho\sigma}F_{\rho\sigma}/2$. In the SM, the additional quark propagator in the two-photon penguin induces an $x^{-1}$ GIM breaking by the loop function (see Fig.~\ref{Fig1}$b$). Hence, the $c$ and $t$-quark contributions are completely negligible compared to the $u$-quark loop. Further, NP effects in these operators should be very suppressed since they are at least of dimension seven. So, whenever it contributes, the two photon penguin represents an irreducible long-distance SM background for the SD processes. The same is true for transitions with more than two photons, with the NP (up-quark loop) even more suppressed (enhanced),
so those will not be considered here.

\subsection{Long-distance effects}

Once QCD is turned back on and with $m_{u}<m_{s,d}<m_{c,t}$, the $c$ and $t$ contributions remain local, but not the up quark loop. At the $K$ mass scale, the former are, together with possible NP, the short-distance (SD) contributions, and the latter are the SM-dominated long-distance (LD)
contributions. Note that the SD contributions are also affected by long-distance effects, since phenomenologically, the matrix elements of the SD operators between low-energy meson states is needed.

\begin{figure}[t]
\centering          \includegraphics[width=16.0cm]{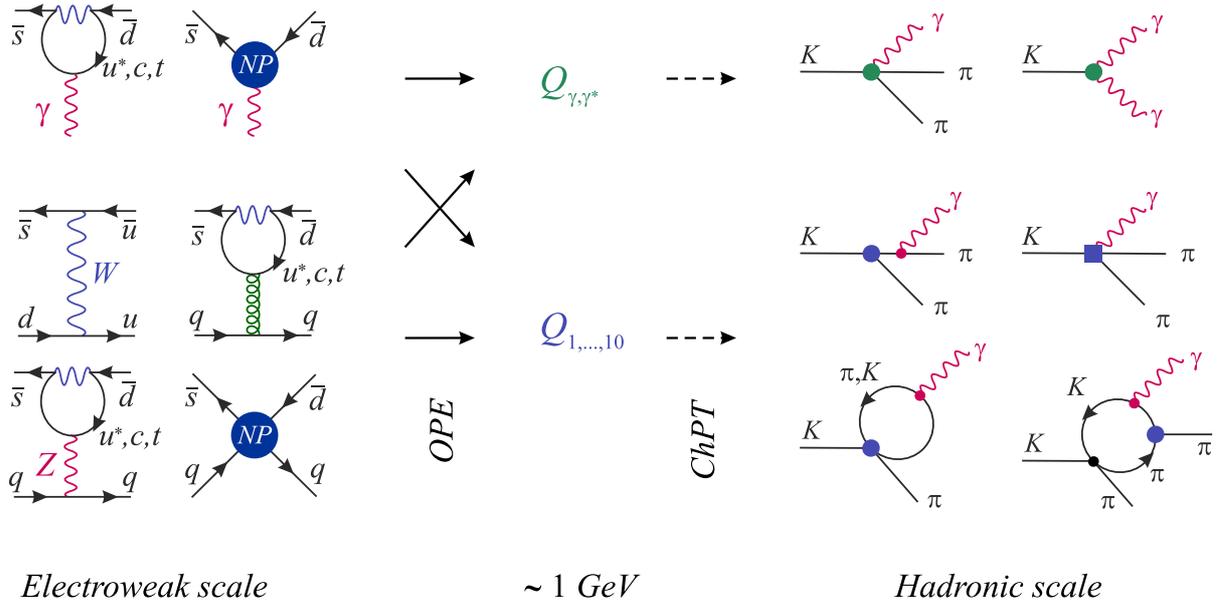}
\caption{Description of the radiative $K$ decays, starting with the
electroweak scale interactions down to chiral perturbation theory, with
illustrative examples of mesonic processes (the photons can be real or
virtual). The green vertices arise from the currents in Eqs.~(\ref{VectorC},~\ref{TensorC}), the blue disks and square from the $\mathcal{O}(p^{2})$ weak Lagrangians Eq.~(\ref{WeakLagr}) and $\mathcal{O}(p^{4})$ weak counterterms Eq.~(\ref{CTp4}), respectively, and finally, the strong (black) and QED (red) vertices from Eq.~(\ref{QCD}).}
\label{Fig2}
\end{figure}

To deal with these LD effects, the first step is to sum up the QCD-corrected interactions among the light quarks into an effective Hamiltonian~\cite{BuchallaBL96}%
\begin{equation}
\mathcal{H}_{eff}(\mu\approx1\text{ GeV})=\sum_{\grave{\imath}=1}^{10}C_{i}\left(  \mu\right)  Q_{i}\left(  \mu\right)  +\mathcal{H}_{eff}^{\gamma}\left(  \mu\right)  +...\;, \label{OPE}
\end{equation}
with the four-quark current-current $(Q_{1,2})$, QCD penguin $(Q_{3,...,6})$, and electroweak penguin $(Q_{7,...,10})$ operators, and $\mathcal{H}_{eff}^{\gamma}\left(  \mu\right)  $ as in Eq.~(\ref{HeffG}). Short-distance physics, including both the SM and NP effects, is encoded into the Wilson coefficients $C_{i}\left(  \mu\right)  $, see Fig.~\ref{Fig2}. The low-virtuality up, down, and strange quarks, i.e. the dynamics going on below the QCD perturbativity frontier $\mu\approx1$ GeV, are dealt through the hadronic matrix elements of the effective operators.

At the hadronic scale, the strong dynamics is represented with chiral perturbation theory (ChPT), the effective theory for QCD with the pseudoscalar mesons as degrees of freedom~\cite{GL}. At $\mathcal{O}(p^{2})$, the strong interaction Lagrangian is
\begin{equation}
\mathcal{L}_{strong}=\dfrac{F^{2}}{4}\langle D^{\mu}UD_{\mu}U^{\dagger}+\chi U^{\dagger}+U\chi^{\dagger}\rangle\;, \label{QCD}
\end{equation}
where $F=F_{\pi}\approx92.4$ MeV, $U$ is a $3\times3$ matrix function of the meson fields, $\chi=2B_{0}\operatorname*{diag}(m_{u},m_{d},m_{s})$ reproduces the explicit chiral symmetry breaking induced by the quark masses, and $\langle...\rangle$ means the flavor trace (we follow the notation of Ref.~\cite{DAmbrosioI96}). The covariant derivative includes external real or virtual photons, $D_{\mu}U=\partial_{\mu}U-ieA_{\mu}[U,Q]$, $Q=\operatorname*{diag}(2/3,-1/3,-1/3)$, as well as static $Z$ or $W$ currents coupled to leptonic states which do not concern us here.

To the strong Lagrangian~(\ref{QCD}), the electroweak operators of $\mathcal{H}_{eff}$ are added as effective interactions among the pseudoscalar mesons. So, the non-local, low-energy tails of the photon penguins of Fig.~\ref{Fig1} are reconstructed using the effective hadronic representations of $Q_{1,...,10}$ to induce the weak transition, and the photon(s) emitted from light charged mesons occurring either as external particles (bremsstrahlung radiation) or inside loops (direct emission radiation), see
Fig.~\ref{Fig2}. Note that the mesonic processes not only represent the $u$ quark loop in Fig.~\ref{Fig1}, but also $d$ and $s$ quark loops since the Fermi interaction is effectively replaced by the whole set of $Q_{1,...,10}$ operators at long-distance. So, let us construct the hadronic representations
of $\mathcal{H}_{eff}$, starting with the electromagnetic operators.

\subsubsection{Electromagnetic operators}

The chiral realization of the $Q_{\gamma^{\ast}}^{\pm}$ operators requires that of the vector and axial-vector quark bilinears. At $\mathcal{O}(p^{2})$, these currents are related by the $SU(3)$ symmetry to the conserved electromagnetic current, and are thus entirely fixed from the
Lagrangian~(\ref{QCD}):%
\begin{equation}
\bar{q}_{L}^{I}\gamma^{\mu}q_{L}^{J}=i\frac{F^{2}}{2}(\partial^{\mu}U^{\dagger}U)^{JI},\;\;\;\;\bar{q}_{R}^{I}\gamma^{\mu}q_{R}^{J}=i\frac{F^{2}}{2}(\partial^{\mu}UU^{\dagger})^{JI}\;. \label{VectorC}
\end{equation}
The $SU(3)$ breaking corrections start at $\mathcal{O}(p^{4})$ and are mild thanks to the Ademollo-Gatto theorem~\cite{AdemolloG64}. They can be precisely estimated from the charged current matrix elements, i.e. from $K_{\ell3}$ decays. See Ref.~\cite{MesciaS07} for a detailed analysis.

The chiral realization of the tensor currents in $Q_{\gamma}^{\pm}$ is more involved and starts at $\mathcal{O}(p^{4})$ since two derivatives are needed to get the correct Lorentz structure. Further, it cannot be entirely fixed but involves specific low-energy constants. By imposing charge conjugation and
parity invariance (valid for QCD), the antisymmetry under $\mu\leftrightarrow \nu$, and the identity $i\varepsilon^{\alpha\beta\mu\nu}\sigma_{\mu\nu}=2\sigma^{\alpha\beta}\gamma_{5}$, only two free real parameters $a_{T}$ and $a_{T}^{\prime}$ remain (parts of these currents were given in Refs.~\cite{Gao02,CDI99})
\begin{subequations}
\label{TensorC}%
\begin{align}
\bar{q}^{I}\sigma_{\mu\nu}P_{L}q^{J}  &  =-i\frac{F^{2}}{2}a_{T}\left(
D_{\mu}U^{\dagger}D_{\nu}UU^{\dagger}-D_{\nu}U^{\dagger}D_{\mu}UU^{\dagger
}-i\varepsilon_{\mu\nu\rho\sigma}D^{\rho}U^{\dagger}D^{\sigma}UU^{\dagger
}\right)  ^{JI}\nonumber\\
&  \;\;\;\;+\frac{F^{2}}{2}a_{T}^{\prime}((F_{\mu\nu}^{L}-i\tilde{F}_{\mu\nu
}^{L})U^{\dagger}+U^{\dagger}(F_{\mu\nu}^{R}-i\tilde{F}_{\mu\nu}^{R}%
))^{JI}\;\;,\\
\bar{q}^{I}\sigma_{\mu\nu}P_{R}q^{J}  &  =-i\frac{F^{2}}{2}a_{T}\left(
D_{\mu}UD_{\nu}U^{\dagger}U-D_{\nu}UD_{\mu}U^{\dagger}U+i\varepsilon_{\mu
\nu\rho\sigma}D^{\rho}UD^{\sigma}U^{\dagger}U\right)  ^{JI}\nonumber\\
&  \;\;\;\;+\frac{F^{2}}{2}a_{T}^{\prime}(U(F_{\mu\nu}^{L}+i\tilde{F}_{\mu\nu
}^{L})+(F_{\mu\nu}^{R}+i\tilde{F}_{\mu\nu}^{R})U)^{JI}\;.
\end{align}
\end{subequations}
Numerically, we will use the lattice estimate~\cite{Mescia00}%
\begin{equation}
B_{T}(2\;\text{GeV})=2m_{K}a_{T}=1.21(12)\;. \label{BT}%
\end{equation}
Being derived from a study of the $\langle\pi|\bar{s}\sigma_{\mu\nu}%
d|K\rangle$ matrix element, $SU(3)$ corrections are under control. A similar
estimate of $B_{T}^{\prime}=2m_{K}a_{T}^{\prime}$ is not available yet.
Instead, we can start from $\langle\gamma|\bar{u}\sigma_{\mu\nu}\gamma
_{5}d|\pi^{-}\rangle$ and invoke the $SU(3)$ symmetry. Ref.~\cite{Mateu07},
through a study of the $VT$ correlator, get $a_{T}^{\prime}=B_{0}/M_{V}^{2}$
and thus $B_{T}^{\prime}=2.7(5)$, assuming the standard ChPT sign conventions
for the matrix elements. Another route is to use the magnetic susceptibility
of the vacuum, $\langle0|\bar{q}\sigma_{\mu\nu}q|0\rangle_{\gamma}$. From the
lattice estimate in Ref.~\cite{MS}, we extract using $a_{T}^{\prime}=-\chi
_{T}B_{0}/2$ the value $B_{T}^{\prime}(2$ GeV$)=2.67(17)$. Both techniques
give similar results though their respective scales do not match. In addition,
sizeable $SU(3)$ breaking effects cannot be ruled out since there is no
Ademollo-Gatto protection for the tensor currents. So, to be conservative, we
shall use%
\begin{equation}
B_{T}^{\prime}(2\;\text{GeV})=2m_{K}a_{T}^{\prime}=3(1)\;. \label{BTP}%
\end{equation}

At $\mathcal{O}(p^{4})$, the magnetic operators contribute to decay modes with
at most two photons. With the chiral suppression expected for higher order
terms, decays with three or more (real or virtual) photons should have a
negligible sensitivity to $Q_{\gamma}^{\pm}$, hence are not included in our study.

In the SM, since the local operators sum up the short-distance part of the
real photon penguins, the factor $m_{s,d}\sim\mathcal{O}(p^{2})$ in
Eq.~(\ref{SMc}) are not included in the bosonization. Instead, they are kept
as perturbative parameters in the Wilson coefficients $C_{\gamma}^{\pm}$, to
be evaluated at the same scale as the form factors $B_{T}$ and $B_{T}^{\prime
}$. Numerically, to account for the large QCD corrections, the Wilson
coefficient of the magnetic operator in $b\rightarrow s\gamma$ can be used for
$\operatorname{Im}C_{\gamma}^{\pm}$, since the CKM elements for the $u$, $c$,
and $t$ contributions scale similarly. With $m_{s}(2$ GeV$)=101_{-21}^{+29}$
MeV~\cite{PDG} and $C_{7\gamma}(2$\thinspace GeV$)\approx-0.36$ from
Ref.~\cite{BuchallaBL96}, we shall use\footnote{For convenience, the same
normalization by $G_{F}m_{K}$ will be adopted throughout the paper. Also, if
not explicitly written, the $C_{\gamma}^{\pm}$ are always understood at the
$\mu=2$ GeV scale.}%
\begin{equation}
\frac{\operatorname{Im}C_{\gamma}^{\pm}(2\;\text{GeV})_{\text{SM}}}{G_{F}%
m_{K}}=\mp\sqrt{2}\frac{C_{7\gamma}(2\,\text{GeV)}}{Q_{d}}\frac{m_{s}%
(2\;\text{GeV})}{m_{K}}\operatorname{Im}\lambda_{t}=\mp0.31(8)\times
\operatorname{Im}\lambda_{t}\;, \label{SM}%
\end{equation}
to be compared to $\mp0.17\operatorname{Im}\lambda_{t}$ with only the top
quark. In view of the large error on $m_{s}$, the LO approximation is
adequate. For $\operatorname{Re}C_{\gamma}^{\pm}$, contrary to the situation
in $b\rightarrow s\gamma$, the top quark is strongly suppressed as
$\operatorname{Re}\lambda_{c}\approx-\operatorname{Re}\lambda_{u}%
\gg\operatorname{Re}\lambda_{t}$. With the light quarks further enhanced by
QCD corrections, an estimate is delicate. Naively rescaling the above result
gives%
\begin{equation}
\frac{\operatorname{Re}C_{\gamma}^{\pm}(2\;\text{GeV})_{\text{SM}}}{G_{F}%
m_{K}}\approx\frac{\operatorname{Re}\lambda_{c}}{\operatorname{Im}\lambda_{c}%
}\times\frac{\operatorname{Im}C_{\gamma}^{\pm}(2\;\text{GeV})_{\text{SM}}%
}{G_{F}m_{K}}\approx\mp0.06\;. \label{ReCSM}%
\end{equation}
Evidently, one should not take this as more than a rough estimate of the order
of magnitude of the $c$ quark and high-virtuality $u$ quark contributions. In
any case, we will be mostly concern by CP-violating observables in the
following, so will not use Eq.~(\ref{ReCSM}).

\subsubsection{Four-quark weak operators}

By matching their chiral structures, the four-quark weak current-current and
penguin operators are represented at $\mathcal{O}(p^{2})$ as~\cite{WeakL}
\begin{subequations}
\label{WeakLagr}%
\begin{align}
\mathcal{L}_{8} &  =F^{4}G_{8}\langle\lambda_{6}L_{\mu}L^{\mu}\rangle\;,\;\;\\
\mathcal{L}_{27} &  =\frac{F^{4}}{18}G_{27}^{1/2}\left(  \langle\lambda
_{1}L_{\mu}\rangle\langle\lambda_{4}L^{\mu}\rangle+\langle\lambda_{2}L_{\mu
}\rangle\langle\lambda_{5}L^{\mu}\rangle-10\langle\lambda_{6}L_{\mu}%
\rangle\langle\lambda_{3}L^{\mu}\rangle+18\langle\lambda_{6}L_{\mu}%
\rangle\langle QL^{\mu}\rangle\right)  \nonumber\\
&  \;\;\;\;+\frac{5F^{4}}{18}G_{27}^{3/2}\left(  \langle\lambda_{1}L_{\mu
}\rangle\langle\lambda_{4}L^{\mu}\rangle+\langle\lambda_{2}L_{\mu}%
\rangle\langle\lambda_{5}L^{\mu}\rangle+2\langle\lambda_{6}L_{\mu}%
\rangle\langle\lambda_{3}L^{\mu}\rangle\right)  \;,\\
\mathcal{L}_{ew} &  =F^{6}e^{2}G_{ew}\langle\lambda_{6}U^{\dagger}QU\rangle\;,
\end{align}
\end{subequations}
where $L^{\mu}\equiv U^{\dagger}D^{\mu}U$, $\lambda_{i}$ are the Gell-Mann
matrices, and $G_{27}\equiv G_{27}^{3/2}=G_{27}^{1/2}$ in the isospin limit.
If QCD was perturbative down to the hadronic scale, the low-energy constants
could be computed from the Wilson coefficients at that scale as
\begin{equation}
\{C_{1}-C_{2},C_{3-6},C_{9},C_{10}\}\rightarrow G_{8}\;\;,\;\;\;\{C_{1}%
+C_{2},C_{9},C_{10}\}\rightarrow G_{27}\;\;,\;\;\;\{C_{7},C_{8}\}\rightarrow
G_{ew}\;.\label{LECs}%
\end{equation}
The ChPT scale is too low for this to be possible however. Instead, the
low-energy constants are fixed from experiment, especially from $K\rightarrow
\pi\pi$. The consequence is that neither the $\Delta I=1/2$ rule, embodied in
their real parts as $\operatorname{Re}G_{27}/\operatorname{Re}G_{8}%
\equiv\omega=1/22.4$, nor the direct CP-violation parameters like
$\varepsilon^{\prime}$ generated from their imaginary parts, can be precisely
computed from first principles.

At tree level, if $\mathcal{L}_{8}$, $\mathcal{L}_{27}$, or $\mathcal{L}_{ew}$ contribute to a radiative decay, it is only through bremsstrahlung amplitudes~\cite{Kpppg,EckerPR88,Rad2}. The dynamics is therefore trivial at $\mathcal{O}(p^{2})$ because Low's theorem~\cite{Low} shows that such emissions are entirely fixed in terms of the non-radiative $K\rightarrow 2\pi,3\pi$ amplitudes. Thus, the non-trivial dynamics corresponding to the low-energy tails of the photon penguins arise at $\mathcal{O}(p^{4})$, where they are represented in terms of non-local meson loops, as well as additional $\mathcal{O}(p^{4})$ local effective interactions, in particular the $\Delta I=1/2$ enhanced $N_{14},...,N_{18}$ octet counterterms~\cite{CT1,CT2}:%
\begin{equation}
\mathcal{L}_{8}^{\text{CT}}=-i\langle\lambda_{6}(N_{14}\{f_{+}^{\mu\nu}%
,L_{\mu}L_{\nu}\}+N_{15}L_{\mu}f_{+}^{\mu\nu}L_{\nu}+N_{16}\{f_{-}^{\mu\nu
},L_{\mu}L_{\nu}\}+N_{17}L_{\mu}f_{-}^{\mu\nu}L_{\nu}+iN_{18}(f_{+\mu\nu}%
^{2}-f_{-\mu\nu}^{2}))\rangle\;, \label{CTp4}%
\end{equation}
with $f_{\pm}^{\mu\nu}\equiv F_{L}^{\mu\nu}\pm U^{\dagger}F_{R}^{\mu\nu}U$,
and $F_{L}^{\mu\nu}=F_{R}^{\mu\nu}=-eQF^{\mu\nu}$ for external photons. There
are also counterterms relevant for the renormalization of the non-radiative
$K\rightarrow n\pi$ amplitudes occurring in the bremsstrahlung contributions,
for the strong structure of the $\pi^{+}\pi^{-}\gamma^{\ast}$ or $K^{+}%
K^{-}\gamma^{\ast}$ vertices, and for the odd-parity sector (proportional to
$\varepsilon$ tensors) which will not concern us here. Note that the need to
compute the $Q_{1,...,10}$ contributions at $\mathcal{O}(p^{4})$ also follows
from the chiral representation~(\ref{TensorC}) of the magnetic operators
starting at that order.

The structure of the effective interactions~(\ref{CTp4}) is dictated by the chiral counting rules and the chiral symmetry properties of the underlying weak operators, but the (renormalized) $N_{i}$ constants cannot be computed from first principles and have to be fixed experimentally, exactly like the $\mathcal{O}(p^{2})$ constants $G_{8,27,ew}$ of Eq.~(\ref{WeakLagr}).

\subsubsection{The hadronic tails of the photon penguins}

The set of interactions included within ChPT is complete, in the sense that all the possible effective interactions with the required symmetries are present at a given order. So, it may appear that at $O(p^{4})$, once the weak interactions~(\ref{WeakLagr}) are added to the strong dynamics~(\ref{QCD}),
and including the counterterms~(\ref{CTp4}), there is no more need to separately include the SD electromagnetic operators through Eq.~(\ref{VectorC}) and~(\ref{TensorC}). All their effects would be accounted for in the values of the low-energy constants. Indeed, these constants should sum up the physics taking place above the mesonic scale, i.e. the hadronic degrees of freedom just above the octet of pseudoscalar mesons~\cite{CT2,Resons} as well as the quark and gluon degrees of freedom above the GeV scale~\cite{QG}.

This actually holds for $Q_{\gamma^{\ast}}^{\pm}$, but not for $Q_{\gamma}^{\pm}$. Indeed, only the former have the same chiral structures as the $N_{i}$ counterterms. Whenever $Q_{\gamma^{\ast}}^{\pm}$ contribute, so do the $N_{i}$, but $Q_{\gamma}^{\pm}$ can contribute to many modes where the $N_{i}$
are absent (see Table~\ref{TableWindows} in the next section) and must therefore appear explicitly in the effective theory. Including the $\Delta I=3/2$ suppressed $\mathcal{L}_{27}^{\text{CT}}$~\cite{CT1,Esposito91} or the $e^{2}$-suppressed $\mathcal{L}_{ew}^{\text{CT}}$~\cite{CTew} counterterms
would not change this picture, so for simplicity we consider only $\mathcal{L}_{8}^{\text{CT}}$.

This mismatch between $\mathcal{L}_{8}^{\text{CT}}$ and $Q_{\gamma}^{\pm}$ has an important dynamical implication since the weak counterterms reflect the chiral structures of the meson loops built on the $Q_{1,...,10}$ operators~(\ref{WeakLagr}) at $\mathcal{O}(p^{4})$. While these meson loops can genuinely represent the low-energy tail of the virtual photon penguin, i.e. the $\log(x_{u})$ singularity of the $D_{0}(x)$ function, they never match the chiral representation of $Q_{\gamma}^{\pm}$. The meson dynamics
lacks the required $m_{s,d}$ chirality flip at $\mathcal{O}(p^{4})$, relying instead on the long-distance dynamics, i.e. momenta. One can understand this phenomenon as the low-energy equivalent of the known importance of the $Q_{2}^{c} = (\bar{s}c)_{V-A} \otimes (\bar{c} b)_{V-A}$ contribution to $b\rightarrow s\gamma$~\cite{D0QCD}. Clearly, $s\rightarrow d\gamma$ has to be even more affected than $b\rightarrow
s\gamma$ by QCD corrections since the photon is never hard ($q_{\gamma}^{2}<m_{K}^{2}$), and an inclusive analysis is not possible. So for $s\rightarrow d\gamma$, the $Q_{2}^{u}= (\bar{s}u)_{V-A} \otimes (\bar{u} d)_{V-A}$ contribution, represented through $Q_{1,...,10}$, corresponds to a whole class of purely long-distance processes, often including IR divergent bremsstrahlung radiations. They are not suppressed at all, contrary to the naive expectation from $D_{0}^{\prime}(x)\rightarrow x$ as $x\rightarrow0$, but instead dominate most of the radiative processes\footnote{By comparison, though the Inami-Lim function
$C_{0}(x)$ for the $Z$ penguin scale like $D_{0}^{\prime}(x)$ in the $x\rightarrow0$ limit, this behavior survives to QCD corrections, and the light-quark contributions are very suppressed, see Ref.~\cite{IsidoriMS05}.}.

With this in mind, we can understand at least qualitatively another striking feature of all the radiative modes where $Q_{\gamma^{\ast}}^{\pm}$ is absent. The meson loops are always finite at $\mathcal{O}(p^{4})$, except for $K_{1}\rightarrow\pi^{+}\pi^{-}\pi^{0}\gamma(\gamma)$~\cite{Kpppg}. This means
that not only the SD part of the magnetic operators decouples, but also to some extent the intermediate QCD degrees of freedom (i.e., the resonances\footnote{Though the counterterms are also scale-independent in the odd-parity sector, driven by the QED anomaly, the resonances are known to be important there~\cite{Anoms}. We will be mostly concerned by the even-parity sector here.}). By contrast, the $N_{i}$ combinations occurring for the modes induced by $Q_{\gamma^{\ast}}^{\pm}$ are always scale dependent, somewhat reminiscent of the factorization of the low-energy part of the virtual photon penguin. So, the behavior of the flavor-changing electromagnetic current is not very different from that of the flavor-conserving one. In that case, being protected by the QED gauge symmetry, the form-factor for $\langle\gamma(q)|\pi^{+}\pi^{-}\rangle$ or $\langle\gamma(q)|K^{+}K^{-}\rangle$ is not renormalized at all at $q^{2}=0$, while vector resonances saturate the off-shell behavior~\cite{CT2,Resons}.

From these observations, we can reasonably expect that whenever a finite
combination of $N_{i}$ occurs for a process with only real photons, it should
be significantly suppressed. Indeed, not only the divergences cancel among the
$N_{i}$, but also the large $Q_{\gamma^{\ast}}^{\pm}$ contribution embedded
into them (this was already noted using large $N_{c}$ arguments in
Ref.~\cite{BrunoP92}), as well as the resonance effects describing the purely
strong structure of the photon. As our analysis of $K^{+}\rightarrow\pi^{+}%
\pi^{0}\gamma$ in Sec. 3 will show, this suppression is supported by the
recent experimental data, see Eq.~(\ref{CT}).

\subsection{Phenomenological windows}

The $K$ decay channels where the electromagnetic operators contribute are
listed in Table~\ref{TableWindows}, together with their CP signatures. For the
electric operators, at least one of the photons needs to be virtual, i.e.
coupled to a Dalitz pair $\ell^{+}\ell^{-}$. In this respect, remark that all
the electromagnetic operators produce the $\ell^{+}\ell^{-}$ pair in the same
$1^{--}$ state, so the electric and magnetic operators can only be
disentangled using real photon decays.

\begin{table}[t] \centering \begin{tabular}[c]{llccccccc}\hline
&  & $\perp$ & $||$ &  &  & $M$ & $E$ & $L$\\\hline
$K_{2}\rightarrow\gamma\gamma$ & \multicolumn{1}{c}{$a_{T}^{\prime}$} &
$\operatorname{Re}C_{\gamma}^{+}$ & $\operatorname{Im}C_{\gamma}^{-}$ & -- &
&  &  & \\
$K_{2}\rightarrow\pi^{0}\gamma\gamma$ & \multicolumn{1}{c}{$a_{T}^{\prime}$} &
$\operatorname{Im}C_{\gamma}^{-}$ & $\operatorname{Re}C_{\gamma}^{+}$ &
$K_{2}\rightarrow\pi^{0}\gamma$ & $a_{T}$ & -- & -- & $\operatorname{Im}%
C_{\gamma^{(\ast)}}^{+}$\\
$K^{+}\rightarrow\pi^{+}\gamma\gamma$ & \multicolumn{1}{c}{$3a_{T}%
+a_{T}^{\prime}$} & $C_{\gamma}^{-}$ & $C_{\gamma}^{+}$ & $K^{+}\rightarrow
\pi^{+}\gamma$ & $a_{T}$ & -- & -- & $C_{\gamma^{(\ast)}}^{+}$\\
$K_{2}\rightarrow\pi^{0}\pi^{0}\gamma\gamma$ & \multicolumn{1}{c}{$a_{T}%
^{\prime}$} & $\operatorname{Re}C_{\gamma}^{+}$ & $\operatorname{Im}C_{\gamma
}^{-}$ & $K_{2}\rightarrow\pi^{0}\pi^{0}\gamma$ & $a_{T}$ & -- & -- &
$\operatorname{Re}C_{\gamma^{(\ast)}}^{-}$\\
$K_{2}\rightarrow\pi^{+}\pi^{-}\gamma\gamma$ & \multicolumn{1}{c}{$a_{T}%
,a_{T}^{\prime}$} & $\operatorname{Re}C_{\gamma}^{+}$ & $\operatorname{Im}%
C_{\gamma}^{-}$ & $K_{2}\rightarrow\pi^{+}\pi^{-}\gamma$ & $a_{T}$ &
$\operatorname{Re}C_{\gamma}^{+}$ & $\operatorname{Im}C_{\gamma}^{-}$ &
$\operatorname{Re}C_{\gamma^{(\ast)}}^{-}$\\
$K^{+}\rightarrow\pi^{+}\pi^{0}\gamma\gamma$ & \multicolumn{1}{c}{$a_{T}%
,a_{T}^{\prime}$} & $C_{\gamma}^{+}$ & $C_{\gamma}^{-}$ & $K^{+}\rightarrow
\pi^{+}\pi^{0}\gamma$ & $a_{T}$ & $C_{\gamma}^{+}$ & $C_{\gamma}^{-}$ &
$C_{\gamma^{(\ast)}}^{-}$\\
$K_{2}\rightarrow3\pi^{0}\gamma\gamma$ & \multicolumn{1}{c}{$a_{T}^{\prime}$}
& $\operatorname{Im}C_{\gamma}^{-}$ & $\operatorname{Re}C_{\gamma}^{+}$ &
$K_{2}\rightarrow3\pi^{0}\gamma$ & $a_{T}$ & -- & -- & $\operatorname{Im}%
C_{\gamma^{(\ast)}}^{+}$\\\hline
\end{tabular}
\caption
{Dominant processes where the electromagnetic operators contribute, omitting the $K\rightarrow(n\pi)\gamma^{\ast}\gamma^{(\ast)}$, $n\geq0$ decays. The $K_1\approx K_{S}$ processes are obtained from $K_2\approx K_{L}$ by inverting real and imaginary parts. The symbol $\perp$ ($||$) means the photon pair in an odd (even) parity state, i.e. a $F_{\mu\nu}\tilde{F}^{\mu\nu}$ ($F_{\mu\nu}F^{\mu\nu}$) coupling, and
similarly, $M$ ($E$) means odd (even) parity magnetic (electric) emissions. For $\pi\pi$ modes, the lowest multipole is understood (i.e., $\pi\pi$ in a $S$ wave for $\gamma\gamma$ modes, and a $P$ wave for $\gamma$ modes). The last column denotes longitudinal off-shell photon emissions, proportional to
$q^{2}g^{\alpha\beta}-q^{\alpha}q^{\beta}$ with $q$ the photon momentum, for which the $Q_{\gamma^{\ast}}^{\pm}$ operators also enters. The $K\rightarrow3\pi\gamma(\gamma)$ decays with charged pions are not included since dominated by bremsstrahlung radiations off $K\rightarrow3\pi$~\protect
\cite{Kpppg}. Finally, $a_{T}$ and $a_{T}^{\prime}$ are the low-energy constants entering the tensor current~(\ref{TensorC}).}
\label{TableWindows}
\end{table}

For most of the decays in Table~\ref{TableWindows}, the LD contributions are
dominant, obscuring the SD parts where NP could be evidenced. The situation is
thus very different than in $b\rightarrow s\gamma$, where the $u$ quark
contribution is suppressed by $V_{ub}\ll1$. However, in $K$ physics, the
long-distance contributions are essentially CP-conserving. Indeed,
CP-violation from the four-quark operators is known to be small from
$\operatorname{Re}(\varepsilon^{\prime}/\varepsilon)^{\exp}$. In the SM, this
follows from the CKM scalings $\operatorname{Re}\lambda_{u}\gg
\operatorname{Re}\lambda_{t}\sim\operatorname{Im}\lambda_{t}$ and
$\operatorname{Im}\lambda_{u}=0$. So, for CP-violating observables, one
recovers a situation reminiscent of $b\rightarrow s\gamma$, with the dominant
SM contributions arising from the charm and top quarks, both of similar size a
priori. Only for such observables can we hope that the interesting
short-distance physics in $Q_{\gamma}^{\pm}$ and $Q_{\gamma^{\ast}}^{\pm}$
emerges from the long-distance SM background.

All the decays in Table~\ref{TableWindows} have a CP-conserving contribution,
and thus in most cases the best available CP-violating observables are
CP-asymmetries. Since they arise from CP-odd interferences between the various
decay mechanisms, the dominant CP-conserving processes must be under
sufficiently good theoretical control. In addition, these CP-asymmetries being
usually small, the decay rates should be sufficiently large, and not
completely dominated by bremsstrahlung radiations. Indeed, even though these
radiations are under excellent theoretical control thanks to Low's
theorem~\cite{Low}, they would render the short-distance physics too difficult
to access experimentally.

Imposing these conditions on the modes in Table~\ref{TableWindows}, the best
windows for the electromagnetic operators are:

\begin{itemize}
\item \textit{Real photons: }Since the branching ratios decrease as the number
of pions increases, the best candidates to constrain $Q_{\gamma}^{\pm}$ are
the $K_{L,S}\rightarrow\gamma\gamma$ decays for two real photons and the
$K\rightarrow\pi\pi\gamma$ decays for a single real photon. All the other
modes with real photons are either significantly more suppressed (see e.g.
Ref.~\cite{EckerPR88,Gao02} for a study of $K\rightarrow\pi\gamma\gamma$), or
dominated by bremsstrahlung contributions. By contrast, these
radiations are suppressed for $K_{L}\rightarrow\pi^{+}\pi^{-}\gamma$ since
$K_{L}\rightarrow\pi^{+}\pi^{-}$ is CP-violating, and for $K^{+}\rightarrow
\pi^{+}\pi^{0}\gamma$ thanks to the $\Delta I=1/2$ rule. The relevant
CP-violating asymmetries are those either between $K_{L}-K_{S}$ decay
amplitudes, between $K^{+}-K^{-}$ differential decay rates, or in some
phase-space variables. This latter possibility usually requires some
additional information on the photon polarization, accessible e.g. through
Dalitz pairs. But besides the significant suppression of the total rates, this
brings in the electric operators, making the analysis much more involved, so
these observables will not be considered here (see e.g. Ref.~\cite{Kppll}).

\item \textit{Virtual photons: }The best candidates to probe the electric
operators are the $K_{L}\rightarrow\pi^{0}\ell^{+}\ell^{-}$ ($\ell=e,\mu$)
decays, for which $K_{L}\rightarrow\pi^{0}\gamma^{\ast}[\rightarrow\ell
^{+}\ell^{-}]$ is CP-violating hence free of the up-quark contribution (see
e.g. Ref.~\cite{MST06}). As detailed in Sec.~\ref{semilept} (see
Fig.~\ref{Fig7}), there are nevertheless an indirect CP-violating piece from
the small $\varepsilon K_{2}$ component of the $K_{L}$ as well as a
CP-conserving contribution from the four-quark operators with two intermediate
photons, but these are suppressed and under control~\cite{BDI03,ISU04}. The
direct CP-asymmetry in $K^{\pm}\rightarrow\pi^{\pm}\ell^{+}\ell^{-}$ is not
competitive because of its small $\sim10^{-9}$ branching ratio, and of the
hadronic uncertainties in the long-distance contributions~\cite{DEIP98,Gao03}%
.\newline With $K_{L}\rightarrow\pi^{0}\ell^{+}\ell^{-}$ sensitive to
$Q_{\gamma^{\ast}}^{+}$, information on $Q_{\gamma^{\ast}}^{-}$ would also be
needed to disentangle the left and right-handed currents. But since
$\langle\gamma|Q_{\gamma^{\ast}}^{-}|K^{0}(q)\rangle\sim q^{\nu}q^{\mu}%
F_{\mu\nu}=0$, and with $K\rightarrow\pi\gamma^{\ast}\gamma$ sensitive again
to $Q_{\gamma^{\ast}}^{+}$, the simplest observables are the $K\rightarrow
\pi\pi\gamma^{\ast}$ and $K\rightarrow\pi\pi\gamma^{\ast}\gamma^{(\ast)}$
modes, which are suppressed and dominated by LD contributions. For the time
being, we will thus concentrate only on $Q_{\gamma^{\ast}}^{+}$.
\end{itemize}

In summary, the best windows to probe for the electromagnetic operators are
the CP-asymmetries in the $K_{L,S}\rightarrow\gamma\gamma$, $K_{L,S}%
\rightarrow\pi^{+}\pi^{-}\gamma$, and $K^{+}\rightarrow\pi^{+}\pi^{0}\gamma$
decays, and the $K_{L}\rightarrow\pi^{0}\ell^{+}\ell^{-}$ decay rates. For
completeness, it should be mentioned that the magnetic operators also
contributes to radiative hyperon decays~\cite{Singer96} or to the
$B_{s}\rightarrow B_{d}^{\ast}\gamma$ transition~\cite{BBg}, which will not be
analyzed here.

\section{Standard Model predictions}

In order to get clear signals of NP, the SM contributions have to be under good theoretical control. We rely on the available OPE analyses for the Wilson coefficients in the SM~\cite{BuchallaBL96}, and concentrate on the remaining long-distance parts of these contributions. For CP-violating observables, they originate either indirectly from the hadronic penguins $Q_{3}\rightarrow Q_{10}$ or directly from the magnetic operators $Q_{\gamma}^{\pm}$. Since the former indirect contributions are suppressed, while the $C_{\gamma}^{\pm}$ are very small in the SM, both often end up being comparable. These LD
contributions have to be estimated in ChPT. This is rather immediate for $Q_{\gamma}^{\pm}$ given the hadronic representations~(\ref{TensorC}), but significantly more involved for the hadronic penguins, requiring a detailed analysis of the meson dynamics relevant for each process. In addition, some
free low-energy constants necessarily enter, which have to be fixed from other observables.

Thus, the goal of this section is threefold. First, the observables relevant for the study of $Q_{\gamma}^{\pm}$ are presented. This includes the $K\rightarrow\pi\pi\gamma$ rate and CP-asymmetries, the $K_{L,S}\rightarrow\gamma\gamma$ direct CP-violation parameters, the rare semileptonic decays $K\rightarrow\pi\ell^{+}\ell^{-}$, and finally, the hadronic parameter $\varepsilon^{\prime}$. Second, the hadronic penguin contributions to the radiative decay observables are brought under control by relating them to well-measured parameters like $\varepsilon^{\prime}$. In doing this, special care is paid on the possible impacts of NP in $Q_{3}\rightarrow Q_{10}$, which have to be separately parametrized. This is crucial to confidently extract the contributions from $Q_{\gamma}^{\pm}$, where NP could also be present. This constitutes the third goal of the section: To establish the master formulas for all the observables relevant in the study of $Q_{\gamma}^{\pm}$, which will form the basis of the NP analysis of the next section.

\subsection{$K\rightarrow\pi\pi\gamma$}

From Lorentz and gauge invariance, the general decomposition of the $K\left(P\right)  \rightarrow\pi_{1}\left(  K_{1}\right)  \pi_{2}\left(  K_{2}\right)\gamma\left(  q\right) $ amplitude is~\cite{EPRKppg,DambrosioMS93,DAmbrosioI94}
\begin{equation}
\mathcal{M}\left(  K\rightarrow\pi_{1}\pi_{2}\gamma\right)  =\left[  E\left(
z_{i}\right)  \frac{K_{2}^{\mu}K_{1}\cdot q-K_{1}^{\mu}K_{2}\cdot q}{m_{K}%
^{3}}+M\left(  z_{i}\right)  \frac{i\varepsilon^{\mu\nu\rho\sigma}K_{1,\nu
}K_{2,\rho}q_{\sigma}}{m_{K}^{3}}\right]  \varepsilon_{\mu}^{\ast
}(q)\;.\label{GenAmpli}%
\end{equation}
The reduced kinematical variables $z_{1,2}=K_{1,2}\cdot q/m_{K}^{2}$ are
related to the energies of the two pions which we identify as $\pi_{1}\pi
_{2}=\pi^{+}\pi^{-}$, $\pi^{0}\pi^{0}$, or $\pi^{+}\pi^{0}$, and $z_{3}%
=z_{1}+z_{2}=E_{\gamma}/m_{K}$ is the photon energy in the $K$ rest-frame.

The two terms $E\left(  z_{i}\right)  $ and $M\left(  z_{i}\right)  $ are
respectively the (dimensionless) electric and magnetic
amplitudes~\cite{Good59}, and do not interfere in the rate once summed over
the photon polarizations. The electric part can be further split into a
bremsstrahlung and a direct emission term:
\begin{equation}
E(z_{1},z_{2})=E_{IB}(z_{1},z_{2})+E_{DE}(z_{1},z_{2})\;,
\end{equation}
while the magnetic part is a pure direct emission, $M\equiv M_{DE}$. When the
photon energy goes to zero, only $E_{IB}$ is divergent and, according to Low's
theorem~\cite{Low}, entirely fixed from the non-radiative process
$K\rightarrow\pi_{1}\pi_{2}$.

The direct emission terms $E_{DE}$ and $M_{DE}$ are constant in that limit. In
addition, they can be expanded in multipoles, according to the angular
momentum of the two pions~\cite{Christ67}:
\begin{equation}
E_{DE}(z_{1},z_{2})e^{i\delta_{DE}}=E_{1}(z_{3})e^{i\delta_{1}}+E_{2}%
(z_{3})e^{i\delta_{2}}(z_{1}-z_{2})+E_{3}(z_{3})e^{i\delta_{3}}(z_{1}%
-z_{2})^{2}+...\;,
\end{equation}
and similarly for $M_{DE}$. There are several interesting features in this
expansion~\cite{DAmbrosioI96}: (1) for $K^{0}$ decays, the odd and even
multipoles produce the $\pi\pi$ pair in opposite CP states (2) when
CP-conserving, the dipole emission $E_{1}$ dominates over higher multipoles
which have to overcome the angular momentum barrier ($|z_{1}-z_{2}|<0.2$), (3)
the strong phases can be assigned consistently to each multipole since it
produces the $\pi\pi$ state in a given angular momentum state, (4) the
magnetic operators $Q_{\gamma}^{-(+)}$ contributes to the electric (magnetic)
dipole emission amplitudes when $\pi_{1}\pi_{2}=\pi^{+}\pi^{0}$ or $\pi^{+}%
\pi^{-}$, and (5) the $E_{IB}$ and $E_{DE}$ amplitudes interfere and have
different weak and strong phases, hence generate a CP-asymmetry for both the
neutral $K^{0}\rightarrow\pi^{+}\pi^{-}\gamma$ and charged $K^{+}%
\rightarrow\pi^{+}\pi^{0}\gamma$ modes. That is how we plan to extract the
$Q_{\gamma}^{-}$ contribution, so let us analyze each decay in turn.

\subsubsection{$K^{+}\rightarrow\pi^{+}\pi^{0}\gamma$}

For the $K^{+}\rightarrow\pi^{+}\pi^{0}\gamma$ decay, instead of $z_{1,2}$,
the standard phase-space variables are chosen as the $\pi^{+}$ kinetic energy
$T_{c}^{\ast}$ and $W^{2}\equiv(q\cdot P)(q\cdot K_{1})/m_{\pi^{+}}^{2}%
m_{K}^{2}$~\cite{Christ67}. Indeed, pulling out the bremsstrahlung
contribution, the differential rate can be written%
\begin{equation}
\frac{\partial^{2}\Gamma}{\partial T_{c}^{\ast}\partial W^{2}}=\frac{\partial
^{2}\Gamma_{IB}}{\partial T_{c}^{\ast}\partial W^{2}}\left(  1-2\frac{m_{\pi
^{+}}^{2}}{m_{K}}\operatorname{Re}\left(  \frac{E_{DE}}{eA_{IB}}\right)
W^{2}+\frac{m_{\pi^{+}}^{4}}{m_{K}^{2}}\left(  \left|  \frac{E_{DE}}{eA_{IB}%
}\right|  ^{2}+\left|  \frac{M_{DE}}{eA_{IB}}\right|  ^{2}\right)
W^{4}\right)  \;, \label{DiffRate}%
\end{equation}
where $A_{IB}=A\left(  K^{+}\rightarrow\pi^{+}\pi^{0}\right)  $ is constant
but both $E_{DE}$ and $M_{DE}$ are functions of $W^{2}$ and $T_{c}^{\ast}$.
The main interest of $K^{+}\rightarrow\pi^{+}\pi^{0}\gamma$ is clearly
apparent: $A_{IB}$ is pure $\Delta I=3/2$ hence suppressed, making the direct
emission amplitudes easier to access. Note that the strong phase of $A_{IB}$
is that of the $\pi\pi$ rescattering in the $I=2$, $L=0$ state, as confirmed
by a full $\mathcal{O}(p^{4})$ computation. This is not trivial a priori since
both Watson's and Low's theorem deal with asymptotic states. Actually, Low's
theorem takes place after Watson's theorem, in agreement with the naive
expectation from the relative strength of QED and strong interactions.

\paragraph{Total and differential rates:\newline \newline }

Given its smallness, we can assume the absence of CP-violation when discussing
these observables. Experimentally, the electric and magnetic amplitudes (taken
as constant) have been fitted in the range $T_{c}^{\ast}\leq80$ MeV and
$0.2<W<0.9$ by NA48/2~\cite{ExpKppg}. Using their parametrization,
\begin{subequations}
\label{DEexp}%
\begin{align}
X_{E}  &  =\frac{-\operatorname{Re}\left(  E_{DE}/eA_{IB}\right)  }{m_{K}%
^{3}\cos(\delta_{1}^{1}-\delta_{0}^{2})}=\left(  -24\pm4\pm4\right)
\;\text{GeV}^{-4}\;,\;\\
X_{M}  &  =\frac{\left|  M_{DE}/eA_{IB}\right|  }{m_{K}^{3}}=\left(
254\pm6\pm6\right)  \;\text{GeV}^{-4}\;,
\end{align}
\end{subequations}
with $\delta_{J}^{I}$ the strong $\pi\pi$ rescattering phase in the isospin
$I$ and angular momentum $J$ state. The magnetic amplitude is dominated by the
QED anomaly and will not concern us here (see e.g.
Refs.~\cite{Anoms,CappielloD07}). For the electric amplitude, we obtain at
$\mathcal{O}(p^{4})$:%
\begin{equation}
X_{E}=\frac{3G_{8}/G_{27}}{40\pi^{2}F_{\pi}^{2}m_{K}^{2}}\frac{\cos
(\delta_{DE}-\delta_{0}^{2})}{\cos(\delta_{1}^{1}-\delta_{0}^{2})}\left[
E^{loop}(W^{2},T_{c}^{\ast})-\dfrac{m_{K}^{2}\operatorname{Re}\bar{N}}%
{m_{K}^{2}-m_{\pi}^{2}}\right]  \;, \label{Eloop}%
\end{equation}
with the expression of $E^{loop}$ given in Appendix~\ref{AppA}. The $\bar{N}$
term contains both the $\mathcal{L}_{8}^{\text{CT}}$
counterterms~\cite{EPRKppg} and the $Q_{\gamma}^{-}$ contributions%
\begin{equation}
\operatorname{Re}\bar{N}\equiv(4\pi)^{2}\operatorname{Re}(N_{14}-N_{15}%
-N_{16}-N_{17})-\frac{2G_{F}}{3G_{8}}B_{T}\frac{\operatorname{Re}C_{\gamma
}^{-}}{G_{F}m_{K}}\;,
\end{equation}
when $27$-plet counterterms are neglected (or rather parametrically included
into the $N_{i}$, together with higher order momentum-independent chiral
corrections). To a good approximation, the loop contribution $E^{loop}%
(W^{2},T_{c}^{\ast})$ is dominated by the leading multipole $E_{1}%
^{loop}(z_{3})$, in which case $\delta_{DE}=\delta_{1}^{1}$. Note that
$E_{1}^{loop}(z_{3})$ is still a function of the photon energy, hence
indirectly of $W^{2}$ and $T_{c}^{\ast}$.

\begin{figure}[t]
\centering    \includegraphics[width=15.0cm]{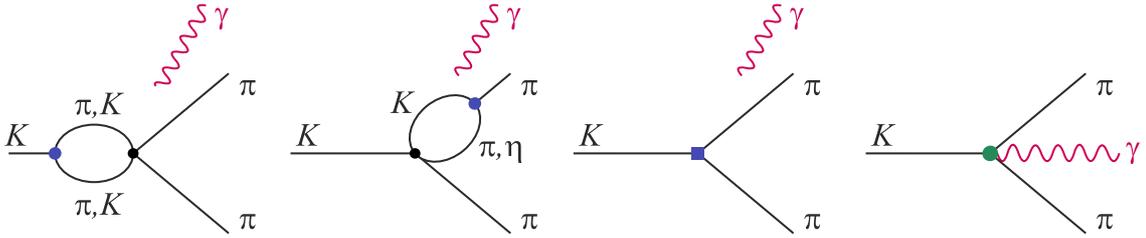}  \caption{Basic
topologies for the $K\rightarrow\pi\pi\gamma$ loops, with the vertices colored
according to the conventions of Fig.~\ref{Fig2}. The photon is to be attached
in all possible ways. However, in accordance with Low's theorem, most of these
diagrams renormalize the $\mathcal{O}(p^{2})$ bremsstrahlung process, leaving
only genuine substracted three-point loops (thus involving at least one
charged meson) for the direct emission amplitudes. The transition is $\Delta
I=1/2$ ($3/2$) when the weak vertex is $K^{+}\pi^{-}\eta$ or $K^{0}\pi^{+}%
\pi^{-}$ ($K^{+}\pi^{-}\pi^{0}$). The counterterms and $Q_{\gamma}^{-}$
contribute only to $K^{+}\rightarrow\pi^{+}\pi^{0}\gamma$ and $K^{0}%
\rightarrow\pi^{+}\pi^{-}\gamma$.}%
\label{Fig3}%
\end{figure}

In our computation of $E_{1}^{loop}$, we include both the $\mathcal{L}_{8}$
and $\mathcal{L}_{27}$ contributions. Indeed, as shown in Fig.~\ref{Fig3}, the
large $\pi\pi$ loop occurs only for the $\Delta I=3/2$ channel, making it
competitive with the $\Delta I=1/2$ contributions arising entirely from the
small $\pi K$ and $\eta K$ loops. As a result, we find $E_{1}^{loop}%
(0)=-0.25$, to be compared to $-0.16$ in Ref.~\cite{DAmbrosioI94}. In
addition, the $\pi\pi$ loop generates a significant slope. Though this
momentum dependence over the experimental phase-space is mild, these cuts are
far from the $z_{3}=0$ point, resulting in a further enhancement. Indeed, over
the experimental range (but not outside of it), $E_{1}^{loop}$ is well
described by%
\begin{equation}
\left[  E_{1}^{loop}(W,T_{c}^{\ast})\right]  _{T_{c}^{\ast}\leq
80MeV,0.2<W<0.9}\approx-0.260-0.051W+0.089\frac{T_{c}^{\ast}}{m_{K}}\;.
\end{equation}
Since experimentally, no slope were included, we average $E_{1}^{loop}$ over
the experimental range (using the $dT_{c}^{\ast}dW$ measure to match the
binning procedure of Ref.~\cite{ExpKppg}), and find%
\begin{equation}
\left\langle E_{1}^{loop}(W,T_{c}^{\ast})\right\rangle _{T_{c}^{\ast}%
\leq80MeV,0.2<W<0.9}=-0.280\;\;\rightarrow\;X_{E}^{loop}=-17.6\;GeV^{-4}\;.
\end{equation}
Note that we checked that in the presence of the slopes as predicted at
$\mathcal{O}(p^{4})$ that the fitted values of $X_{E}$ and $X_{M}$ are not
altered significantly.

Once $E_{1}^{loop}$ is known, we can constrain the local term $\bar{N}$ using
the experimental measurement of $X_{E}$:%
\begin{equation}
\operatorname{Re}\bar{N}=0.095\pm0.083\;. \label{CT}%
\end{equation}
This is much smaller than the $\mathcal{O}(1)$ expected for the $N_{i}$ on
dimensional grounds or from factorization~\cite{EPRKppg}, but confirms the
picture described in Sec.~2.1.3. Evidently, so long as the $N_{i}$ are not
better known, we cannot get an unambiguous bound on $\operatorname{Re}%
C_{\gamma}^{-}$. Still, barring a large fortuitous cancellation,%
\begin{equation}
\frac{|\operatorname{Re}C_{\gamma}^{-}|}{G_{F}m_{K}}\lesssim0.1\;.
\label{ReCm}%
\end{equation}
Note that this bound is rather close to our naive estimate~(\ref{ReCSM}) of
the charm-quark contribution to the real photon penguin in the SM.

\paragraph{Direct CP-violating asymmetries:\newline \newline }

CP-violation in $K^{+}\rightarrow\pi^{+}\pi^{0}\gamma$ is quantified by the
parameter $\varepsilon_{+0\gamma}^{\prime}$, defined from%
\begin{equation}
\operatorname{Re}\left(  \frac{E_{DE}}{eA_{IB}}\right)  \left(  K^{\pm
}\rightarrow\pi^{\pm}\pi^{0}\gamma\right)  \approx\frac{\operatorname{Re}%
E_{DE}}{e\operatorname{Re}A_{IB}}\left[  \cos(\delta_{DE}-\delta_{0}^{2}%
)\mp\sin(\delta_{DE}-\delta_{0}^{2})\varepsilon_{+0\gamma}^{\prime}\right]
\;, \label{Asym1}%
\end{equation}
as~\cite{DAmbrosioI96}%
\begin{equation}
\varepsilon_{+0\gamma}^{\prime}\equiv\frac{\operatorname{Im}E_{DE}%
}{\operatorname{Re}E_{DE}}-\frac{\operatorname{Im}A_{IB}}{\operatorname{Re}%
A_{IB}}\;.
\end{equation}
To reach this form, we use the fact that both $\operatorname{Im}E_{DE}$ and
$\operatorname{Im}A_{IB}$ change sign under $CP$, but not the strong phase
$\delta_{DE}$ and $\delta_{0}^{2}$, and work to first order in
$\operatorname{Im}A_{IB}/\operatorname{Re}A_{IB}$. Since $E_{2}$ has the same
strong phase as $A_{IB}$, and higher multipoles are completely negligible, we
can replace $E_{DE}$ by the dipole emission $E_{1}$ to an excellent
approximation, so that $\delta_{DE}=\delta_{1}^{1}$.

Plugging Eq.~(\ref{Asym1}) in Eq.~(\ref{DiffRate}), we get the differential
asymmetry, which can be integrated over phase-space according to various
definitions. Still, no matter the choice, these phase-space integrations tend
to strongly suppress the overall sensitivity to $\varepsilon_{+0\gamma
}^{\prime}$ since the rate is dominantly CP-conserving~\cite{DAmbrosioI96}.
For example, NA48/2~\cite{ExpKppg} use the partially integrated asymmetry%
\begin{equation}
a_{CP}(W^{2})=\frac{\partial\Gamma^{+}/\partial W^{2}-\partial\Gamma
^{-}/\partial W^{2}}{\partial\Gamma^{+}/\partial W^{2}+\partial\Gamma
^{-}/\partial W^{2}}=\frac{-2m_{\pi^{+}}^{2}m_{K}^{2}X_{E}W^{2}\;\sin
(\delta_{DE}-\delta_{0}^{2})\;\varepsilon_{+0\gamma}^{\prime}}{1+2m_{\pi^{+}%
}^{2}m_{K}^{2}X_{E}W^{2}+m_{\pi^{+}}^{4}m_{K}^{4}(\left|  X_{E}\right|
^{2}+\left|  X_{M}\right|  ^{2})W^{4}}\;,
\end{equation}
where the dependences of $X_{E}$ and $X_{M}$ on $T_{c}^{\ast}$ are dropped,
which is a reasonable approximation within the considered phase-space. Given
the experimental values for $X_{E}$ and $X_{M}$, and combined with
$\sin(\delta_{1}^{1}-\delta_{0}^{2})\approx\sin(7^\circ)\approx0.12$~\cite{ExpKppg,ACGH01}, $a_{CP}(W^{2})\lesssim0.01\varepsilon
_{+0\gamma}^{\prime}$ over the whole $W^{2}$ range. Clearly, integrating over
$W^{2}$ to get the total rate charge asymmetry (or the induced direct
CP-asymmetry in $K^{\pm}\rightarrow\pi^{\pm}\pi^{0}$~\cite{DibP90}) would
suppress the sensitivity even more. Because of this, the current bound is
rather weak~\cite{ExpKppg}%
\begin{equation}
\sin(\delta_{DE}-\delta_{2})\varepsilon_{+0\gamma}^{\prime}=\left(
-2.5\pm4.2\right)  \times10^{-2}\;. \label{ep0gExp}%
\end{equation}

Actually, thanks to the fact that $X_{E}<0$, there is an alternative
observable which is not phase-space suppressed. Defining $\partial^{2}%
\Gamma_{DE}^{\pm}=\partial^{2}\Gamma^{\pm}-\partial^{2}\Gamma_{IB}^{\pm}$, and
integrating over $T_{c}^{\ast}$, the direct emission differential rates
$\partial\Gamma_{DE}^{+}/\partial W^{2}$ and $\partial\Gamma_{DE}^{-}/\partial
W^{2}$ vanish at slightly different values of $W^{2}$, so we can construct the
asymmetry,%
\begin{equation}
a_{CP}^{0}=\frac{W_{\partial\Gamma_{DE}^{+}/\partial W^{2}=0}^{2}%
-W_{\partial\Gamma_{DE}^{-}/\partial W^{2}=0}^{2}}{W_{\partial\Gamma_{DE}%
^{+}/\partial W^{2}=0}^{2}+W_{\partial\Gamma_{DE}^{-}/\partial W^{2}=0}^{2}%
}=-\tan(\delta_{DE}-\delta_{2})\varepsilon_{+0\gamma}^{\prime}\;. \label{a0CP}%
\end{equation}
The zeros are around $W^{2}\approx0.16$, i.e. within the experimental range
$0.2<W<0.9$. Of course, it remains to be seen whether the experimental
precision needed to perform significant fits to the zeros of $\partial
\Gamma_{DE}^{\pm}/\partial W^{2}$ is not prohibitive.

\begin{figure}[t]
\centering           \includegraphics[width=8.0cm]{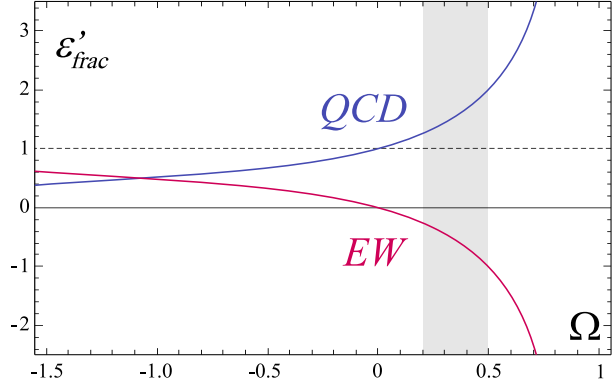}
\caption{Fractions of QCD and electroweak penguins in $\varepsilon^{\prime}$.
The absence of electroweak penguins corresponds to $\Omega=0$. Destructive
interference occurs for values between $0$ and $1$ (with a singularity at $1$
since it corresponds to a complete cancellation between both types of
penguins). Current analyses in the SM favor a limited destructive
interference, i.e. $\Omega\in\lbrack+0.2,+0.5]$ (see e.g.
Ref.~\protect
\cite{BurasG01,BurasJ04,Guadagnoli}).}%
\label{Fig5}%
\end{figure}

Let us analyze the prediction for $\varepsilon_{+0\gamma}^{\prime}$ in the SM.
At $\mathcal{O}(p^{4})$, discarding for now the counterterms and the
electromagnetic operators, we obtain (see Appendix~\ref{AppA})%
\begin{equation}
\varepsilon_{+0\gamma}^{\prime}(z_{3})=\frac{\sqrt{2}|\varepsilon^{\prime}%
|}{\omega}f(z_{3},\Omega)\;,\;\;f(z_{3},\Omega)=\frac{-1}{1+\omega
h_{20}(z_{3})}-\frac{\Omega}{1-\Omega}\frac{\omega\delta h_{20}(z_{3}%
)}{1+\omega h_{20}(z_{3})}\;,\label{ep0g}%
\end{equation}
where $\omega=1/22.4$, $h_{20}(z_{3})$ is the ratio of the $G_{27}$ and
$G_{8}$ loop functions, enhanced by the $\pi\pi$ contributions to the former,
while $\delta h_{20}(z_{3})$ is the ratio of the $G_{ew}$ and $G_{8}$ loop
functions and is $\mathcal{O}(1)$. The parameter $\Omega$ is defined as%
\begin{equation}
\frac{\operatorname{Im}A_{2}}{\operatorname{Im}A_{0}}\equiv\omega
\Omega\;.\label{xparam}%
\end{equation}
It represents the fraction of electroweak versus QCD penguins in
$\varepsilon^{\prime}$,%
\begin{equation}
\varepsilon^{\prime}=i\frac{e^{i(\delta_{0}^{2}-\delta_{0}^{0})}}{\sqrt{2}%
}\omega\left(  \frac{\operatorname{Im}A_{2}}{\operatorname{Re}A_{2}%
}-\frac{\operatorname{Im}A_{0}}{\operatorname{Re}A_{0}}\right)
=i\frac{e^{i(\delta_{0}^{2}-\delta_{0}^{0})}}{\sqrt{2}}\frac{\operatorname{Im}%
A_{0}}{\operatorname{Re}A_{0}}\omega(\Omega-1)\;.
\end{equation}
As shown in Fig.~\ref{Fig5}, a conservative range is $\Omega\in\lbrack
-1,+0.8]$. Values between $[+0.2,+0.5]$ are favored by current analyses in the
SM, but large NP cannot be ruled out.

A crucial observation is that $\varepsilon_{+0\gamma}^{\prime}$ is rather
insensitive to $\Omega$, because $\omega\delta h_{20}(z_{3})$ is suppressed by
$\omega$, so that $f(z_{3},\Omega)\approx-2/3$. Varying $\Omega$ in the large
range $[-1,+0.8]$, as well as including the potential impact of the
$\mathcal{L}_{8}^{\text{CT}}$ counterterms (subject to the constraint
Eq.~(\ref{CT})) does not affect $\varepsilon_{+0\gamma}^{\prime}$ much (see
Appendix~\ref{AppA}), and we conservatively obtain%
\begin{equation}
\varepsilon_{+0\gamma}^{\prime}(Q_{3,...,10})=-0.55(25)\times\frac{\sqrt
{2}|\varepsilon^{\prime}|}{\omega}=-0.64(31)\times10^{-4}\;,\label{Finalep0g}%
\end{equation}
using $\operatorname{Re}(\varepsilon^{\prime}/\varepsilon)^{\exp}%
=(1.65\pm26)\times10^{-3}$~\cite{PDG}. The slight growth of $\varepsilon
_{+0\gamma}^{\prime}$ with $z_{3}$ is negligible compared to its error. Since
it is based on the experimental value of $|\varepsilon^{\prime}|$, and given
the large range allowed for $\Omega$, this estimate is valid even in the
presence of NP in the four-quark operators.

The stability of this prediction actually means that even a precise
measurement of $\varepsilon_{+0\gamma}^{\prime}$ would not help to understand
the physical content of $\varepsilon^{\prime}$, which would require measuring
$\Omega$. On the other hand, it may help to unambiguously distinguish a
contribution from $Q_{\gamma}^{-}$,
\begin{equation}
\varepsilon_{+0\gamma}^{\prime}(Q_{\gamma}^{-})=\frac{\operatorname{Im}%
E_{DE}(Q_{\gamma}^{-})}{\operatorname{Re}E_{DE}}=\frac{B_{T}}{20\pi^{2}%
}\frac{G_{F}/G_{27}}{F_{\pi}^{2}(m_{K}^{2}-m_{\pi}^{2})X_{E}}%
\frac{\operatorname{Im}C_{\gamma}^{-}}{G_{F}m_{K}}%
=+2.8(7)\frac{\operatorname{Im}C_{\gamma}^{-}}{G_{F}m_{K}}\;,
\end{equation}
where we used the experimental determination~(\ref{DEexp}) of
$\operatorname{Re}E_{DE}$. So, the magnetic operator is competitive with the
four-quark operators already in the SM, where we find from Eq.~(\ref{SM}),
\begin{equation}
\varepsilon_{+0\gamma}^{\prime}(Q_{\gamma}^{-})|_{\text{SM}}=+1.2(4)\times
10^{-4}\;. \label{ep0gQgSM}%
\end{equation}
Hence, summing Eq.~(\ref{Finalep0g}) and~(\ref{ep0gQgSM}), there is a
significant cancellation at play and $\varepsilon_{+0\gamma}^{\prime
}|_{\text{SM}}=0.5(5)\times10^{-4}$. This is still far below the current bound
on $\varepsilon_{+0\gamma}^{\prime}$ derived from Eq.~(\ref{ep0gExp}), which
translates as%
\begin{equation}
\frac{\operatorname{Im}C_{\gamma}^{-}}{G_{F}m_{K}}=-0.08\pm0.13\;,
\label{ImCg}%
\end{equation}
thus leaving ample room for NP effects.

\subsubsection{$K_{L}\rightarrow\pi^{+}\pi^{-}\gamma$}

For this mode, the large $\pi\pi$ loop is present in both the $\Delta I=1/2$
and $\Delta I=3/2$ channel, see Fig.~\ref{Fig3}, so including the latter does
not change the picture for the total rate. On the other hand, the situation
for the CP-violating parameter $\bar{\varepsilon}_{+-\gamma}^{\prime}$,
defined from~\cite{DAmbrosioI96}%
\begin{equation}
\bar{\varepsilon}_{+-\gamma}^{\prime}\equiv\eta_{+-\gamma}-\eta_{+-}%
\;,\;\;\eta_{+-\gamma}\equiv\frac{A(K_{L}\rightarrow\pi^{+}\pi^{-}%
\gamma)_{E_{IB}+E_{1}}}{A(K_{S}\rightarrow\pi^{+}\pi^{-}\gamma)_{E_{IB}+E_{1}%
}}\;,\;\;\eta_{+-}\equiv\frac{A(K_{L}\rightarrow\pi^{+}\pi^{-})}%
{A(K_{S}\rightarrow\pi^{+}\pi^{-})}\;,\label{etapm0}%
\end{equation}
is altered significantly. The restriction to the dipole terms originates in
their dominance in the $K_{S}$ decay. The parameter $\eta_{+-\gamma}$ is then
purely CP-violating since the $K_{L}\rightarrow\pi^{+}\pi^{-}\gamma$ dipole
emissions violate CP. The direct dipole emission amplitudes $E_{1}^{L,S}$ for
$K_{L,S}\rightarrow\pi^{+}\pi^{-}\gamma$ are functions of the photon energy
$z_{3}$ only, and can be written as%
\begin{equation}
E_{1}^{S}=\operatorname{Re}E_{+-}\;,\;\;E_{1}^{L}=i\operatorname{Im}%
E_{+-}+\bar{\varepsilon}\operatorname{Re}E_{+-}\;.\label{hpm}%
\end{equation}
Parametrizing the CP-violating IB amplitude as $E_{IB}^{L}=\eta_{+-}E_{IB}%
^{S}$, including the strong phases but working to leading order in $\omega$
and in the CP-violating quantities~\cite{DAmbrosioI96},%
\begin{equation}
\bar{\varepsilon}_{+-\gamma}^{\prime}=e^{i(\delta_{1}^{1}-\delta_{0}^{0}%
)}\frac{m_{K}z_{1}z_{2}}{e\sqrt{2}}\frac{\operatorname{Re}E_{+-}%
}{\operatorname{Re}A_{0}}\left(  \varepsilon^{\prime}+i\left(
\frac{\operatorname{Im}A_{0}}{\operatorname{Re}A_{0}}-\frac{\operatorname{Im}%
E_{+-}}{\operatorname{Re}E_{+-}}\right)  \right)  \;.
\end{equation}
As stated in Ref.~\cite{DAmbrosioI96}, $\bar{\varepsilon}_{+-\gamma}^{\prime}$
is a measure of direct CP-violation. The $z_{1}z_{2}$ momentum dependence
comes from the bremsstrahlung amplitude $E_{IB}^{S}$, which we write in terms
of the $K\rightarrow\pi\pi$ isospin amplitudes using $A(K_{S}\rightarrow
\pi^{+}\pi^{-})=\sqrt{2}A_{0}+A_{2}$. Over the $K^{0}\rightarrow\pi^{+}\pi
^{-}\gamma$ phase-space, $z_{1}z_{2}$ is the largest when $E_{\gamma}^{\ast}$
is at its maximum (and the bremsstrahlung at its minimum), but always strongly
suppresses the asymmetry since $z_{1}z_{2}\lesssim0.030$. Following
Ref.~\cite{TandeanV00}, to avoid dragging along this phase-space factor, we
define the direct CP-violating parameter $\varepsilon_{+-\gamma}^{\prime}$%
\begin{equation}
\varepsilon_{+-\gamma}^{\prime}\equiv\frac{\bar{\varepsilon}_{+-\gamma
}^{\prime}}{z_{1}z_{2}}=\frac{\eta_{+-\gamma}-\eta_{+-}}{z_{1}z_{2}}\;.
\end{equation}

Experimentally, this parameter has been studied indirectly through the
time-dependence observed in the $\pi^{+}\pi^{-}\gamma$ decay
channel~\cite{KppgKTeV} (using material in the beam to regenerate $K_{S}$
states), which is sensitive to the interference between the $K_{L}%
\rightarrow\pi^{+}\pi^{-}\gamma$ and $K_{S}\rightarrow\pi^{+}\pi^{-}\gamma$
decay amplitudes. Importantly, the experimental parameter $\eta_{+-\gamma}$
used in Ref.~\cite{KppgKTeV} (also quoted by the PDG~\cite{PDG}) is not the
same as the one in Eq.~(\ref{etapm0}) but requires additional phase-space
integrations. Following Ref.~\cite{TandeanV00} to pull these out, the
experimental measurement $\tilde{\eta}_{+-\gamma}=(2.35\pm0.07)\times10^{-3}$
translates as%
\begin{equation}
|\varepsilon_{+-\gamma}^{\prime}|<0.06\;. \label{PMGexp}%
\end{equation}

The $E_{+-}$ amplitude can be predicted at $\mathcal{O}(p^{4})$ in ChPT, with
the result (neglecting the counterterms and electromagnetic operators for now)%
\begin{equation}
\frac{\operatorname{Im}E_{+-}}{\operatorname{Re}E_{+-}}%
=\frac{\operatorname{Im}A_{0}}{\operatorname{Re}A_{0}}\frac{1+\omega
\Omega(h_{20}^{\prime}(z_{3})+\delta h_{20}^{\prime}(z_{3}))}{1+\omega
h_{20}^{\prime}(z_{3})}\;,\label{epmg}%
\end{equation}
where $\Omega$ is defined in Eq.~(\ref{xparam}), and $h_{20}^{\prime}(z_{3})$,
$\delta h_{20}^{\prime}(z_{3})$ are ratios of loop functions (see
Appendix~\ref{AppA}). Because the $\pi\pi$ loop is allowed in the
$\Delta I=1/2$ channel, $h_{20}^{\prime}(z_{3})\approx1/\sqrt{2}\ll\omega
^{-1}$ while $\delta h_{20}^{\prime}(z_{3})$ is tiny and can be safely
neglected. Plugging this in $\varepsilon_{+-\gamma}^{\prime}$, the sensitivity
to $\Omega$ disappears completely%
\begin{equation}
\varepsilon_{+-\gamma}^{\prime}(Q_{3,...,10})=ie^{i(\delta_{1}^{1}-\delta
_{0}^{0})}\frac{m_{K}}{e\sqrt{2}}\frac{\operatorname{Re}E_{+-}}%
{\operatorname{Re}A_{0}}|\varepsilon^{\prime}|\left(  e^{i(\delta_{0}%
^{2}-\delta_{0}^{0})}-1\right)  \;.
\end{equation}
As for $\varepsilon_{+0\gamma}^{\prime}$, there is no way to learn something
about $\varepsilon^{\prime}$ by measuring $\varepsilon_{+-\gamma}^{\prime}$.
Also, remark that $\varepsilon_{+-\gamma}^{\prime}$ is suppressed by the
$\Delta I=1/2$ rule through its proportionality to $|\varepsilon^{\prime}|$,
contrary to $\varepsilon_{+0\gamma}^{\prime}$ in Eq.~(\ref{Finalep0g}).

The same combination of counterterms occur for $K^{0}\rightarrow\pi^{+}\pi
^{-}\gamma$ and $K^{+}\rightarrow\pi^{+}\pi^{0}\gamma$. The bound in
Eq.~(\ref{CT}) shows that this combination is of the order of the $\pi K$ and
$\eta K$ loops, which are much smaller than the $\pi\pi$ loop. So, they can be
safely neglected and we finally predict%
\begin{equation}
\varepsilon_{+-\gamma}^{\prime}(Q_{3,...,10})\approx\dfrac{m_{K}^{2}}{(4\pi
F_{\pi})^{2}}h_{0}(z_{3}/2)\times|\varepsilon^{\prime}|\times e^{-i\pi
/3}=-1.5(5)\times10^{-6}\times e^{-i\pi/3}\;,
\end{equation}
with $h_{0}(z_{3}/2)\approx-4\sqrt{2}\operatorname{Re}h_{\pi\pi}\left( -z_{3}\right)  \approx-2.2$, $\delta_{0}^{2}-\delta_{0}^{0}\approx-45^\circ$, and $\delta_{1}^{1}-\delta_{0}^{2}\approx7^\circ$. We conservatively add by hand a 30\% error to account for the chiral corrections to the loop functions. This result is an order of magnitude below the bound derived in Ref.~\cite{DAmbrosioI96} because having kept track of the $G_{8}$, $G_{27}$, and $G_{ew}$ contributions, we could prove that $\varepsilon_{+-\gamma}^{\prime}(Q_{3,...,10})$ is suppressed by the $\Delta I=1/2$ rule. As for $\varepsilon_{+0\gamma}^{\prime}$, this estimate is valid even in the presence of NP in the four-quark operators since it is independent
of $\Omega$ and takes $\operatorname{Re}(\varepsilon^{\prime}/\varepsilon)^{\exp}$ as input.

With $\varepsilon_{+-\gamma}^{\prime}(Q_{3,...,10})$ extremely suppressed,
$\varepsilon_{+-\gamma}^{\prime}$ becomes sensitive to the presence of the
$Q_{\gamma}^{-}$ operator, even in the SM. Its impact on $E_{DE}^{S}$ is
negligible given the bound~(\ref{ReCm}) but $E_{DE}^{L}$ receives an extra
contribution (see Appendix~\ref{AppA}), so that%
\begin{equation}
\varepsilon_{+-\gamma}^{\prime}(Q_{\gamma}^{-})=\frac{-G_{F}/G_{8}}%
{6(2\pi)^{2}}B_{T}\frac{m_{K}^{4}}{F_{\pi}^{2}(m_{K}^{2}-m_{\pi}^{2}%
)}\frac{\operatorname{Im}C_{\gamma}^{-}}{G_{F}m_{K}}e^{i\phi_{\gamma}}%
\approx0.2\frac{\operatorname{Im}C_{\gamma}^{-}}{G_{F}m_{K}}e^{i\phi_{\gamma}%
}\;,
\end{equation}
with $\phi_{\gamma}\equiv\delta_{1}^{1}-\delta_{0}^{0}+\pi/2\approx52^\circ$ and $G_{8}<0$ in our conventions. With the SM value~(\ref{SM}) for $\operatorname{Im}C_{\gamma}^{-}$, this gives%
\begin{equation}
\varepsilon_{+-\gamma}^{\prime}(Q_{\gamma}^{-})_{\text{SM}}=+8(3)\times
10^{-6}\times e^{i\phi_{\gamma}}\;,
\end{equation}
which is about five times larger than $\varepsilon_{+-\gamma}^{\prime
}(Q_{3,...,10})$, but still very small compared to $\varepsilon_{+0\gamma
}^{\prime}$. The current measurement~(\ref{PMGexp}) requires%
\begin{equation}
\frac{|\operatorname{Im}C_{\gamma}^{-}|}{G_{F}m_{K}}<0.3\;, \label{PMGbound}%
\end{equation}
which is slightly looser than the bound~(\ref{ImCg}) obtained from the direct
CP-asymmetry in $K^{+}\rightarrow\pi^{+}\pi^{0}\gamma$.

\subsection{$K_{L,S}\rightarrow\gamma\gamma$}

CP-violating asymmetries for $K\rightarrow\gamma\gamma$ can be defined through
the parameters (adopting the notation of Ref.~\cite{DAmbrosioI96})%
\begin{equation}
\eta_{\gamma\gamma}^{\perp}=\frac{A(K_{S}\rightarrow(\gamma\gamma)_{\perp}%
)}{A(K_{L}\rightarrow(\gamma\gamma)_{\perp})}=\varepsilon+\varepsilon_{\perp
}^{\prime}\;,\;\;\eta_{\gamma\gamma}^{||}=\frac{A(K_{L}\rightarrow
(\gamma\gamma)_{||})}{A(K_{S}\rightarrow(\gamma\gamma)_{||})}=\varepsilon
+\varepsilon_{||}^{\prime}\;.
\end{equation}
Experimentally, these CP-violating parameters could be accessed through
time-dependent interference experiments, i.e. with $K^{0}$ or $\bar{K}^{0}$
beams~\cite{Kgg}, so the photon polarization need not be measured using the
suppressed decays with Dalitz pairs.

\begin{figure}[t]
\centering
\includegraphics[width=13.0cm]{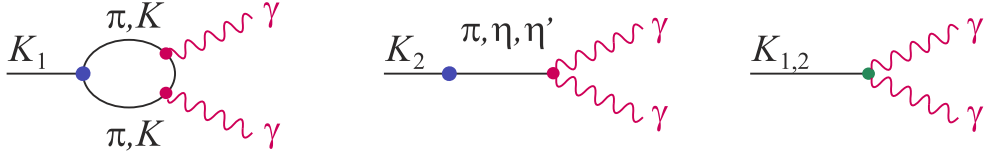}  \caption{The transition
$K\rightarrow\gamma\gamma$ in the SM, with the vertices colored according to
the conventions of Fig.~\ref{Fig2}. The meson loop produces the $\gamma
\gamma_{||}$ state, while the meson poles produce the $\gamma\gamma_{\perp}$
state thanks to the QED anomaly. The direct $Q_{\gamma}^{\pm}$ contributions
produces both the $\gamma\gamma_{||}$ and $\gamma\gamma_{\perp}$ states.}%
\label{Fig4}%
\end{figure}

Let us parametrize the $K^{0}\rightarrow\gamma(k_{1},\mu)\gamma(k_{2},\nu)$
amplitudes as
\begin{subequations}
\label{Agg}%
\begin{align}
A(K^{0}  &  \rightarrow(\gamma\gamma)_{||})=\frac{1}{\sqrt{2}}A_{\gamma\gamma
}^{||}\times(\alpha G_{F}m_{K})\times(k_{1}^{\nu}k_{2}^{\mu}-k_{1}\cdot
k_{2}g^{\mu\nu})\;,\\
A(K^{0}  &  \rightarrow(\gamma\gamma)_{\perp})=\frac{1}{\sqrt{2}}%
A_{\gamma\gamma}^{\perp}\times(\alpha G_{F}m_{K})\times i\varepsilon^{\mu
\nu\rho\sigma}k_{1,\rho}k_{2\sigma}\;,\;\;
\end{align}
\end{subequations}
so that the direct CP-violating parameters are expressed as%
\begin{equation}
\varepsilon_{||,\perp}^{\prime}=i\left(  \frac{\operatorname{Im}%
A_{\gamma\gamma}^{||,\perp}}{\operatorname{Re}A_{\gamma\gamma}^{||,\perp}%
}-\frac{\operatorname{Im}A_{0}}{\operatorname{Re}A_{0}}\right)  \;.
\label{epsigg}%
\end{equation}
We can fix $|A_{\gamma\gamma}^{||}|=0.133(4)$ and $|A_{\gamma\gamma}^{\perp
}|=0.0800(3)$ from the $K_{L,S}\rightarrow\gamma\gamma$ decay rates~\cite{PDG}%
, which are dominantly CP-conserving. In ChPT, $A_{\gamma\gamma}^{||}$
originates from a $\pi^{+}\pi^{-}$ loop and $A_{\gamma\gamma}^{\perp}$ is
induced by the $\pi^{0}$, $\eta$, $\eta^{\prime}$ meson poles together with
the QED anomaly, see Fig.~\ref{Fig4}.

\subsubsection{Two-photon penguin contributions}

In the absence of the electromagnetic operators, $K^{0}\rightarrow\gamma
\gamma$ is induced by the two-photon penguin. The parameters $\varepsilon
_{||,\perp}^{\prime}$ are then generated indirectly by the $Q_{3,...,10}$
contributions to the weak vertices in Fig.~\ref{Fig4}, and directly by the two
photon penguins with $c$ and $t$ quarks (see Eq.~(\ref{Qgg})). However, as
said in Sec.~2, these short-distance contributions are suppressed by the
quadratic decoupling of the heavy modes in the two-photon penguin
loop~\cite{DAmbrosioI96}:%
\begin{equation}
\frac{|\operatorname{Re}A_{\gamma\gamma}^{||,\perp}|_{c,t}}{|\operatorname{Re}%
A_{\gamma\gamma}^{||,\perp}|_{u}}<10^{-4}\;\;\rightarrow|\varepsilon
_{||,\perp}^{\prime}|_{c,t}\approx\frac{|\operatorname{Im}A_{\gamma\gamma
}^{||,\perp}|_{c}}{|\operatorname{Re}A_{\gamma\gamma}^{||,\perp}|_{u}%
}<\frac{\operatorname{Im}\lambda_{c}}{\operatorname{Re}\lambda_{c}}%
\times10^{-4}\approx10^{-7}\;.
\end{equation}
This contribution will turn out to be negligible both for $\varepsilon_{\perp
}^{\prime}$ and $\varepsilon_{||}^{\prime}$.

Concerning the long-distance contribution, let us start with $\varepsilon
_{||}^{\prime}$. Since $A_{\gamma\gamma}^{||}$ is induced by a $\pi\pi$ loop,
CP-violation comes entirely from the $K^{0}\rightarrow\pi^{+}\pi^{-}$ vertex,
as is obvious adopting a dispersive approach. By using $A\left(
K_{S}\rightarrow\pi^{+}\pi^{-}\right)  =\sqrt{2}A_{0}+A_{2}$ (without strong
phases), we recover the result of Ref.~\cite{BuccellaDM91}%
\begin{equation}
\varepsilon_{||}^{\prime}(Q_{3,...,10})=i\frac{\operatorname{Im}A_{0}%
}{\operatorname{Re}A_{0}}\left(  \frac{\sqrt{2}+\omega\Omega}{\sqrt{2}+\omega
}-1\right)  =\frac{\varepsilon^{\prime}e^{-i(\delta_{0}^{2}-\delta_{0}^{0})}%
}{1+\omega/\sqrt{2}}\;. \label{epsiPARA}%
\end{equation}
As for $\varepsilon_{+0\gamma}^{\prime}$ and $\varepsilon_{+-\gamma}^{\prime}%
$, $\varepsilon_{||}^{\prime}$ is insensitive to $\Omega$, so this expression
remains valid in the presence of NP. Also, being suppressed by the $\Delta
I=1/2$ rule, the tiny value $|\varepsilon_{||}^{\prime}(Q_{3,...,10}%
)|\approx4\times10^{-6}$ is obtained.

The situation is different for $\varepsilon_{\perp}^{\prime}$. It was
demonstrated in Ref.~\cite{GerardST} that only the $Q_{1}$ operator has the
right structure to generate $A_{\gamma\gamma}^{\perp}$ through the QED
anomaly. Then, $\operatorname{Im}A_{\gamma\gamma}^{\perp}=0$ since
current-current operators are CP-conserving (proportional to $\lambda
_{u}=V_{us}^{\ast}V_{ud}$), leaving $\varepsilon_{\perp}^{\prime}$ as a pure
and $\Delta I=1/2$ enhanced measure of the QCD penguins%
\begin{equation}
\varepsilon_{\perp}^{\prime}(Q_{3,...,10})=-i\frac{\operatorname{Im}A_{0}%
}{\operatorname{Re}A_{0}}=i\frac{\sqrt{2}|\varepsilon^{\prime}|}%
{\omega(1-\Omega)}\;. \label{epsiPERP}%
\end{equation}

One may be a bit puzzled by the appearance of $\operatorname{Im}A_{0}$ in this
$K\rightarrow\gamma\gamma$ observable. Actually, this originates from the very
definition of $\varepsilon$ in the $K\rightarrow\pi\pi$ system. It is the
choice made there to define a convention-independent physical parameter which
renders it implicitly dependent on $K\rightarrow\pi\pi$ amplitudes. Besides,
Eq.~(\ref{epsiPERP}) is clearly only valid in the usual CKM phase-convention,
contrary to Eq.~(\ref{epsigg}) which is convention-independent. For example,
if the Wu-Yang phase convention $\operatorname{Im}A_{0}=0$ is
adopted~\cite{WuYang}, then $\langle\gamma\gamma|Q_{1}|K_{L}\rangle$ gets a
non-zero weak phase since $\operatorname{Im}\lambda_{u}\neq0$, and
$\varepsilon_{\perp}^{\prime}$ stays the same.

Evidently, given the current information on the $Q_{6}$ contribution to
$\varepsilon^{\prime}$, it is not possible to give a precise prediction for
$\varepsilon_{\perp}^{\prime}$. With $\Omega\in\lbrack-1,+0.8]$,
$\varepsilon_{\perp}^{\prime}\ $spans an order of magnitude:
\begin{equation}
5\times10^{-5}<-i\varepsilon_{\perp}^{\prime}(Q_{3,...,10})<7\times10^{-4}\;.
\label{epsiperpnum}%
\end{equation}
A value of a few $10^{-4}$ is likely as $\Omega\in\lbrack+0.2,+0.5]$ is
favored in the SM, see Fig.~\ref{Fig5}.

This result is different from earlier estimates~\cite{BuccellaDM91}, obtained
before the structure of the $K_{L}\rightarrow\gamma\gamma$ amplitude was
elucidated Ref.~\cite{GerardST}. Further, from that analysis, we do not expect
that the residual $Q_{6}$ contributions in $K_{2}\rightarrow\gamma\gamma$
could alter Eq.~(\ref{epsiPERP}), especially given its large $\Delta I=1/2$
enhanced value (\ref{epsiperpnum}). Indeed, the origin of the vanishing of the
$K_{2}\rightarrow\gamma\gamma$ amplitude at $\mathcal{O}(p^{4})$ is now
understood as the inability of $SU(3)$ ChPT to catch the $Q_{1}$ contribution
at leading order. But once accounted for either through higher order
counterterms or by first working within $U(3)$ ChPT, this $Q_{1}$ contribution
is seen to dominate the $K_{2}\rightarrow\gamma\gamma$ amplitude.

Though only ten times smaller than $\varepsilon$, measuring $\varepsilon
_{\perp}^{\prime}$ would be very challenging. Still, any information would be
very rewarding: with its unique sensitivity to the QCD penguins, it could be
used to finally resolve the physics content of $\varepsilon^{\prime}$.
Further, it would also help in estimating $\varepsilon$ precisely, since the
term $i\operatorname{Im}A_{0}/\operatorname{Re}A_{0}$ enters directly
there~\cite{Guadagnoli,BurasGI10}.

\subsubsection{Electromagnetic operator contributions}

The magnetic operators $Q_{\gamma}^{\pm}$ contribute to $K\rightarrow
\gamma\gamma$ as%
\begin{equation}
A_{\gamma\gamma}^{||,\perp}\rightarrow A_{\gamma\gamma}^{||,\perp
}+\frac{2F_{\pi}}{9\pi m_{K}}B_{T}^{\prime}\frac{C_{\gamma}^{-,+}}{G_{F}m_{K}%
}\;. \label{ampligg}%
\end{equation}
Given the good agreement between theory and experiment for the $K_{S,L}%
\rightarrow\gamma\gamma$ rate, we require that their contributions is less
than $10\%$ of the full amplitude, giving%
\begin{equation}
\frac{|\operatorname{Re}C_{\gamma}^{\pm}|}{G_{F}m_{K}}\lesssim0.3\;.
\label{KggBD}%
\end{equation}
The stronger bound~(\ref{ReCm}) from $K^{+}\rightarrow\pi^{+}\pi^{0}\gamma$
thus shows that the impact of $Q_{\gamma}^{\pm}$ on the total rates is
negligible (assuming $|\operatorname{Re}C_{\gamma}^{+}|\approx
|\operatorname{Re}C_{\gamma}^{-}|)$.

Plugging Eq.~(\ref{ampligg}) in Eq.~(\ref{epsigg}), the $Q_{\gamma}^{\pm}$
contribution to the direct CP-violation parameters are
\begin{equation}
|\varepsilon_{||}^{\prime}(Q_{\gamma}^{-})|\approx\frac{1}{3}%
\frac{|\operatorname{Im}C_{\gamma}^{-}|}{G_{F}m_{K}}\;,\;\;|\varepsilon
_{\perp}^{\prime}(Q_{\gamma}^{+})|\approx\frac{1}{2}\frac{|\operatorname{Im}%
C_{\gamma}^{+}|}{G_{F}m_{K}}\;.
\end{equation}
In the SM, $|\varepsilon_{||}^{\prime}(Q_{\gamma}^{-})|\approx1.4\times
10^{-5}$ is nearly an order of magnitude larger than $\varepsilon_{||}%
^{\prime}(Q_{3,...,10})$, Eq.~(\ref{epsiPARA}). On the contrary, the SM
contribution $|\varepsilon_{\perp}^{\prime}(Q_{\gamma}^{+})|\approx
2\times10^{-5}$ is too small to compete with $\varepsilon_{\perp}^{\prime
}(Q_{3,...,10})$, Eq.~(\ref{epsiPERP}). In the absence of a significant NP
enhancement, $\varepsilon_{\perp}^{\prime}$ thus remains a pure measure of the QCD penguins.

\subsection{Rare semileptonic decays\label{semilept}}

\begin{figure}[t]
\centering           \includegraphics[width=16.0cm]{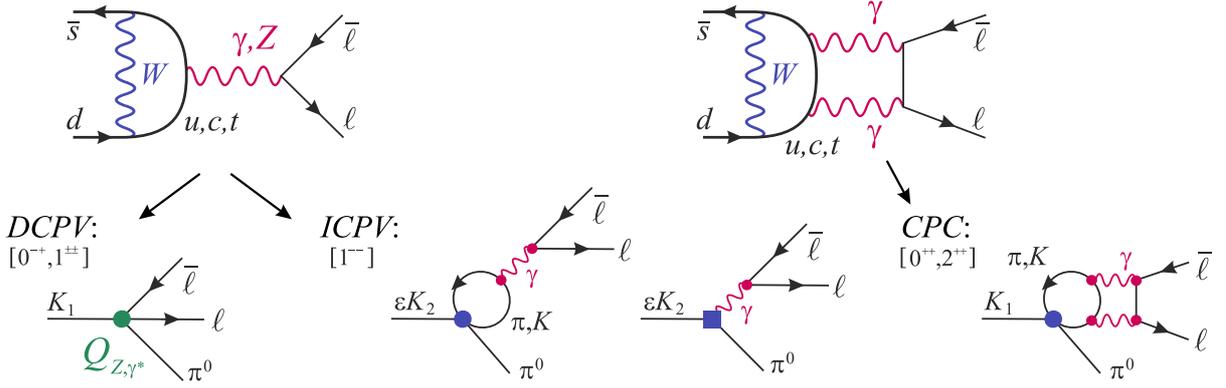}
\caption{The anatomy of the rare semileptonic decays, following the color
coding defined in Fig.~\ref{Fig2}. For $K\rightarrow\pi\nu\bar{\nu}$, only the
$Z$ penguin contributes. For $K_{L}\rightarrow\pi^{0}\ell^{+}\ell^{-}$, in
addition to the direct CP-violating contributions (DCPV) from the $Z$ and
$\gamma^{\ast}$ penguins, the long-distance dominated indirect CP-violating
contribution (ICPV) and the CP-conserving two-photon penguin contribution
(CPC) also enter. The $J^{PC}$ state of the lepton pair is indicated, showing
that only the DCPV and ICPV processes can interfere in the $1^{--}$ channel.}%
\label{Fig7}%
\end{figure}

The $K_{L}\rightarrow\pi^{0}\ell^{+}\ell^{-}$ decays are sensitive to several
FCNC currents. In the SM, both the virtual and real photon penguins, as well
as the $Z$ penguins can contribute (together with their associated $W$ boxes),
see Fig.~\ref{Fig7}. Since NP could a priori affect all these FCNC in a
coherent way, they have to be accounted for. Further, to separately constrain
the $Z$ penguins, we include the rare $K\rightarrow\pi\nu\bar{\nu}$ decays in
the analysis. So, in the present section, we collect the master formula for
the $K_{L}\rightarrow\pi^{0}e^{+}e^{-}$, $K_{L}\rightarrow\pi^{0}\mu^{+}%
\mu^{-}$, $K^{+}\rightarrow\pi^{+}\nu\bar{\nu}$ and $K_{L}\rightarrow\pi
^{0}\nu\bar{\nu}$ decay rates, starting from the effective Hamiltonian%
\begin{gather}
\mathcal{H}_{\text{eff}}=-\frac{G_{F}\alpha}{\sqrt{2}}\sum_{\ell=e,\mu,\tau
}(C_{\nu,\ell}\;Q_{\nu,\ell}+C_{V,\ell}\;Q_{V,\ell}+C_{A,\ell}\;Q_{A,\ell
})+h.c.\;,\label{HFF}\\
Q_{V,\ell}=\bar{s}\gamma^{\mu}d\otimes\bar{\ell}\gamma_{\mu}\ell
\;,\;\;Q_{A,\ell}=\bar{s}\gamma^{\mu}d\otimes\bar{\ell}\gamma_{\mu}\gamma
_{5}\ell\;,\;Q_{\nu,\ell}=\bar{s}\gamma^{\mu}d\otimes\bar{\nu}_{\ell}%
\gamma_{\mu}(1-\gamma_{5})\nu_{\ell}\;,\nonumber
\end{gather}
to which only the magnetic operators $Q_{\gamma}^{\pm}$ should be added, since
$Q_{\gamma^{\ast}}^{\pm}$ are implicitly included in $Q_{V,\ell}$.

\subsubsection{Electric operators and SM predictions}

Thanks to the excellent control on the vector currents~(\ref{VectorC}), the
branching ratios for $K\rightarrow\pi\nu\bar{\nu}$ are predicted very
precisely:%
\begin{subequations}
\begin{align}
\mathcal{B}\left(  K^{+}\rightarrow\pi^{+}\nu_{\ell}\bar{\nu}_{\ell}\right)
&  =0.1092(5)\cdot10^{-11}\times r_{us}^{2}\times|\omega_{\nu,\ell}%
|^{2}\;,\label{Kpnn}\\
\mathcal{B}\left(  K_{L}\rightarrow\pi^{0}\nu_{\ell}\bar{\nu}_{\ell}\right)
&  =0.471(3)\cdot10^{-11}\times r_{us}^{2}\times(\operatorname{Im}\omega
_{\nu,\ell})^{2}\;,
\end{align}
\end{subequations}
with $r_{us}=0.225/|V_{us}|$ and $\omega_{\nu,\ell}=C_{\nu,\ell}/10^{-4}$.
Since experimentally, the neutrino flavors are not detected, the
$K\rightarrow\pi\nu\bar{\nu}$ rate is the sum of the rates into $\nu
_{e,\mu,\tau}$.

As shown in Fig.~\ref{Fig7}, the situation for $K_{L}\rightarrow\pi^{0}%
\ell^{+}\ell^{-}$ is more complex as the indirect CP-violation $K_{L}%
=\varepsilon K_{1}\rightarrow\pi^{0}\gamma^{\ast}[\rightarrow\ell^{+}\ell
^{-}]$~\cite{DEIP98} and the CP-conserving contribution $K_{L}\rightarrow
\pi^{0}\gamma\gamma\lbrack\rightarrow\ell^{+}\ell^{-}]$~\cite{BDI03,ISU04}
have to be included (see Appendix~\ref{AppKpll} for an updated error
analysis):%
\begin{equation}%
\begin{tabular}
[c]{ll}%
\multicolumn{2}{l}{$\mathcal{B}(K_{L}\rightarrow\pi^{0}\ell^{+}\ell
^{-})=\left(  C_{dir}^{\ell}r_{us}^{2}+C_{int}^{\ell}\bar{a}_{S}r_{us}%
+C_{mix}^{\ell}\bar{a}_{S}^{2}+C_{\gamma\gamma}^{\ell}\right)  \cdot
10^{-12}\;,\smallskip\smallskip$}\\
$C_{dir}^{e}=2.355(13)\;(\omega_{V,e}^{2}+\omega_{A,e}^{2})\;,$ &
$C_{dir}^{\mu}=0.553(3)\omega_{V,\mu}^{2}+1.266(12)\omega_{A,\mu}%
^{2}\;,\smallskip$\\
$C_{int}^{e}=7.3(2)\;[-7.0(2)]\;\omega_{V,e}\;,$ & $C_{int}^{\mu
}=1.73(4)\;[-1.74(4)]\;\omega_{V,\mu}\;,\smallskip$\\
$C_{mix}^{e}=12.2(4)\;[11.5(5)]\;,$ & $C_{mix}^{\mu}=2.81(6)\;,\smallskip$\\
$C_{\gamma\gamma}^{e}\approx0\;,$ & $C_{\gamma\gamma}^{\mu}=4.7(1.3)\;,$%
\end{tabular}
\label{MasterKL}%
\end{equation}
with $\bar{a}_{S}=1.25(22)$, $\omega_{X,\ell}=\operatorname{Im}C_{X,\ell
}/10^{-4}$. Importantly, if there is some NP, it would enter through
$\omega_{i}$ only because all the rest is fixed from experimental
data~\cite{MST06}. The theoretically disfavored case of destructive
interference between the direct and indirect CP-violating contributions is
indicated in square brackets~\cite{BrunoP92,BDI03}.

In the SM, the QCD corrected Wilson coefficients $\omega_{\nu,\ell}%
^{\text{SM}}$ are known very precisely. Though $\omega_{\nu,\tau}^{\text{SM}}$
is slightly different than $\omega_{\nu,e(\mu)}^{\text{SM}}$ owing to the
large $\tau$ mass, the standard phenomenological parametrization employs a
unique coefficient,
\begin{equation}
\omega_{\nu}^{\text{SM}}=-\frac{\lambda_{t}X_{t}+\bar{\lambda}^{4}%
\operatorname{Re}\lambda_{c}(P_{c}+\delta P_{u,c})}{2\pi\sin^{2}\theta
_{W}\times10^{-4}}=4.84(22)-i1.359(96)\;, \label{CnSM}%
\end{equation}
valid for $\ell=e,\mu,\tau$, with $X_{t}=1.465(16)$~\cite{BrodGS10},
$P_{c}=0.372(15)$~\cite{BurasGHN05}, $\delta P_{u,c}=0.04(2)$%
~\cite{IsidoriMS05} (with $\bar{\lambda}=0.2255$). The difference $\omega
_{\nu,e(\mu)}^{\text{SM}}-\omega_{\nu,\tau}^{\text{SM}}$ is implicitly
embedded into the definition of $P_{c}$, up to a negligible $0.2\%$
effect~\cite{BuchallaBL96}. With the CKM coefficients from
Ref.~\cite{CKMfitter}, the rates in the SM are thus%
\begin{equation}
\mathcal{B}(K^{+}\rightarrow\pi^{+}\nu\bar{\nu})^{\text{SM}}=8.25(64)\cdot
10^{-11}\;,\;\mathcal{B}(K_{L}\rightarrow\pi^{0}\nu\bar{\nu})^{\text{SM}%
}=2.60(37)\cdot10^{-11}\;.
\end{equation}

For $K_{L}\rightarrow\pi^{0}\ell^{+}\ell^{-}$, the Wilson coefficients are
$\operatorname{Im}C_{i}=\operatorname{Im}\lambda_{t}y_{i}$ with $y_{A,\ell
}^{\text{SM}}(M_{W})=-0.68(3)$ and $y_{V,\ell}^{\text{SM}}(\mu\approx
1\;$GeV$)=0.73(4)$~\cite{BuchallaBL96}. Using again the CKM elements from
Ref.~\cite{CKMfitter} gives the rate%
\begin{subequations}
\begin{align}
\mathcal{B}(K_{L}  &  \rightarrow\pi^{0}e^{+}e^{-})^{\text{SM}}=3.23_{-0.79}%
^{+0.91}\cdot10^{-11}\;[1.37_{-0.43}^{+0.55}\cdot10^{-11}]\;,\\
\mathcal{B}(K_{L}  &  \rightarrow\pi^{0}\mu^{+}\mu^{-})^{\text{SM}%
}=1.29_{-0.23}^{+0.24}\cdot10^{-11}\;\,[0.86_{-0.17}^{+0.18}\cdot10^{-11}]\;.
\end{align}
\end{subequations}
The errors are currently dominated by that on $\bar{a}_{S}$.

These predictions can be compared to the current experimental results
\begin{equation}%
\begin{array}
[c]{llll}%
\mathcal{B}(K^{+}\rightarrow\pi^{+}\nu\bar{\nu})^{\text{exp}}=1.73_{-1.05}%
^{+1.15}\times10^{-10} & \text{\cite{KPpnunu}\ }, & \mathcal{B}(K_{L}%
\rightarrow\pi^{0}e^{+}e^{-})^{\text{exp}}<2.8\times10^{-10} &
\text{\cite{KTeVelec}\ },\\
\,\mathcal{B}(K_{L}\rightarrow\pi^{0}\nu\bar{\nu})^{\text{exp}}<2.6\times
10^{-8} & \text{\cite{K0pnunu}\ }, & \mathcal{B}(K_{L}\rightarrow\pi^{0}%
\mu^{+}\mu^{-})^{\text{exp}}<3.8\times10^{-10} & \text{\cite{KTeVmuon}\ }.
\end{array}
\label{rareKexp}%
\end{equation}
At $90\%$ CL, this measurement of $\mathcal{B}(K^{+}\rightarrow\pi^{+}\nu
\bar{\nu})$ becomes an upper limit at $3.35\times10^{-10}$~\cite{KPpnunu}.
Improvements are expected in the future, with J-Parc aiming at a hundred SM
events for $K_{L}\rightarrow\pi^{0}\nu\bar{\nu}$, and NA62 at a similar amount
of $K^{+}\rightarrow\pi^{+}\nu\bar{\nu}$ events. The $K_{L}\rightarrow\pi
^{0}\ell^{+}\ell^{-}$ modes are not yet included in the program of these
experiments, but should be tackled in a second phase.

\subsubsection{Magnetic operators in $K^{0}\rightarrow\pi^{0}\ell^{+}\ell^{-}$}

Only the $Q_{\gamma}^{+}$ operator occurs in the $K^{0}\rightarrow\pi^{0}%
\ell^{+}\ell^{-}$ decays:%
\begin{equation}
A(K^{0}(P)\rightarrow\pi^{0}\gamma^{\ast}(q))_{Q_{\gamma}^{+}}=-\frac{eG_{F}%
}{24\sqrt{2}\pi^{2}}B_{T}\frac{C_{\gamma}^{+}}{G_{F}m_{K}}\left(  q^{2}P^{\mu
}-q^{\mu}P\cdot q\right)  \;.
\end{equation}
For $K_{S}\rightarrow\pi^{0}\ell^{+}\ell^{-}$, this contribution is
CP-conserving and parametrically included in $a_{S}$ since it is fixed from
experiment. If we require that there is no large cancellations, i.e. that the
$Q_{\gamma}^{+}$ operator at most accounts for half of $|a_{S}|\approx1.2$, we
get from Eq.~(\ref{KSAS}) in Appendix~\ref{AppKpll},%
\begin{equation}
\frac{|\operatorname{Re}C_{\gamma}^{+}|}{G_{F}m_{K}}\lesssim
\frac{3|\bar{a}_{S}|}{2B_{T}}\approx1.5\;. \label{BoundaS}%
\end{equation}
This bound is nearly an order of magnitude looser than the one derived from
$K_{L}\rightarrow\gamma\gamma$ in Eq.~(\ref{KggBD}).

For $K_{L}\rightarrow\pi^{0}\ell^{+}\ell^{-}$, the whole effect of $Q_{\gamma
}^{+}$ is to shift the value of the vector current~\cite{BurasCIRS99,MST06}:%
\begin{equation}
\omega_{V,\ell}\times10^{-4}=\operatorname{Im}C_{V,\ell}+\frac{Q_{d}}%
{2\sqrt{2}\pi}\frac{B_{T}\left(  0\right)  }{f_{+}\left(  0\right)
}\frac{\operatorname{Im}C_{\gamma}^{+}}{G_{F}m_{K}}\approx\operatorname{Im}%
C_{V,\ell}-\frac{1}{21.3}\frac{\operatorname{Im}C_{\gamma}^{+}}{G_{F}m_{K}}\;,
\label{shift}%
\end{equation}
where we assume the slopes of $B_{T}(z)$ and $f_{+}\left(  z\right)  $ are
both saturated by the same resonance (which is a valid first order
approximation). The relative sign between the $Q_{\gamma}^{+}$ and $Q_{V,\ell
}$ contributions agrees with Ref.~\cite{BurasCIRS99}.

In the SM, $\operatorname{Im}C_{V,\ell}\approx0.99\times10^{-4}$ and
$|\operatorname{Im}C_{\gamma}^{+}|/G_{F}m_{K}\approx4\times10^{-5}$, so the
shift is negligible. However, in case there is some NP, it quickly becomes
visible. In the absence of any other NP effects (which is a strong assumption,
as we will see in the next section), the current experimental
bounds~(\ref{rareKexp}) imply%
\begin{subequations}
\begin{align}
K_{L}\overset{}{\rightarrow}\pi^{0}e^{+}e^{-}  &  \Rightarrow
\;-0.018<\frac{\operatorname{Im}C_{\gamma}^{+}}{G_{F}m_{K}}<+0.030\;,\\
K_{L}\overset{}{\rightarrow}\pi^{0}\mu^{+}\mu^{-}  &  \Rightarrow
\;-0.050<\frac{\operatorname{Im}C_{\gamma}^{+}}{G_{F}m_{K}}<+0.063\;,
\end{align}
\end{subequations}
at $90\%$ confidence and treating all theory errors as Gaussian. This is about
an order of magnitude tighter than the bound (\ref{ImCg}) on
$\operatorname{Im}C_{\gamma}^{-}$ derived from $K^{+}\rightarrow\pi^{+}\pi
^{0}\gamma$.

\subsection{Virtual effects in $\varepsilon^{\prime}/\varepsilon$}

Up to now, the photon produced by the electromagnetic operators was either
real or coupled to a Dalitz pair, but it could also couple to quarks. At the
level of the OPE, such effects are dealt with as $\mathcal{O}(\alpha)$ mixing
among the four-quark operators, and sum up at $\mu\approx1$ GeV in the Wilson
coefficients of Eq.~(\ref{OPE}). The non-perturbative tail of these mixings
are computed as QED corrections to the matrix elements of the effective
operators between hadron states. Currently, only the left-handed electric
operator (i.e., the virtual photon penguin) is included in the
OPE~\cite{BuchallaBL96} and in the $K\rightarrow\pi\pi$ matrix elements and
observables~\cite{Cirigliano03}. The magnetic operators are left aside given
their strong suppression in the SM.

\subsubsection{Magnetic operators in hadronic observables}

In the presence of NP, the magnetic operators could be much more enhanced than
the electric operators, so their impact on hadronic observables must be
quantified. Though in principle we should amend the whole OPE (i.e., initial
conditions and running), we will instead compute only the low-energy part of
these corrections. Indeed, the photon produced by $Q_{\gamma}^{\pm}$ can be
on-shell, so the dominant part of the mixing $Q_{\gamma}^{\pm}\rightarrow
Q_{1,...,10}$ is likely to arise at the matrix-element level. In any case, the
missing SD contributions do not represent the main source of uncertainty.
Indeed, the meson-photon loops induced by $Q_{\gamma}^{\pm}$ are UV-divergent,
requiring specific but unknown counterterms. So, at best, the order of
magnitude of the LD mixing effects can be estimated. To this end, the loops
are computed in dimensional regularization and only the leading $\log
(\mu/m_{\pi})$ or $\log(\mu/m_{K})$ is kept, with $\mu\approx m_{\rho}$.
Parametrizing the momentum dependences of the $B_{T}$, $B_{T}^{\prime}$
form-factors and of the electromagnetic form-factors of the $\pi$ and $K$
mesons using vector-meson dominance would lead to similar results.%

\begin{figure}[t]
\centering\includegraphics[width=14.0cm]{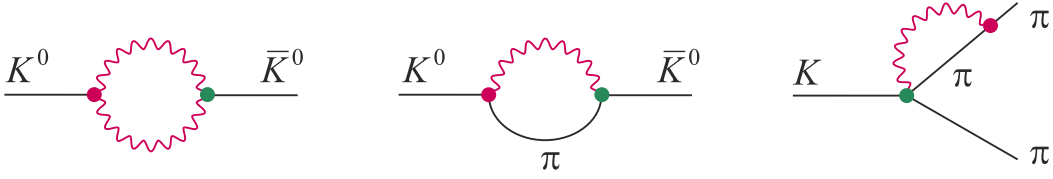}
\caption{The virtual effects from $Q_{\gamma}^{\pm}$ on $\Delta S=2$ observables (reversed diagrams are understood) and on $\varepsilon^{\prime}$ from $K^{0}\rightarrow\pi^{+}\pi^{-}$. Red vertices stand for the SM transitions (which are not necessarily local, see for example Fig.~\ref{Fig4}
), while green vertices are induced by $Q_{\gamma}^{\pm}$.}
\label{Fig7b}
\end{figure}

Let us start with the impact of $Q_{\gamma}^{\pm}$ on $\varepsilon^{\prime}$.
The diagram of Fig.~\ref{Fig7b} induces a correction to $\eta_{+-}%
=A(K_{L}\rightarrow\pi^{+}\pi^{-})/A(K_{S}\rightarrow\pi^{+}\pi^{-})$ and
thereby, discarding strong phases for simplicity
\begin{equation}
\frac{|\operatorname{Re}(\varepsilon^{\prime}/\varepsilon)|_{\gamma}%
}{\operatorname{Re}(\varepsilon^{\prime}/\varepsilon)^{\exp}}\approx
\frac{3\alpha}{256\pi^{3}}B_{T}\frac{G_{F}}{|G_{8}|}\frac{\log(m_{\rho}%
/m_{\pi})}{|\varepsilon|\operatorname{Re}(\varepsilon^{\prime}/\varepsilon
)^{\exp}}\frac{|\operatorname{Im}C_{\gamma}^{-}|}{G_{F}m_{K}}\approx
2\frac{|\operatorname{Im}C_{\gamma}^{-}|}{G_{F}m_{K}}\;.\label{AmpliVirt}%
\end{equation}
The photon loop is IR safe since $Q_{\gamma}^{-}$ does not contribute to the
bremsstrahlung amplitude in $K^{0}\rightarrow\pi^{+}\pi^{-}\gamma$. Let us
stress again that this is only an order of magnitude estimate. Besides the
neglected SD mixings, unknown effects of similar size as Eq.~(\ref{AmpliVirt})
are necessarily present to absorb the divergence. Plugging in the bound on
$\operatorname{Im}C_{\gamma}^{-}$ obtained from the measured $K^{+}%
\rightarrow\pi^{+}\pi^{0}\gamma$ direct CP-asymmetry, Eq.~(\ref{ImCg}),%
\begin{equation}
(\varepsilon_{+0\gamma}^{\prime})^{\exp}\;\Rightarrow
\;\frac{|\operatorname{Re}(\varepsilon^{\prime}/\varepsilon)|_{\gamma}%
}{\operatorname{Re}(\varepsilon^{\prime}/\varepsilon)^{\exp}}=(16\pm
26)\%\;.\label{Bound}%
\end{equation}
So, even in the presence of a large NP contribution to $Q_{\gamma}^{-}$, the
impact on $\varepsilon^{\prime}$ remains smaller than its current theoretical
error in the SM.

For completeness, let us also compute the contribution of the magnetic
operators to the $\Delta S=2$ observables, for which perturbative QED
corrections are significantly suppressed. At long distance, the magnetic
operators contribute to $\langle\bar{K}^{0}|H_{W}|K^{0}\rangle$ through the
transitions $K^{0}\rightarrow\pi\gamma^{\ast}\rightarrow\bar{K}^{0}$ and
$K^{0}\rightarrow\gamma\gamma\rightarrow\bar{K}^{0}$, see Fig.~\ref{Fig7b}.
Neglecting the momentum dependences of the $K\rightarrow\gamma\gamma$ and
$K\rightarrow\pi\gamma^{\ast}$ vertices and keeping only the leading
$\log(m_{\rho}/m_{\pi})$, we obtain
\begin{equation}
\mu_{12}\equiv\frac{\langle\bar{K}^{0}|Q_{\gamma}^{\pm}|K^{0}\rangle}%
{M_{K}\Delta M_{K}^{\exp}}=(a_{\gamma\gamma}^{\perp}+a_{\pi\gamma
})\frac{C_{\gamma}^{+}}{G_{F}m_{K}}+a_{\gamma\gamma}^{||}\frac{C_{\gamma}^{-}%
}{G_{F}m_{K}}\;,
\end{equation}
with (see Eq.~(\ref{Agg}) for the definition of $A_{\gamma\gamma}^{i}$ and
Eq.~(\ref{KSAS}) for that of $a_{S}$)
\begin{subequations}
\label{ds2}%
\begin{align}
|a_{\gamma\gamma}^{i}| &  \approx\frac{\alpha^{2}}{72\pi^{3}}B_{T}^{\prime
}\frac{G_{F}^{2}m_{K}^{4}F_{\pi}}{\Delta M_{K}^{\exp}}|A_{\gamma\gamma}%
^{i}|\log(m_{\rho}/m_{K})\approx7\times10^{-6}\,|A_{\gamma\gamma}^{i}|\;,\\
|a_{\pi\gamma}| &  \approx\frac{\alpha}{512\pi^{5}}B_{T}|a_{S}|\frac{G_{F}%
^{2}m_{\pi}^{4}m_{K}}{\Delta M_{K}^{\exp}}\log(m_{\rho}/m_{\pi})\approx
8\times10^{-7}\;.
\end{align}
\end{subequations}
Numerically, $a_{\pi\gamma}\sim a_{\pi\gamma}$, even though they are not of
the same order in $\alpha$, because of the absence of a $K^{0}\rightarrow
\pi^{0}\gamma^{\ast}$ vertex at leading order (see Eq.~(\ref{KSAS}) in
Appendix~\ref{AppKpll}), and because the momentum scale in the $a_{\pi\gamma}$
loop is entirely set by the pion mass instead of the transferred momentum of
$\mathcal{O}(m_{K})$, as in $a_{\gamma\gamma}$. With such small values for
$a_{\gamma\gamma}$ and $a_{\pi\gamma}$, neither $\Delta M_{K}(Q_{\gamma}^{\pm
})\sim\operatorname{Re}\mu_{12}$ nor $\varepsilon_{K}(Q_{\gamma}%
)\sim\operatorname{Im}\mu_{12}$ can compete with the non-radiative $\Delta
S=2$ processes, even in the presence of NP in $Q_{\gamma}^{\pm}$.

\subsubsection{Gluonic penguin operators}

In complete analogy with the electromagnetic operators, gluonic FCNC are
described by effective operators of dimensions greater than four. For
instance, the chromomagnetic operators producing either a real or a virtual
gluon are%
\begin{equation}
\mathcal{H}_{eff}^{\gamma}=C_{g}^{\pm}Q_{g}^{\pm}+h.c.\;,\;\;\;Q_{g}^{\pm
}=\frac{g}{16\pi^{2}}(\bar{s}_{L}\sigma^{\alpha\beta}t^{a}d_{R}\pm\bar{s}%
_{R}\sigma^{\alpha\beta}t^{a}d_{L})G_{\alpha\beta}^{a}\;. \label{ChromoMag}%
\end{equation}
The chromoelectric operators $Q_{g^{\ast}}^{\pm}$, whose form can easily be
deduced from Eq.~(\ref{HeffG2}), contribute only for a virtual gluon.

\begin{figure}[t]
\centering       \includegraphics[width=3.5cm]{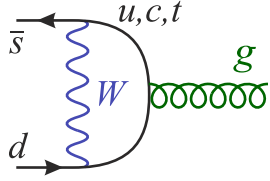}  \caption{The
gluonic penguin in the SM.}%
\label{Fig8}%
\end{figure}

In the SM, both $Q_{g}^{\pm}$ and $Q_{g^{\ast}}^{\pm}$ arise from the diagram
shown in Fig.~\ref{Fig8}. As for $Q_{\gamma}^{\pm}$, the former are suppressed
by the light-quark chirality flips hence completely negligible, but the
chromoelectric operators are sizeable and enter into the initial conditions
for the four-quark operators~\cite{BuchallaBL96}. They are thus hidden inside
the weak low-energy constants in Eq.~(\ref{LECs}), together with the hadronic
virtual photon and $Z$ penguins (see Fig.~\ref{Fig2}).

The chromomagnetic operators are not included in the standard OPE, since they
are negligible in the SM. But being of dimension-five, they could get
significantly enhanced by NP. This would have two main effects. First, through
the OPE mixing\footnote{The $Q_{\gamma}^{\pm}\rightarrow Q_{g}^{\pm}$ mixings
are not included in Eq.~(\ref{OPEgg}), even though they become relevant if
$C_{\gamma}^{\pm}\gg C_{g}^{\pm}$. However, such effects are presumably
LD-dominated, and thus were already included in Eq.~(\ref{AmpliVirt}) together
with $Q_{\gamma}^{\pm}\rightarrow Q_{1,...,10}$.}, $Q_{g}^{\pm}$ generate
$Q_{\gamma}^{\pm}$. When both arise at a high-scale $\mu_{NP}\gtrsim M_{W}$,
assuming only the SM colored particle content, neglecting the mixings with the
four-quark operators, and working to LO~\cite{BurasCIRS99}:%
\begin{align}
C_{\gamma}^{\pm}(\mu_{c}) &  =\eta^{2}\left[  C_{\gamma}^{\pm}(\mu
_{NP})+8(1-\eta^{-1})C_{g}^{\pm}(\mu_{NP})\right]  \;,\;\;C_{g}^{\pm}(\mu
_{c})=\eta C_{g}^{\pm}(\mu_{NP})\;,\nonumber\\
\eta &  \equiv\eta(\mu_{NP})=\left(  \frac{\alpha_{S}(\mu_{NP})}{\alpha
_{S}(m_{t})}\right)  ^{2/21}\left(  \frac{\alpha_{S}(m_{t})}{\alpha_{S}%
(m_{b})}\right)  ^{2/23}\left(  \frac{\alpha_{S}(m_{b})}{\alpha_{S}(\mu_{c}%
)}\right)  ^{2/25}\;.\label{OPEgg}%
\end{align}
Numerically, $\eta(\mu)=0.90,0.89,0.88$ for $\mu=0.1,0.5,1$ TeV, respectively.
Indirectly, all the bounds on $C_{\gamma}^{\pm}$ can thus be translated as
bounds on $C_{g}^{\pm}$.

However, there is another more direct impact of $Q_{g}^{\pm}$ on phenomenology
since it contributes to $K\rightarrow\pi\pi$, hence to $\varepsilon^{\prime}%
$~\cite{BurasCIRS99}%
\begin{equation}
\operatorname{Re}(\varepsilon^{\prime}/\varepsilon)_{g}=\frac{11}{64\pi^{2}%
}\frac{\omega}{|\varepsilon||\operatorname{Re}A_{0}|}\frac{m_{\pi}^{2}%
m_{K}^{2}}{F_{\pi}(m_{s}+m_{d})}\eta B_{G}\operatorname{Im}C_{g}^{-}%
\approx3B_{G}\frac{\operatorname{Im}C_{g}^{-}}{G_{F}m_{K}}\;, \label{boundCg}%
\end{equation}
with, neglecting $\Delta I=3/2$ contributions, $|\operatorname{Re}A_{0}%
|=\sqrt{2}F_{\pi}(m_{K}^{2}-m_{\pi}^{2})|\operatorname{Re}G_{8}|$ and $F_{\pi
}=92.4$ MeV. The hadronic parameter $B_{G}$ parametrizes the departure of
$\langle(\pi\pi)_{0}|Q_{g}^{-}|K^{0}\rangle$ from the chiral quark model, and
lies presumably in the range $1\rightarrow4$~\cite{BurasCIRS99}. Given that
the SM prediction for $\operatorname{Re}(\varepsilon^{\prime}/\varepsilon)$ is
rather close to $\operatorname{Re}(\varepsilon^{\prime}/\varepsilon)^{\exp}%
$~\cite{BurasJ04}, but its uncertainty is itself of the order of
$\operatorname{Re}(\varepsilon^{\prime}/\varepsilon)^{\exp}$, we simply impose
that $|\operatorname{Re}(\varepsilon^{\prime}/\varepsilon)_{g}|\leq
\operatorname{Re}(\varepsilon^{\prime}/\varepsilon)^{\exp}$, which gives,%
\begin{equation}
\frac{|\operatorname{Im}C_{g}^{-}|}{G_{F}m_{K}}\lesssim5\times10^{-4}\;.
\label{BdCg}%
\end{equation}
For comparison, imposing that $\left|  \operatorname{Re}A_{0}\right|  _{g}$ is
at most of the order of $\left|  \operatorname{Re}A_{0}\right|  ^{\exp}$ gives
the much looser constraint $|\operatorname{Re}C_{g}^{-}|/G_{F}m_{K}\lesssim
10$. Note, however, that the bound~(\ref{BdCg}) is not to be taken too
strictly. First, the $B_{G}$ parameter is set to $1$, but could be slightly
smaller or bigger. Second, $Q_{g}^{\pm}$ is not the only FCNC affecting
$\operatorname{Re}(\varepsilon^{\prime}/\varepsilon)$ (see Fig.~\ref{Fig2}).
This bound could get relaxed in the presence of NP in the other penguins. This
will be analyzed in more details in the next section.

\section{New Physics effects}

In most models of New Physics, new degrees of freedom and additional sources
of flavor breaking offer alternative mechanisms to induce the FCNC
transitions. The goal of the present section is to quantify the possible
phenomenological impacts of NP in the dimension-five magnetic operators
$Q_{\gamma}^{\pm}$ of Eq.~(\ref{HeffG}). As discussed in details in the
previous sections, CP-conserving processes are fully dominated by the SM
long-distance contributions. So, throughout this section, we concentrate
exclusively on CP-violating observables, from which the short-distance physics
can be more readily accessed along with possible signals of NP.

The cleanest observables to identify a large enhancement of $Q_{\gamma}^{\pm}$
are the direct CP-asymmetries in $K\rightarrow\pi\pi\gamma$ and $K\rightarrow
(\gamma\gamma)_{||}$, which would then satisfy%
\begin{equation}
\frac{1}{3}|\varepsilon_{+0\gamma}^{\prime}(Q_{\gamma}^{-})|\approx
5|\varepsilon_{+-\gamma}^{\prime}(Q_{\gamma}^{-})|\approx3|\varepsilon
_{||}^{\prime}(Q_{\gamma}^{-})|\approx\frac{|\operatorname{Im}C_{\gamma}^{-}%
|}{G_{F}m_{K}}\;. \label{CmAsym}%
\end{equation}
Indeed, the contributions from the four-quark operators (QCD and electroweak
penguins) is small and under control,
\begin{equation}
\frac{3\omega}{2\sqrt{2}}|\varepsilon_{+0\gamma}^{\prime}(Q_{3,...,10}%
)|\approx\frac{5}{2}|\varepsilon_{+-\gamma}^{\prime}(Q_{3,...,10}%
)|\approx|\varepsilon_{||}^{\prime}(Q_{3,...,10})|\approx|\varepsilon^{\prime
}|\;,
\end{equation}
with $\omega=1/22.4$. By using the experimental $\varepsilon^{\prime}$ value,
these estimates are independent of the presence of NP in $Q_{3,...,10}$. On
the other hand, the $K_{S,L}\rightarrow(\gamma\gamma)_{\perp}$ asymmetry is
very sensitive to $\Omega$, representing the ratio of the electroweak to the
QCD penguin contributions in $\varepsilon^{\prime}$:%
\begin{equation}
\varepsilon_{\perp}^{\prime}(Q_{3,...,10})=-i\frac{\operatorname{Im}A_{0}%
}{\operatorname{Re}A_{0}}=i\frac{\sqrt{2}|\varepsilon^{\prime}|}%
{\omega(1-\Omega)}\;\;,\;\;\;\;|\varepsilon_{\perp}^{\prime}(Q_{\gamma}%
^{+})|\approx\frac{1}{2}\frac{|\operatorname{Im}C_{\gamma}^{+}|}{G_{F}m_{K}%
}\;. \label{CpAsym}%
\end{equation}
So, knowing the impact of $Q_{\gamma}^{+}$, the asymmetry $\varepsilon_{\perp
}^{\prime}$ can be used to extract the otherwise inaccessible QCD penguin
contributions to $\varepsilon^{\prime}$.

The experimental information on these four asymmetries is however limited,
with only the loose bound (\ref{ep0gExp}) on $\varepsilon_{+0\gamma}^{\prime}$
and (\ref{PMGexp}) on $\varepsilon_{+-\gamma}^{\prime}$ currently available.
So, to get some information on $Q_{\gamma}^{\pm}$, two routes will be explored.

\begin{figure}[t]
\centering                   \includegraphics[width=9.5cm]{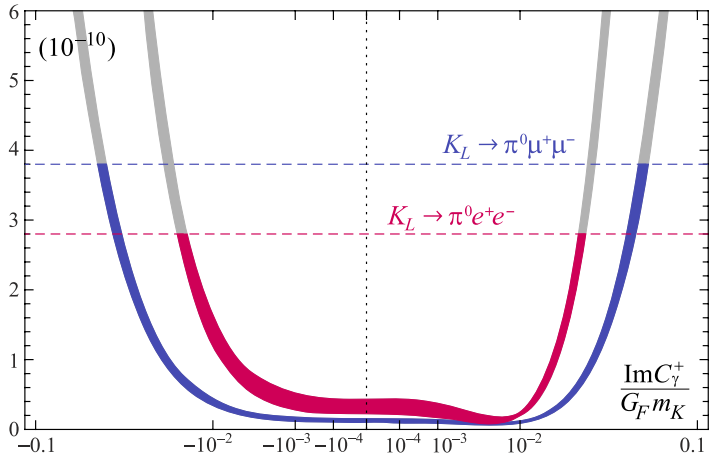}
\caption{The sensitivity of the $K_{L}\rightarrow\pi^{0}\ell^{+}\ell^{-}$
decays to the magnetic penguin operator $Q_{\gamma}^{+}$, in the absence of
any other source of NP. These curves are actually parabolas, but blown out to
emphasize the small $\operatorname{Im}C_{\gamma}^{+}/G_{F}m_{K}$ region (whose
SM value is in the $10^{-5}$ range). The horizontal lines signal the
experimental bounds on $K_{L}\rightarrow\pi^{0}\ell^{+}\ell^{-}$. The contours
stand for $90\%$ confidence regions given the current theoretical errors in
Eq.~(\ref{MasterKL}). Their apparent thinning as $|\operatorname{Im}C_{\gamma
}^{+}|$ increases is purely optical, except just below $10^{-2}$ where the
$Q_{\gamma}^{+}$ contribution precisely cancel out with the SM one in the
vector current (positive DCPV--ICPV interference is assumed).}%
\label{FigKpll}%
\end{figure}

First, we can use the $K_{L}\rightarrow\pi^{0}\ell^{+}\ell^{-}$ decay rates,
for which the experimental bounds are currently in the $10^{-10}$ range. As
shown in Fig.~\ref{FigKpll}, these modes are rather sensitive to $Q_{\gamma
}^{+}$ once $|\operatorname{Im}C_{\gamma}^{+}|/G_{F}m_{K}$ is above a few
$10^{-3}$. In the absence of any other source of NP, the experimental
bounds~(\ref{rareKexp}) give
\begin{subequations}
\label{BDnaive}%
\begin{align}
K_{L}\overset{}{\rightarrow}\pi^{0}e^{+}e^{-} &  \Rightarrow
\;-0.018<\frac{\operatorname{Im}C_{\gamma}^{+}}{G_{F}m_{K}}<+0.030\;,\\
K_{L}\overset{}{\rightarrow}\pi^{0}\mu^{+}\mu^{-} &  \Rightarrow
\;-0.050<\frac{\operatorname{Im}C_{\gamma}^{+}}{G_{F}m_{K}}<+0.063\;.
\end{align}
\end{subequations}
To compare with the direct CP-asymmetries~(\ref{CmAsym}), sensitive to
$Q_{\gamma}^{-}$, we first need to study how NP could affect the relationship
between $Q_{\gamma}^{+}$ and $Q_{\gamma}^{-}$. If the SM relation $C_{\gamma
}^{+}\approx-C_{\gamma}^{-}$ survives, the direct CP-asymmetries could be
relatively large, with for example $-8\%<\varepsilon_{+0\gamma}^{\prime}<5\%$
from $K_{L}\rightarrow\pi^{0}e^{+}e^{-}$. Then, since NP can enter in
$K_{L}\rightarrow\pi^{0}\ell^{+}\ell^{-}$ through other FCNC, for example by
affecting the electroweak penguins, we must also study their possible
interferences with $Q_{\gamma}^{+}$, and quantify how broadly the
bounds~(\ref{BDnaive}) could get relaxed.

A second route is to use $\varepsilon^{\prime}$. Indeed, in many NP models,
the magnetic operators $Q_{\gamma}^{\pm}$ are accompanied by chromomagnetic
operators $Q_{g}^{\pm}$, which contribute directly to $\varepsilon^{\prime}$,
\begin{equation}
\operatorname{Re}(\varepsilon^{\prime}/\varepsilon)_{g}\approx3B_{G}%
\frac{\operatorname{Im}C_{g}^{-}}{G_{F}m_{K}}\;, \label{QgE}%
\end{equation}
with $B_{G}$ the hadronic bag parameter a priori of $\mathcal{O}(1)$. If the
Wilson coefficients of $Q_{\gamma}^{\pm}$ and $Q_{g}^{\pm}$ are similar, the
current measurement $\operatorname{Re}(\varepsilon^{\prime}/\varepsilon
)^{\exp}=(1.65\pm26)\times10^{-3}$~\cite{PDG} imposes strong constraints, and
would naively imply that the direct CP-asymmetries in Eq.~(\ref{CmAsym}) are
at most of $\mathcal{O}(10^{-3})$. However, not only the relationship between
$Q_{g}^{\pm}$ and $Q_{\gamma}^{\pm}$ is model-dependent, but as for
$K_{L}\rightarrow\pi^{0}\ell^{+}\ell^{-}$, many other FCNC enter in
$\varepsilon^{\prime}$ and their possible correlations with $Q_{g}^{\pm}$ must
be analyzed.

The only way to relate the NP occurring in the various FCNC is to adopt a
specific picture for the NP dynamics. Evidently, this cannot be done
model-independently. Instead, the strategy will be to classify the models into
broad classes, and within each class, to stay as model-independent as
possible. In practice, these classes are in one-to-one correspondence with the
choice of basis made for the effective semileptonic FCNC operators. Once a
basis is chosen, bounds on the Wilson coefficients of these operators are
derived by turning them on one at a time. In this way, fine-tunings between
the chosen operators are explicitly ruled out. This is where the
model-dependence enters~\cite{Basis}. On the other hand, the magnetic
operators are kept on at all times, since it is precisely their interference
with the semileptonic FCNC which we want to resolve. Note that the alternative
procedure of performing a full scan over parameter space is (usually) basis
independent, but we prefer to avoid that method as the many possible
fine-tuning among the semileptonic operators would obscure those with the
magnetic ones. Further, we will see that with our method, it is possible to
get additional insight because the bounds do depend on the basis, and thus
allow discriminating among the NP scenarios.

\subsection{Model-independent analysis}

The most model-independent operator basis is the one minimizing the
interferences between the NP contributions in physical
observables~\cite{Basis}. It is the one in Eq.~(\ref{HFF}), which we reproduce
here for convenience:
\begin{gather}
\mathcal{H}_{\text{Pheno}}=-\frac{G_{F}\alpha}{\sqrt{2}}\sum_{\ell=e,\mu,\tau
}(C_{\nu,\ell}\;Q_{\nu,\ell}+C_{V,\ell}\;Q_{V,\ell}+C_{A,\ell}\;Q_{A,\ell
})+C_{\gamma}^{\pm}Q_{\gamma}^{\pm}+h.c.\;,\label{BasisPheno}\\
Q_{V,\ell}=\bar{s}\gamma^{\mu}d\otimes\bar{\ell}\gamma_{\mu}\ell
\;,\;\;Q_{A,\ell}=\bar{s}\gamma^{\mu}d\otimes\bar{\ell}\gamma_{\mu}\gamma
_{5}\ell\;,\;\;Q_{\nu,\ell}=\bar{s}\gamma^{\mu}d\otimes\bar{\nu}_{\ell}%
\gamma_{\mu}(1-\gamma_{5})\nu_{\ell}\;,\nonumber\\
Q_{\gamma}^{\pm}=\frac{Q_{d}e}{16\pi^{2}}\;(\bar{s}_{L}\sigma^{\mu\nu}d_{R}%
\pm\bar{s}_{R}\sigma^{\mu\nu}d_{L})\,F_{\mu\nu}\;.\nonumber
\end{gather}
The four-fermion operators do not interfere in the rates since they produce
different final states, while $Q_{\gamma}^{+}$ and $Q_{\gamma}^{-}$ have
opposite CP-properties (see Table~\ref{TableWindows}). On the other hand,
$Q_{\gamma}^{\pm}$ and $Q_{V,\ell}\ni Q_{\gamma^{\ast}}^{\pm}$ involve an
intermediate photon hence necessarily interfere. Note that the coefficients in
Eq.~(\ref{BasisPheno}) are understood to be purely induced by the NP: the SM
contributions have to be added separately.

Given the current data, the bounds on the CP-violating parts of the Wilson
coefficients are (we define $\rho^{-1}\equiv21.3G_{F}m_{K}$ from
Eq.~(\ref{shift}))%
\begin{equation}%
\begin{array}
[c]{rc}%
K^{+}\rightarrow\pi^{+}\pi^{0}\gamma\Rightarrow & -160<\rho\operatorname{Im}%
C_{\gamma}^{-}<80\;,\smallskip\\
K_{L}\rightarrow\pi^{0}e^{+}e^{-}\Rightarrow & -14<\operatorname{Im}%
C_{V,e}-\rho\operatorname{Im}C_{\gamma}^{+}<\;\;8\oplus\lbrack
-10<\operatorname{Im}C_{A,e}<11\wedge\;\;-8<\rho\operatorname{Im}C_{\gamma
}^{+}<14]\;,\smallskip\\
K_{L}\rightarrow\pi^{0}\mu^{+}\mu^{-}\Rightarrow & -29<\operatorname{Im}%
C_{V,\mu}-\rho\operatorname{Im}C_{\gamma}^{+}<24\oplus\lbrack
-16<\operatorname{Im}C_{A,\mu}<18\wedge-24<\rho\operatorname{Im}C_{\gamma}%
^{+}<29]\;,\smallskip\\
K^{+}\rightarrow\pi^{+}\nu\bar{\nu}\;\Rightarrow & -14<\operatorname{Im}%
C_{\nu,\ell}<17\;\;(\ell=e\oplus\mu\oplus\tau)\;.
\end{array}
\label{MIbnd}%
\end{equation}
All the numbers are in unit of $10^{-4}$. The symbol ``$\oplus$'' stands for
the exclusive alternative, since e.g. $C_{A,\ell}$ and $C_{V,\ell}$ are not
turned on simultaneously, while ``$\Lambda$'' means that the bounds are
correlated, i.e. the coefficients fall within an elliptical contour in the
corresponding plane. For comparison, $\operatorname{Im}C_{V,\ell}^{\text{SM}}%
$, $\operatorname{Im}C_{A,\ell}^{\text{SM}}$ and $\operatorname{Im}C_{\nu
,\ell}^{\text{SM}}$ are all around $10^{-4}$. For the magnetic operators, the
SM value in Eq.~(\ref{SM}) implies $\rho\operatorname{Im}C_{\gamma}^{\pm
,\text{SM}}\approx\mp0.015\operatorname{Im}\lambda_{t}\sim\mathcal{O}%
(10^{-6})$.

For the neutrino modes, NP is separately turned on in each $\operatorname{Im}%
C_{\nu,\ell}$, $\ell=e,\mu,\tau$. Assuming leptonic universality would
decrease the bound by about $\sqrt{3}$ since then all three $C_{\nu,e}%
=C_{\nu,\mu}=C_{\nu,\tau}$ would simultaneously contribute. The direct bounds
on $\operatorname{Im}C_{\nu,\ell}$ from $K_{L}\rightarrow\pi^{0}\nu\bar{\nu}$
are currently not competitive, so the experimental bound on the $K^{+}%
\rightarrow\pi^{+}\nu\bar{\nu}$ mode is used setting $\operatorname{Re}%
C_{\nu,\ell}=0$. The maximal value for $K_{L}\rightarrow\pi^{0}\nu\bar{\nu}$
can then be predicted%
\begin{equation}
\mathcal{B}(K_{L}\rightarrow\pi^{0}\nu\bar{\nu})<1.2\times10^{-9}\;,
\label{KLnn}%
\end{equation}
which corresponds to a saturation of the Grossman-Nir Bound \cite{GrossmanN97}
(including the isospin breaking effects in the vector form-factor, but
forbidding a destructive interference between the CP-conserving SM and NP
contributions since $\operatorname{Re}C_{\nu,\ell}=0$). This is more than an
order of magnitude below the current experimental limit, but about 50 times
larger than the SM prediction.

For $K_{L}\rightarrow\pi^{0}\ell^{+}\ell^{-}$, the bound on the vector current is less strict than on the axial-vector current because of the interference with the indirect CP-violating contribution. The theoretically favored case of positive DCPV-ICPV interference is assumed (relaxing this would not change much the numbers). Finally, the impact of $Q_{\gamma}^{-}$ on $\varepsilon^{\prime}$ is estimated to be below $30\%$ of its experimental value given the bound from $K^{+}\rightarrow\pi^{+}\pi^{0}\gamma$, see Eq.~(\ref{Bound}), hence is neglected.

\begin{figure}[t]
\centering       \includegraphics[width=9.5cm]{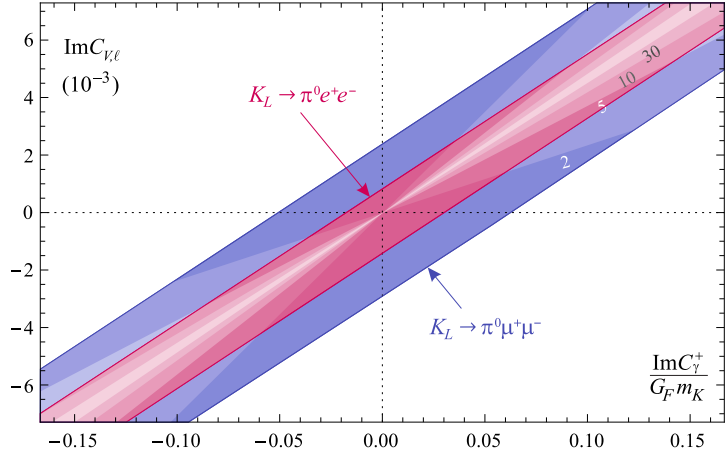}  \caption{The band in
the $\operatorname{Im}C_{V,\ell}-\operatorname{Im}C_{\gamma}^{+}$ plane
allowed by the $K_{L}\rightarrow\pi^{0}\ell^{+}\ell^{-}$ experimental bounds.
The degree of fine-tuning is represented by the lighter areas, where
$|\operatorname{Im}C_{V,\ell}-\rho\operatorname{Im}C_{\gamma}^{+}%
|/|\rho\operatorname{Im}C_{\gamma}^{+}|<1/r$, $r=2,5,10,30$. Assuming
$\operatorname{Im}C_{\gamma}^{+}=-\operatorname{Im}C_{\gamma}^{-}$,
$\varepsilon_{+0\gamma}^{\prime}$ could thus reach its $K^{+}\rightarrow
\pi^{+}\pi^{0}\gamma$ experimental bound for $r\gtrsim5$.}%
\label{Fig11}%
\end{figure}

To resolve the bound in the vector current and thereby disentangle $C_{\gamma
}^{+}$ and $C_{V,\ell}$, one is forced to specify at which level a destructive
interference becomes a fine-tuning, see Fig.~\ref{Fig11}. This introduces some
model-dependence since a specific NP model could generate $Q_{\gamma}^{\pm}$
and $Q_{V,\ell}$ (or $Q_{\gamma^{\ast}}^{\pm}$) coherently. In this respect,
it should be noted that the basis of four-fermion operators in
Eq.~(\ref{BasisPheno}) is not complete. It lacks the scalar, pseudoscalar,
tensor and pseudotensor four-fermion operators. Naively, all these operators
produce the lepton pair in different states and do not interfere in the
rate~\cite{MST06}. Introducing large NP in any of them would thus render the
bounds~(\ref{MIbnd}) weaker. There is however one exception. In $K_{L}%
\rightarrow\pi^{0}\ell^{+}\ell^{-}$, the tensor operators,%
\begin{equation}
Q_{T,\ell}=\bar{s}\sigma^{\mu\nu}d\otimes\bar{\ell}\sigma_{\mu\nu}\ell\;,
\label{TensorOP}%
\end{equation}
do produce the leptons in the same $1^{--}$ state as $Q_{V,\ell}$ and
$Q_{\gamma}^{+}$~\cite{MST06}. So, effectively, $Q_{T,\ell}$ can be absorbed
into $Q_{V,\ell}$. But then, owing to their similar structures, it is not
impossible that $Q_{\gamma}^{\pm}$ and $Q_{T,\ell}$ are generated
simultaneously, and thus that $Q_{\gamma}^{\pm}$ is tightly correlated to this
effective $Q_{V,\ell}$.

In the next two sections, several NP scenarios are considered, in order to
investigate under which circumstances the bounds on $C_{\gamma}^{+}$ and
$C_{V,\ell}$ can be resolved. Of course, ultimately, better measurements of
the direct CP-asymmetries are the cleanest option to get to $C_{\gamma}^{\pm}%
$. But before pushing for an experimental effort in that direction, it is
essential to have a more precise idea of their maximal sizes under a large
spectrum of NP scenarios.

\subsubsection{Hadronic current and Minimal Flavor Violation}

The NP scenarios are organized into two broad classes according to the way the
leptonic currents of the effective operators are parametrized. So, before
entering that discussion, let us consider here their hadronic parts, whose
generic features transcend the various scenarios.

Only the vector current $\bar{s}\gamma_{\mu}d$ enters in Eq.~(\ref{BasisPheno}%
) because the axial-vector current $\bar{s}\gamma_{\mu}\gamma_{5}d$ drops out
of the $K\rightarrow\pi\nu\bar{\nu}$ and $K_{L}\rightarrow\pi^{0}\ell^{+}%
\ell^{-}$ matrix elements. It would thus be equivalent to replace $\bar
{s}\gamma_{\mu}d$ by the $SU(2)_{L}\otimes U(1)_{Y}$ invariant forms $\bar
{Q}\gamma_{\mu}Q$ and $\bar{D}\gamma_{\mu}D$, with $Q^{T}=(u,d)_{L}$ and
$D=d_{R}$. By contrast, the magnetic operators require an extra Higgs doublet
field to reach an $SU(2)_{L}$ invariant form:
\begin{equation}
Q_{\gamma}^{\pm}\sim(\bar{Q}\sigma^{\mu\nu}DH\pm\bar{D}\sigma^{\mu\nu}%
QH^{\ast})\,F_{\mu\nu}\;. \label{Higgs}%
\end{equation}
After electroweak symmetry breaking, this operator collapses to that in
Eq.~(\ref{HeffG2}). Consequently, if the NP respects the $SU(2)_{L}\otimes
U(1)_{Y}$ symmetry, $Q_{\gamma}^{\pm}$ and semileptonic operators are equally
suppressed by the NP scale since they are all of dimension six. However, the
magnetic operators are a priori much more sensitive to the electroweak
symmetry breaking mechanism, so that the scaling between the two types of
operators cannot be assessed model-independently. Its phenomenological
extraction is thus important, and could help discriminate among models.

The effective operators in Eq.~(\ref{BasisPheno}) induce the $s\rightarrow d$
flavor transition, while the leptonic currents (or the photon) are flavor
diagonal. Model-independently, the underlying gauge symmetry properties of an
operator does not preclude anything about its flavor-breaking capabilities.
However, the situation changes if we ask for the NP to have no more sources of
flavor breaking than the SM. This is the Minimal Flavor Violation
hypothesis~\cite{MFV}. For the operators at hand, it implies that the hadronic
currents scale as%
\begin{equation}
\bar{Q}^{I}\gamma_{\mu}(\mathbf{Y}_{u}^{\dagger}\mathbf{Y}_{u})^{IJ}%
Q^{J}\;,\;\;\bar{D}^{I}\gamma_{\mu}(\mathbf{Y}_{d}\mathbf{Y}_{u}^{\dagger
}\mathbf{Y}_{u}\mathbf{Y}_{d}^{\dagger})^{IJ}D^{J}\;,\;\;\bar{Q}^{I}%
\sigma^{\mu\nu}(\mathbf{Y}_{u}^{\dagger}\mathbf{Y}_{u}\mathbf{Y}_{d}%
)^{IJ}D^{J}\;, \label{MFV0}%
\end{equation}
with $v\mathbf{Y}_{d}=\mathbf{m}_{d}$, $v\mathbf{Y}_{u}=\mathbf{m}_{u}V$, $\mathbf{m}_{u,d}$ the diagonal quark mass matrices, and $v$ the Higgs vacuum expectation value. The CKM matrix $V$ is put in
$\mathbf{Y}_{u}$ so that the down-quark fields in the operators of Eq.~(\ref{BasisPheno}) are mass eigenstates. Also, we limit the MFV expansions to the leading sources of flavor-breaking (i.e., minimal number of $\mathbf{Y}_{u,d}$) for simplicity.

Under MFV, the NP operators acquire many SM-like properties. First, $\bar
{D}\gamma_{\mu}D$ is doubly suppressed by the light quark Yukawa couplings,
and is thus not competitive with $\bar{Q}\gamma_{\mu}Q$. Second, the chirality
flip in $\bar{Q}^{I}\sigma^{\mu\nu}D^{J}$ comes from the external light quark
masses, and are thus significantly suppressed. Finally, the $s\rightarrow d$
transitions become correlated to the $b\rightarrow d$ and $b\rightarrow s$
transitions since
\begin{equation}
v^{2}(\mathbf{Y}_{u}^{\dagger}\mathbf{Y}_{u})^{IJ}\approx m_{t}^{2}%
V_{3I}^{\dagger}V_{3J}^{\,}\;. \label{MFV1}%
\end{equation}
Of course, this correlation is not always strict as additional terms in the
MFV expansion can be relevant. Still, it drives the overall scale of the
observables in each sector.

We do not intend to perform a full MFV analysis here. Instead, our goal is to
quantify, under the MFV ansatz, the maximal NP effects $Q_{\gamma}^{\pm}$
could induce given the current situation in $b\rightarrow s\gamma$. From
Eqs.~(\ref{Higgs},~\ref{MFV0},~\ref{MFV1}), discarding $m_{s(d)}$ against
$m_{b(s)}$,%
\begin{equation}
Q_{\gamma}^{\pm}|_{d_{R}^{I}\rightarrow d_{L}^{J}}\sim C_{7\gamma}(\mu
_{EW})\;(\bar{Q}^{J}\sigma^{\mu\nu}(\mathbf{Y}_{u}^{\dagger}\mathbf{Y}%
_{u}\mathbf{Y}_{d})^{JI}D^{I})\;H\;F_{\mu\nu}\;\;\;\Rightarrow\frac{Q_{\gamma
}^{\pm}|_{s\rightarrow d}}{Q_{\gamma}^{\pm}|_{b\rightarrow s}}\sim
\frac{V_{ts}^{\dagger}V_{td}\;m_{s}}{V_{ts}^{\dagger}V_{tb}\;m_{b}}\;.
\label{MFV}%
\end{equation}
The flavor-universality of the Wilson coefficient $C_{7\gamma}(\mu_{EW})$
embodies the MFV hypothesis. The NP shift still allowed by $b\rightarrow
s\gamma$ is~\cite{HurthIKM08}
\begin{equation}
\delta C_{7\gamma}(\mu_{EW})=[-0.14,0.06]\cup\lbrack1.42,1.62]\;,
\label{Shift}%
\end{equation}
for constructive and destructive interference with the SM contributions. The
latter has a lower probability, and would require significant cancellations
among the NP effects in $B\rightarrow X_{s}\ell^{+}\ell^{-}$. From
Eq.~(\ref{SM}), and including the LO QCD reduction~\cite{BuchallaBL96}, such a
shift can be written in our conventions as%
\begin{equation}
\frac{\left.  \operatorname{Im}C_{\gamma}^{\pm}\right|  _{\text{MFV}}}%
{G_{F}m_{K}}-\frac{\left.  \operatorname{Im}C_{\gamma}^{\pm}\right|
_{\text{SM}}}{G_{F}m_{K}}\approx\pm\frac{2}{3}\,\operatorname{Im}\lambda
_{t}\,\delta C_{7\gamma}(\mu_{EW})\;.
\end{equation}
For comparison, the SM prediction is $\mp0.31(8)\times\operatorname{Im}%
\lambda_{t}$. So, there would be no visible effects for $\delta C_{7\gamma
}(\mu_{EW})\in\lbrack-0.14,0.06]$, and at most a factor four enhancement for
$\delta C_{7\gamma}(\mu_{EW})\in\lbrack1.42,1.62]$.

This is hardly sufficient to push any of the asymmetries within the
experimentally accessible range, while the impact on $K_{L}\rightarrow\pi
^{0}\ell^{+}\ell^{-}$ would be buried in the theoretical errors, see
Fig.~\ref{FigKpll}. However, it is well-known that MFV is particularly effective
for $K$ physics since it suppresses the NP contributions by the small
$V_{ts}^{\ast}V_{td}\sim10^{-4}$. So, this is the best place to test MFV. A
deviation with respect to the strict\ ansatz~(\ref{MFV}) could lead to visible effects.

\subsection{Tree-level FCNC}

The basis of operators in Eq.~(\ref{BasisPheno}) maximally breaks the
$SU(2)_{L}\otimes U(1)_{Y}$ symmetry. Neutrinos are completely decoupled from
the charged leptons, and the vector and axial-vector operators (as well as
$Q_{\gamma}^{+}$ and $Q_{\gamma}^{-}$) maximally mix currents of opposite
chiralities. To be specific, the $SU(2)_{L}\otimes U(1)_{Y}$ invariant basis
\cite{BuchmullerW86} is, after projecting the hadronic currents of
semileptonic operators on their vector components,%
\begin{gather}
\mathcal{H}_{\text{Gauge}}=-\frac{G_{F}\alpha}{\sqrt{2}}\sum_{\ell=e,\mu,\tau
}(C_{L,\ell}\;Q_{L,\ell}+C_{L,\ell}^{\prime}\;Q_{L,\ell}^{\prime}+C_{R,\ell
}\;Q_{R,\ell})+C_{\gamma}^{L,R}Q_{\gamma}^{L.R}+h.c.\;,\label{LLRbasis}\\
Q_{L}\equiv\bar{s}\gamma^{\mu}d\otimes\bar{L}\gamma_{\mu}L\;,\;\;Q_{L}%
^{\prime}\equiv\bar{s}\gamma^{\mu}d\otimes\bar{L}\gamma_{\mu}\sigma
^{3}L\;,\;\;Q_{R}\equiv\bar{s}\gamma^{\mu}d\otimes\bar{E}\gamma_{\mu
}E\;\;,\nonumber\\
Q_{\gamma}^{L}=\frac{Q_{d}e}{16\pi^{2}v}\;\bar{s}_{R}\sigma^{\mu\nu}%
d_{L}\,H^{\ast}\,F_{\mu\nu}\;,\;\;\,Q_{\gamma}^{R}=\frac{Q_{d}e}{16\pi^{2}%
v}\;\bar{s}_{L}\sigma^{\mu\nu}d_{R}\,H\,F_{\mu\nu}\;,\nonumber
\end{gather}
with $L^{T}=(\nu_{\ell},\ell)_{L}$ and $E=\ell_{R}$. It is related to the
phenomenological basis (\ref{BasisPheno}) through nearly democratic
transformations%
\begin{equation}
\left(
\begin{array}
[c]{c}%
C_{\nu,\ell}\\
C_{V,\ell}\\
C_{A,\ell}%
\end{array}
\right)  =\frac{1}{2}\left(
\begin{array}
[c]{ccc}%
1 & 1 & 0\\
1 & -1 & 1\\
-1 & 1 & 1
\end{array}
\right)  \left(
\begin{array}
[c]{c}%
C_{L,\ell}\\
C_{L,\ell}^{\prime}\\
C_{R,\ell}%
\end{array}
\right)  \;,\;\left(
\begin{array}
[c]{c}%
C_{\gamma}^{-}\\
C_{\gamma}^{+}%
\end{array}
\right)  =\frac{1}{2}\left(
\begin{array}
[c]{cc}%
1 & -1\\
1 & 1
\end{array}
\right)  \left(
\begin{array}
[c]{c}%
C_{\gamma}^{R}\\
C_{\gamma}^{L}%
\end{array}
\right)  \;, \label{LLR}%
\end{equation}
for each $\ell=e,\mu,\tau$. As in Eq.~(\ref{BasisPheno}), the SM contributions
are not encoded into $\mathcal{H}_{\text{Gauge}}$, and have to be added separately.

The $\mathcal{H}_{\text{Gauge}}$ basis represents a class of models where the
four-fermion effective operators arise entirely from some high-scale
$SU(2)_{L}\otimes U(1)_{Y}$ invariant tree-level interactions. It is
characterized by the correlations it imposes among the phenomenologically non
interfering operators in $\mathcal{H}_{\text{Pheno}}$. A well-known example of
model within this class is the MSSM with R-parity violating
couplings~\cite{RPV}, but more generic leptoquark models are also of this
form~\cite{LQ}. Note that in these two cases, the $Q_{\gamma}^{R,L}$ operators
nevertheless arise only at the loop level since both the photon and the Higgs
(see Eq.~(\ref{Higgs})) have flavor-diagonal couplings at tree-level.

The $\mathcal{H}_{\text{Gauge}}$ basis completely decouples the three leptonic
flavors. This is adequate since generic leptoquark couplings do not respect
leptonic universality. Actually, one would expect that lepton-flavor violating
(LFV) operators should arise, inducing in particular $K\rightarrow(\pi)e\mu$
which corresponds to an $s+\mu\rightarrow d+e$ transition. Those modes are
very constrained experimentally, with bounds often lower that for
lepton-flavor conserving (LFC) modes. So, if LFV and LFC couplings have
similar sizes, there can be no large effects in the LFC modes. However, to
relate the LFC and LFV couplings is far from immediate, and requires some
additional inputs on the dynamics (see e.g. Ref.\cite{LQMFV} for studies
within MFV). So in the present work, we concentrate exclusively on LFC decay
channels. Still, let us emphasize again that leptonic universality is not
expected to hold in the present scenario.

\begin{figure}[t]
\centering       \includegraphics[width=9.5cm]{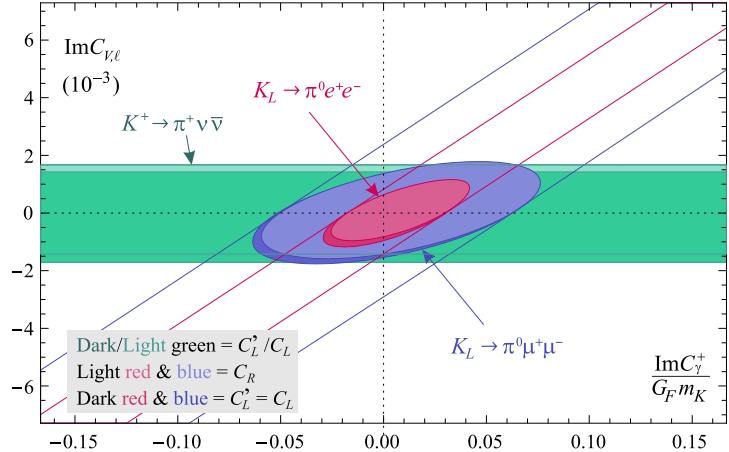}  \caption{Tree-level
FCNC scenario, with $C_{\gamma}^{L,R}$ together with either $C_{L}^{\prime}$,
$C_{L}$, or $C_{R}$ turned on. The diagonal bands show the model-independent
limits of Fig.~\ref{Fig11}.}%
\label{Fig12}%
\end{figure}

Adopting the $SU(2)_{L}\otimes U(1)_{Y}$ invariant basis, the Wilson
coefficients of the semileptonic operators in Eq.~(\ref{LLRbasis}) are turned
on one at a time while either $C_{\gamma}^{L}$ or $C_{\gamma}^{R}$ is kept on.
The bounds are then completely resolved and rather strict (all numbers in
units of $10^{-4}$)%
\begin{equation}%
\begin{array}
[c]{rc}%
K_{L}\rightarrow\pi^{0}e^{+}e^{-}\Rightarrow & -20<(-\operatorname{Im}%
C_{L,e}\oplus\operatorname{Im}C_{L,e}^{\prime}\oplus\operatorname{Im}%
C_{R,e})<24\;\;\wedge\;\;-14<\rho\operatorname{Im}C_{\gamma}^{+}%
<19\;,\smallskip\\
K_{L}\rightarrow\pi^{0}\mu^{+}\mu^{-}\Rightarrow & -33<(-\operatorname{Im}%
C_{L,\mu}\oplus\operatorname{Im}C_{L,\mu}^{\prime}\oplus\operatorname{Im}%
C_{R,\mu})<37\;\;\wedge\;\;-30<\rho\operatorname{Im}C_{\gamma}^{+}%
<36\;,\smallskip\\
K^{+}\rightarrow\pi^{+}\nu\bar{\nu}\;\Rightarrow & -28<(\operatorname{Im}%
C_{L,\ell}\oplus\operatorname{Im}C_{L,\ell}^{\prime})<34\;\;(\ell=e\oplus
\mu\oplus\tau)\;.
\end{array}
\label{BoundLLR}%
\end{equation}
Indeed, $C_{\gamma}^{L}$ and $C_{\gamma}^{R}$ cannot grow unchecked since the
bounds from $K_{L}\rightarrow\pi^{0}(\ell^{+}\ell^{-})_{1^{--}}$ would then
require a large interference with $C_{L}$, $C_{L}^{\prime}$, or $C_{R}$ . But
these Wilson coefficients also contribute either to the neutrino modes (via
$Q_{\nu,\ell}$) or to the axial-vector current (via $Q_{A,\ell}$), which are
separately bounded since non-interfering. So, $C_{L}$, $C_{L}^{\prime}$, or
$C_{R}$ have maximal allowed values, and so have $C_{\gamma}^{L}$ and
$C_{\gamma}^{R}$. The slight asymmetries between minimal and maximal values
are due to the SM contributions. As in Eq.~(\ref{MIbnd}), ``$\oplus$'' denotes
exclusive alternatives and ``$\wedge$'' means that the bounds are correlated.
For example, both $\operatorname{Im}C_{L,\ell}$ and $\operatorname{Im}%
C_{\gamma}^{+}$ cannot reach their maximal values simultaneously, but rather
should fall within the elliptical contour in the $\operatorname{Im}C_{L,\ell}%
$--$\operatorname{Im}C_{\gamma}$ plane, see Fig.~\ref{Fig12}. Looking at these
contours, the bound from $K_{L}\rightarrow\pi^{0}e^{+}e^{-}$ is clearly
tighter than that from $K^{+}\rightarrow\pi^{+}\nu\bar{\nu}$, but
$K_{L}\rightarrow\pi^{0}\mu^{+}\mu^{-}$ is less constraining (except of course
for $C_{R,\mu}$). Thus, as long as leptonic universality is not imposed,
$C_{L,\mu}$ and $C_{L,\mu}^{\prime}$ are only bounded by $K^{+}\rightarrow
\pi^{+}\nu\bar{\nu}$, and $K_{L}\rightarrow\pi^{0}\nu\bar{\nu}$ can reach is
maximal model-independent bound~(\ref{KLnn}). Still, even if $K^{+}%
\rightarrow\pi^{+}\nu\bar{\nu}$ limits $C_{L,\mu}^{(\prime)}$, the
$K_{L}\rightarrow\pi^{0}\mu^{+}\mu^{-}$ rate can always reach its current
experimental limit either through $C_{R,\mu}$ or with the help of $Q_{\gamma
}^{+}$.

The comparison of these bounds with Eq.~(\ref{MIbnd}) illustrates the
consequence of introducing some model-dependence. A scenario with tree-level
FCNC is completely bounded by the data. Further, both $Q_{\gamma}^{L,R}$
contribute to all the decays in Table~\ref{TableWindows}, since $C_{\gamma
}^{-}=+(-)C_{\gamma}^{+}$ when $C_{\gamma}^{R(L)}$ is turned on. Thus, we give
in Eq.~(\ref{BoundLLR}) the bounds on $\operatorname{Im}C_{\gamma}^{+}$, which
directly translates as maximal values for all the direct
CP-asymmetries~(\ref{CmAsym},~\ref{CpAsym}). Since leptonic universality holds
for $Q_{\gamma}^{\pm}$, the tightest bound from $K_{L}\rightarrow\pi^{0}%
e^{+}e^{-}$ must be satisfied, i.e.%
\begin{equation}
-0.03<\frac{\operatorname{Im}C_{\gamma}^{+}}{G_{F}m_{K}}<0.04\;.
\label{ImCgLLR}%
\end{equation}
This represents only a slight extension of the range~(\ref{BDnaive}), obtained
in the absence of NP but in $Q_{\gamma}^{\pm}$.

Scalar or tensor four-fermion operators are not included in
Eq.~(\ref{LLRbasis}), even though they could arise from leptoquark exchanges.
The reason is that they cannot alter the bounds~(\ref{BoundLLR}) if we write
them in $SU(2)_{L}\otimes U(1)_{Y}$ invariant forms. The only four-fermion
operators able to interfere with the vector ones is $Q_{T,\ell}$ of
Eq.~(\ref{TensorOP}), but it must here be replaced by%
\begin{equation}
Q_{T,\ell}^{L}=\bar{s}\sigma^{\mu\nu}d\otimes\bar{L}\sigma_{\mu\nu
}E,\;\;Q_{T,\ell}^{R}=\bar{s}\sigma^{\mu\nu}d\otimes\bar{E}\sigma_{\mu\nu}L\;.
\end{equation}
Each of these operators has a pseudotensor piece $\bar{s}\sigma^{\mu\nu
}d\otimes\bar{\ell}\sigma_{\mu\nu}\gamma_{5}\ell$ which is the only current
able to produce the lepton pair in a $1^{+-}$ state~\cite{MST06}. There is
thus no entanglement, and $Q_{T,\ell}^{L}$ and $Q_{T,\ell}^{R}$ are both
directly bounded by the total $K_{L}\rightarrow\pi^{0}\ell^{+}\ell^{-}$ rate.
Hence numerically, the bounds are similar to those in Eq.~(\ref{BoundLLR}),
and Eq.~(\ref{ImCgLLR}) is not affected.

\subsection{Loop-level FCNC}

For a given lepton flavor, the $\mathcal{H}_{\text{Gauge}}$ basis maximally
couples the semileptonic operators, while the $\mathcal{H}_{\text{Pheno}}$
basis maximally decouples them. An intermediate picture emerges if the NP
generates FCNC only at the loop level. This can be due to some discrete
symmetries (like $R$-parity) or to some generalized GIM mechanism. By
construction, most NP models are of this type, for example the MSSM (see
Sec.~\ref{MSSM}), little Higgs~\cite{LH}, left-right symmetry
\cite{TandeanV00,LR}, fourth generation~\cite{4G}, some extra dimension models
\cite{UED},..., because the loop suppression of the FCNC naturally allows for
the NP particles to be lighter, hopefully within range of the LHC.

An appropriate basis to study this scenario is derived from the situation in
the SM. Indeed, the NP should induce the quark flavor transition $s\rightarrow
d$, but the lepton pair is flavor-diagonal and could still be produced by SM
currents, i.e., $\gamma$ and/or $Z$ bosons. So, in the absence of new vector
interactions, the SM basis is adequate:%
\begin{equation}
\mathcal{H}_{\text{PB}}=-\frac{G_{F}\alpha}{\sqrt{2}}(C_{Z}\;Q_{Z}%
+C_{A}\;Q_{A}+C_{B}\;Q_{B})+C_{\gamma}^{L,R}Q_{\gamma}^{L,R}+h.c.\;,
\label{HPB}%
\end{equation}
with ($s_{W}^{2}\equiv\sin^{2}\theta_{W}=0.231$)
\begin{subequations}
\label{QZAB}%
\begin{align}
Z\text{ penguin}  &  :Q_{Z}\equiv s_{W}^{2}Q_{L}+(1-s_{W}^{2})Q_{L}^{\prime
}+2s_{W}^{2}Q_{R}\;,\\
\gamma^{\ast}\text{ penguin}  &  :Q_{A}\equiv\frac{s_{W}^{2}}{4}(Q_{L}%
-Q_{L}^{\prime}+2Q_{R})\;,\\
W\text{ boxes}  &  :Q_{B}\equiv-\frac{3}{2}Q_{L}-\frac{5}{2}Q_{L}^{\prime}\;.
\end{align}
\end{subequations}
In the presence of NP at the loop-level, it is natural to use the SM-like
$Q_{\gamma}^{L,R}$ operators of Eq.~(\ref{LLRbasis}) since the chirality flip
is a priori different for the $L\rightarrow R$ and $R\rightarrow L$
transitions. Indeed, even though the drastic SM scaling $C_{\gamma}^{L}\sim
m_{s}\gg C_{\gamma}^{R}\sim m_{d}$ needs not survive in the presence of NP, it
is nevertheless expected that $(C_{\gamma}^{L}+C_{\gamma}^{R})/(C_{\gamma}%
^{L}-C_{\gamma}^{R})$ is of $\mathcal{O}(1)$.

The $Q_{L}$, $Q_{L}^{\prime}$ and $Q_{R}$ operators are never independent in
this scenario, even before the electroweak symmetry breaking takes place.
Indeed, though there is a one-to-one correspondence between the $W_{3}^{\mu}$
penguin and $Q_{L}^{\prime}$, the $B^{\mu}$ penguin generates both $Q_{L}$ and
$Q_{R}$ with a fixed (``fine-tuned'') relative coefficient. Combined with
Eq.~(\ref{LLR}), the transformation back to the phenomenological basis is%
\begin{equation}
\left(
\begin{array}
[c]{c}%
C_{\nu,\ell}\\
C_{V,\ell}\\
C_{A,\ell}%
\end{array}
\right)  =\frac{1}{2}\left(
\begin{array}
[c]{ccc}%
1 & 0 & -4\\
4s_{W}^{2}-1 & s_{W}^{2} & 1\\
1 & 0 & -1
\end{array}
\right)  \left(
\begin{array}
[c]{c}%
C_{Z}\\
C_{A}\\
C_{B}%
\end{array}
\right)  \;, \label{AZB}%
\end{equation}
while the $Q_{\gamma}^{L,R}$ operators are related to the $Q_{\gamma}^{\pm}$
as in Eq.~(\ref{LLRbasis}). In the SM without QCD, the semileptonic
coefficients are directly given in terms of the Inami-Lim functions as (beware
that the SM contributions are not included in $\mathcal{H}_{\text{PB}}$, which
parametrizes only the NP contributions) \cite{BuchallaBL96}%
\begin{equation}
C_{A}^{\text{SM}}=-\lambda_{t}D_{0}(x_{t})/\pi s_{W}^{2}\;,\;\;C_{Z}%
^{\text{SM}}=-\lambda_{t}C_{0}(x_{t})/\pi s_{W}^{2}\;,\;\;C_{B}^{\text{SM}%
}=-\lambda_{t}B_{0}(x_{t})/\pi s_{W}^{2}\;, \label{SMZAB}%
\end{equation}
so the $\mathcal{H}_{\text{PB}}$ basis coincides with Penguin-Box expansion of
Ref.~\cite{PB}. Remark that lepton universality is strictly enforced to match
the physical picture of NP entering only for the $s\rightarrow d$ penguins,
but this can easily be lifted. Also, (pseudo)scalar or (pseudo)tensor
operators are not introduced, as none of the SM penguins can produce them.

In the SM, only specific combinations of the electroweak penguins and boxes
are gauge invariant~\cite{PB}. Those combinations are precisely those entering
into $C_{\nu,\ell}$, $C_{V,\ell}$, and $C_{A,\ell}$, since their operators are
directly producing different physical states. Of course, by construction, the
$\mathcal{H}_{\text{Gauge}}$ basis~(\ref{LLRbasis}) is also gauge invariant.
To check this starting with the SM expressions (\ref{SMZAB}) requires first
extending the basis~(\ref{HPB}) to differentiate the boxes according to the
weak isospin state of the lepton pairs~\cite{PB}%
\begin{equation}
Q_{B,\pm1/2}\equiv\frac{1}{2}(Q_{L}\pm Q_{L}^{\prime})\;\;\Leftrightarrow
\;\;\left(
\begin{array}
[c]{c}%
Q_{B}\\
Q_{B}^{\prime}%
\end{array}
\right)  =\left(
\begin{array}
[c]{cc}%
-4 & 1\\
-1 & 1
\end{array}
\right)  \left(
\begin{array}
[c]{c}%
Q_{B,+1/2}\\
Q_{B,-1/2}%
\end{array}
\right)  \;.
\end{equation}
The combination $Q_{B}$ occurs in Eq.~(\ref{QZAB}) because its Wilson
coefficient is separately gauge invariant, see Ref.~\cite{PB}, while
$Q_{B}^{\prime}$ is redundant once the gauge is fixed (we work in the
t'Hooft-Feynman gauge).

So, if one insists on gauge invariance, the $\mathcal{H}_{\text{PB}}$ basis
collapses either onto the $\mathcal{H}_{\text{Pheno}}$ basis or the
$\mathcal{H}_{\text{Gauge}}$ basis. Still, using directly the $\mathcal{H}%
_{\text{PB}}$ basis for parametrizing NP makes sense because its operators
encode different physics~\cite{PB,BurasS98}. Indeed, the dominant NP
contribution in the $Z$ penguin effectively comes from a dimension-four
operator after electroweak symmetry breaking~\cite{NirW97}, while the
$\gamma^{\ast}$ penguin is of dimension six. The box operator $Q_{B}$ is there
to complete the basis, but is rather suppressed in general. Finally, the
magnetic operators $Q_{\gamma}^{L,R}$ are separately gauge-invariant, of
dimension five after the electroweak symmetry breaking, and require a
chirality flip mechanism. So, it is only if there is a new gauge boson, and a
corresponding new penguin not necessarily aligned with the SM structures, that
significant fine-tunings between the $\mathcal{H}_{\text{PB}}$ operators could
arise. This will be dealt with in the next section.

Coincidentally, the $\mathcal{H}_{\text{PB}}$ basis is rather close to the
model-independent basis $\mathcal{H}_{\text{Pheno}}$ because $4s_{W}%
^{2}\approx1$. Indeed, $Q_{Z}$ essentially drops out from the vector current,
leaving $Q_{A}$ and $Q_{\gamma}^{+}$ completely entangled in $K_{L}%
\rightarrow\pi^{0}(\ell^{+}\ell^{-})_{1^{--}}$, while the $Q_{B}$ and $Q_{Z}$
pair is fully resolved through the non-interfering $C_{\nu,\ell}$ and
$C_{A,\ell}$ contributions to $K\rightarrow\pi\nu\bar{\nu}$ and $K_{L}%
\rightarrow\pi^{0}(\ell^{+}\ell^{-})_{1^{++},0^{-+}}$. The main difference
between the $\mathcal{H}_{\text{PB}}$ and $\mathcal{H}_{\text{Pheno}}$ bases
is in the magnetic penguins, since the former relates $Q_{\gamma}^{+}$ and
$Q_{\gamma}^{-}$ through $(C_{\gamma}^{L}+C_{\gamma}^{R})/(C_{\gamma}%
^{L}-C_{\gamma}^{R})\sim\mathcal{O}(1)$.

Turning on $C_{Z}$, $C_{A}$, and $C_{B}$ one at a time while keeping
$C_{\gamma}^{R,L}$ on, the bounds are (in units of $10^{-4}$)%
\begin{equation}%
\begin{array}
[c]{cc}%
K_{L}\rightarrow\pi^{0}e^{+}e^{-}\;\Rightarrow & -14<(s_{W}^{2}%
/2)\operatorname{Im}C_{A}-\rho\operatorname{Im}C_{\gamma}^{+}<8\;\;\oplus
\;\smallskip\\
& [\;-20<(\operatorname{Im}C_{Z}\oplus-\operatorname{Im}C_{B})<24\;\wedge
\;-8<\rho\operatorname{Im}C_{\gamma}^{+}<14\;]\;,\smallskip\smallskip\\
K_{L}\rightarrow\pi^{0}\mu^{+}\mu^{-}\;\Rightarrow & -29<(s_{W}^{2}%
/2)\operatorname{Im}C_{A}-\rho\operatorname{Im}C_{\gamma}^{+}<24\;\;\oplus
\;\smallskip\\
& [\;-33<(\operatorname{Im}C_{Z}\oplus-\operatorname{Im}C_{B})<37\;\wedge
\;-24<\rho\operatorname{Im}C_{\gamma}^{+}<29\;]\;,\smallskip\smallskip\\
K^{+}\rightarrow\pi^{+}\nu\bar{\nu}\;\Rightarrow & -15<(\operatorname{Im}%
C_{Z}\oplus-4\operatorname{Im}C_{B})<21\;.
\end{array}
\label{BoundsZAB}%
\end{equation}
As before, ``$\wedge$'' denotes a contour in the corresponding plane within
the quoted extremes, while ``$\oplus$'' is the exclusive alternative.
Comparing with Eq.~(\ref{MIbnd}), the presence of $Q_{Z}$ or $Q_{B}$ in the
vector current has no impact on the range for $\operatorname{Im}C_{\gamma}%
^{+}$. The bound from $K^{+}\rightarrow\pi^{+}\nu\bar{\nu}$ are stricter
because leptonic universality is now imposed. This actually permits to combine
all the modes, so that $\operatorname{Im}C_{Z}$ is best constrained by
$K_{L}\rightarrow\pi^{0}e^{+}e^{-}$ together with $K^{+}\rightarrow\pi^{+}%
\nu\bar{\nu}$, and $\operatorname{Im}C_{B}$ entirely by $K^{+}\rightarrow
\pi^{+}\nu\bar{\nu}$ thanks to the factor $-4$ in Eq.~(\ref{AZB}). The photon
operators $Q_{A}$ and $Q_{\gamma}^{\pm}$ are unconstrained at this level, so
let us investigate how to resolve this ambiguity within the present scenario.

\subsubsection{Hadronic Electroweak penguins\label{EWloop}}

The photon and the $Z$ boson are also coupled to quarks, and thus affect
$\varepsilon^{\prime}$. So, if NP generates the $Q_{Z}$ and $Q_{A}$ operators
entirely through these SM gauge interactions, we must impose%
\begin{equation}
\operatorname{Re}(\varepsilon^{\prime}/\varepsilon)^{\text{NP}}\approx\pi
s_{W}^{2}\operatorname{Im}\left[  11.3\times C_{Z}+3.1\times C_{A}+2.9\times
C_{B}\right]  \;.\;\label{epeNP}%
\end{equation}
This simplified formula is obtained from Ref.~\cite{BurasJ04} by parametrizing
the NP contributions to the OPE initial conditions at $M_{W}$ in terms of
$C_{Z,A,B}$, setting the bag factors to their large $N_{c}$ values, and taking
$m_{s}(m_{c})=121$ MeV. We do not include the $Q_{\gamma}^{-}$ contribution to
$\varepsilon^{\prime}$ since the experimental bound~(\ref{ImCg}) implies that
it is below 30\% of $\operatorname{Re}(\varepsilon^{\prime}/\varepsilon
)^{\text{exp}}$, see Eq.~(\ref{Bound}). It should be clear that this formula
is only a rough estimate. Deviations with respect to the strict large $N_{c}$
limits are likely, even though the coefficients of $C_{Z}$ and $C_{A}$ are
most dependent on $B_{8}^{3/2}$ which is better known than $B_{6}^{1/2}$ (see
Ref.~\cite{BurasJ04}). To account simultaneously for this uncertainty and that
on the SM contribution, we conservatively require $|\operatorname{Re}%
(\varepsilon^{\prime}/\varepsilon)^{\text{NP}}|<2\operatorname{Re}%
(\varepsilon^{\prime}/\varepsilon)^{\text{exp}}$.

\begin{figure}[t]
\centering
\begin{tabular}
[b]{ll}%
$a.\;\text{\raisebox{-4.5cm}{\includegraphics[width=7.7cm]{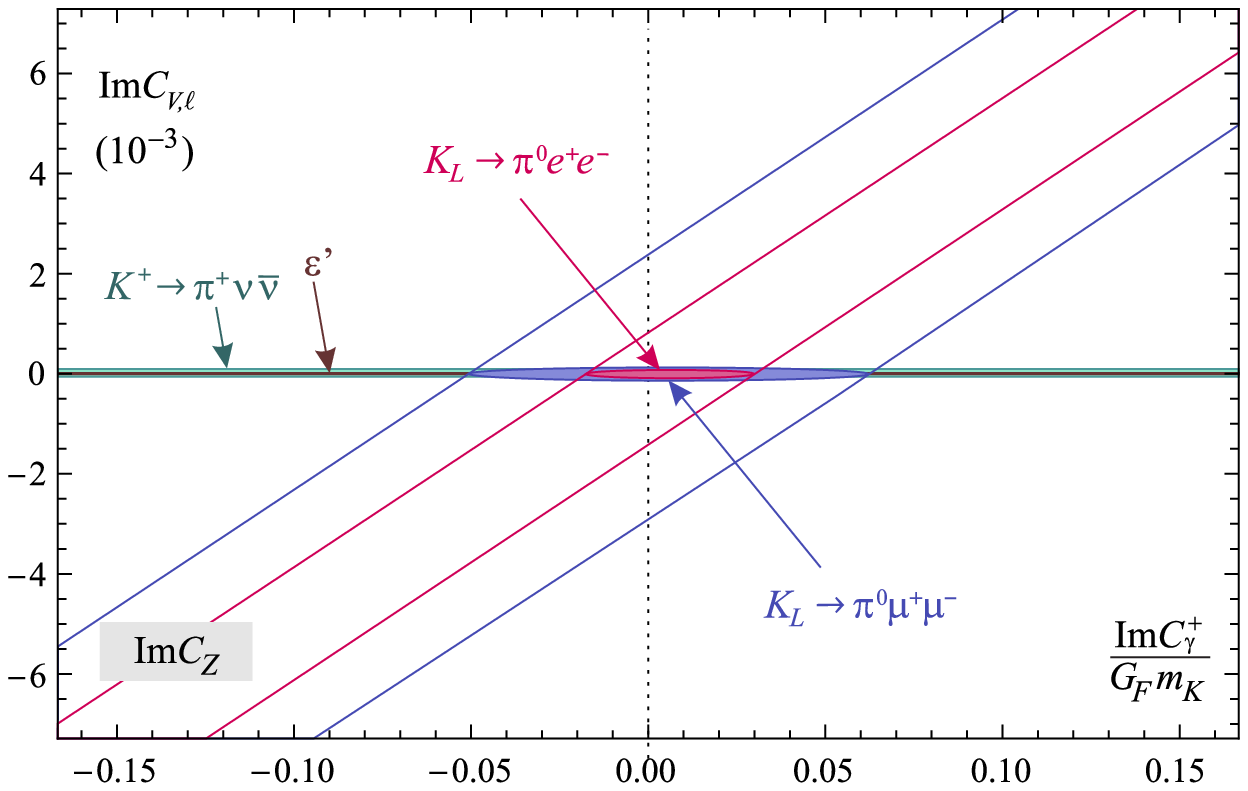}}  }$ &
$b.\;\text{\raisebox{-4.5cm}{\includegraphics[width=7.7cm]{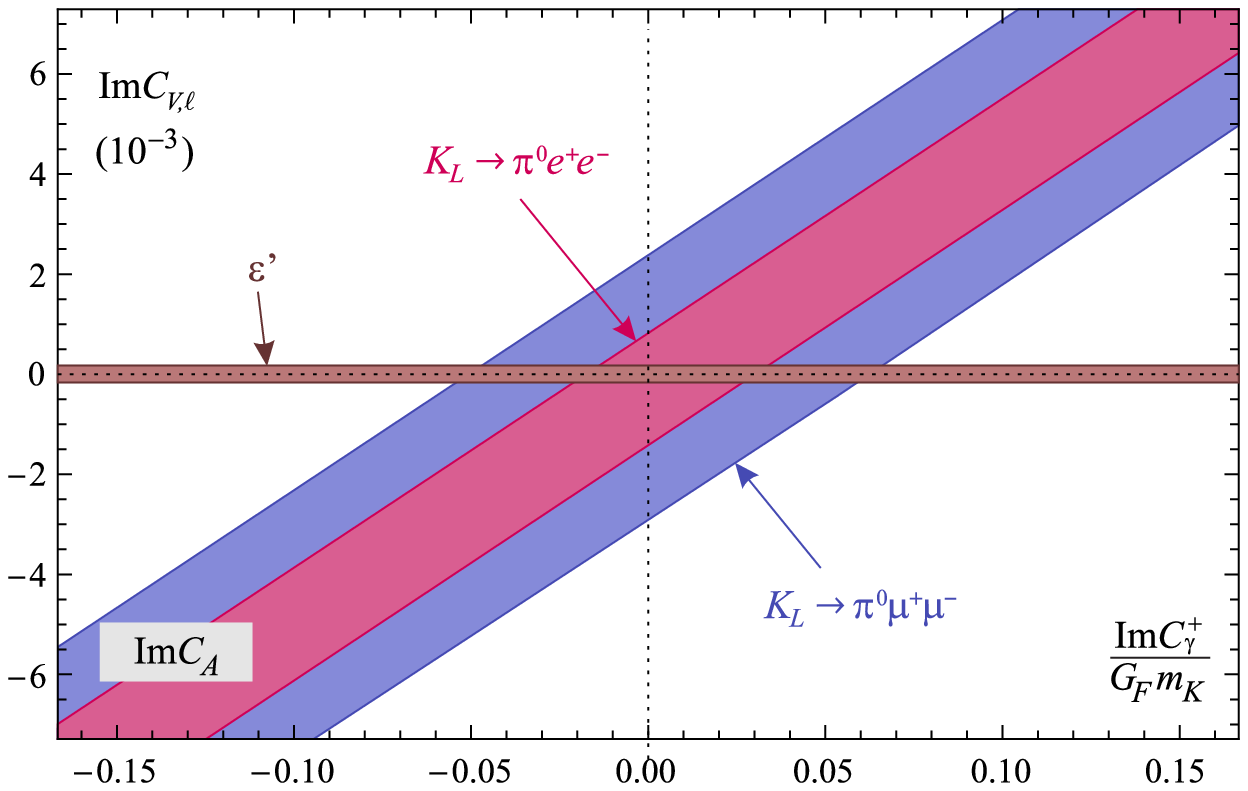}}  }$\\
$c.\;\text{\raisebox{-4.5cm}{\includegraphics[width=7.7cm]{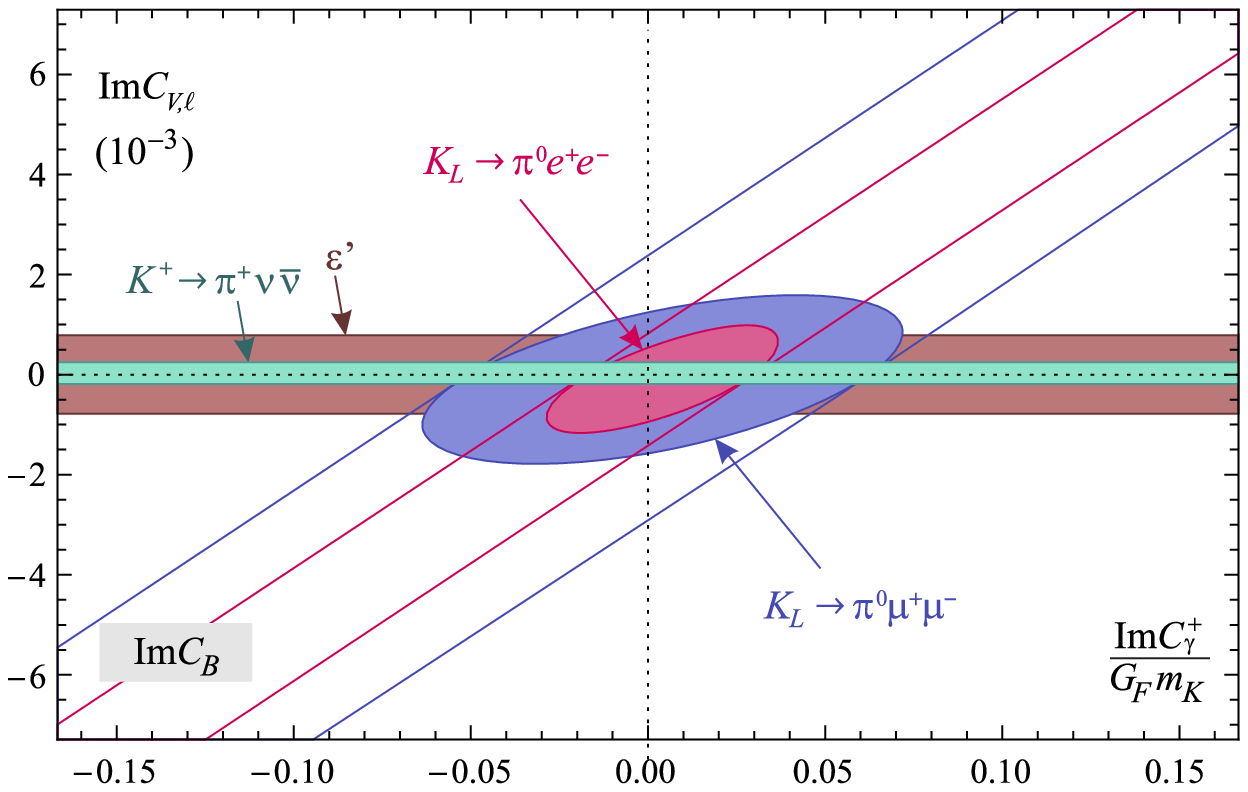}}  }$ &
$d.\;\text{\raisebox{-4.5cm}{\includegraphics[width=7.7cm]{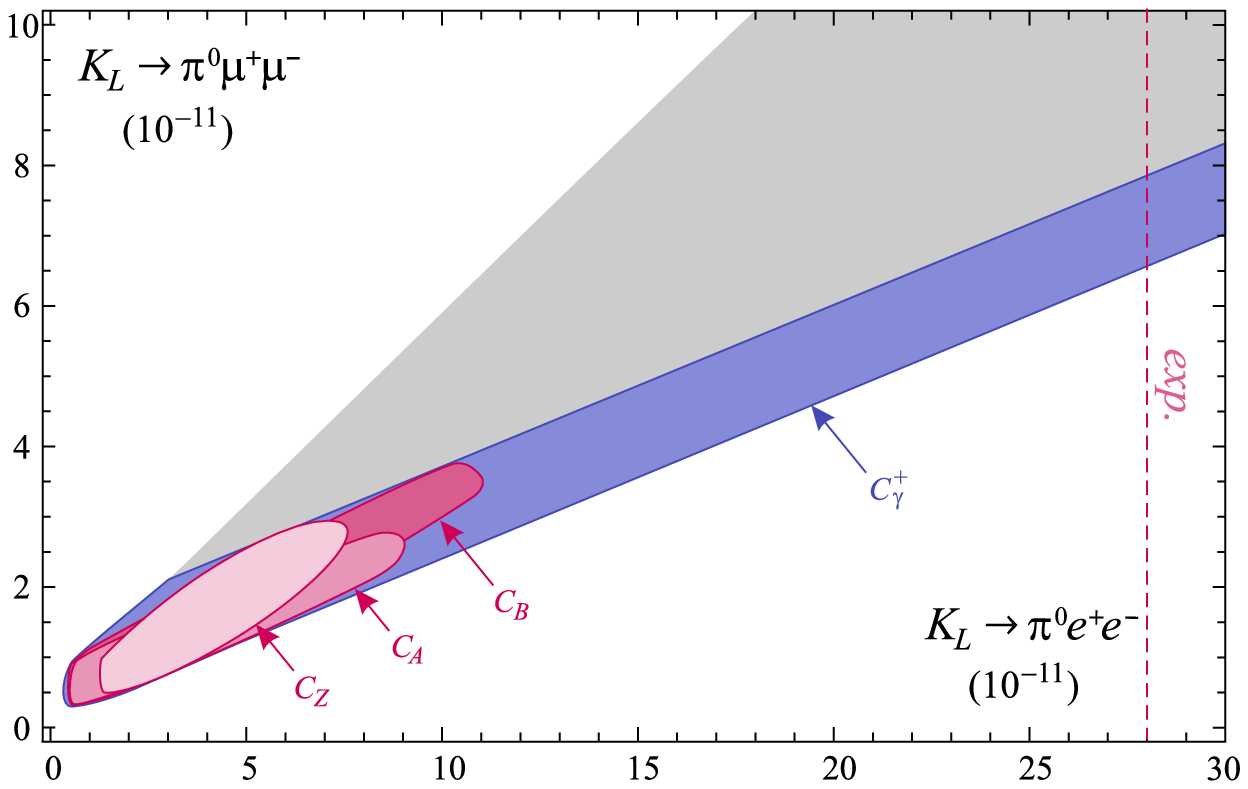}}  }$%
\end{tabular}
\caption{Loop-level FCNC scenario, with each electroweak operator separately
turned on together with $Q_{\gamma}^{\pm}$. ($a-c$) Contours in the
$\operatorname{Im}C_{V,\ell}-\operatorname{Im}C_{\gamma}^{+}$ plane as allowed
by the $K^{+}\rightarrow\pi^{+}\nu\bar{\nu}$, $K_{L}\rightarrow\pi^{0}\ell
^{+}\ell^{-}$, and $\varepsilon^{\prime}$ experimental bounds. ($d$) The
correlation between $K_{L}\rightarrow\pi^{0}e^{+}e^{-}$ and $K_{L}%
\rightarrow\pi^{0}\mu^{+}\mu^{-}$, when generated exclusively by $Q_{Z}$,
$Q_{A}$, or $Q_{B}$ (red), or with one of these together with $Q_{\gamma}^{+}$
(blue). The grey background is the area accessible with uncorrelated vector
and axial-vector currents (assuming leptonic universality). See
Ref.~\protect\cite{MST06} for more information.}%
\label{Fig13}%
\end{figure}

Even if rather imprecise, the constraints from $\operatorname{Re}%
(\varepsilon^{\prime}/\varepsilon)$ are currently tighter than those coming
from rare decays for $C_{Z}$ and $C_{A}$. Numerically, turning on one
semileptonic operator at a time, Eq.~(\ref{epeNP}) imposes (all numbers are in
units of $10^{-4}$)
\begin{equation}
\operatorname{Re}(\varepsilon^{\prime}/\varepsilon)\;\Rightarrow
\;|\operatorname{Im}C_{Z}|<4\;\oplus\;|\operatorname{Im}C_{A}|<15\;\oplus
\;|\operatorname{Im}C_{B}|<16\;. \label{epBound}%
\end{equation}
As shown in Fig.~\ref{Fig13}, for such values, the contributions to
$C_{V,\ell}$ are tiny. Thus, the maximal values for $\operatorname{Im}%
C_{\gamma}^{+}$ are the same as without any other NP sources,
Eq.~(\ref{BDnaive}), which requires that $K_{L}\rightarrow\pi^{0}e^{+}e^{-}$
saturates its current experimental limit. Since lepton universality holds, the
$K_{L}\rightarrow\pi^{0}\mu^{+}\mu^{-}$ rate is smaller but tightly correlated
to $K_{L}\rightarrow\pi^{0}e^{+}e^{-}$, see Fig.~\ref{Fig13}. Concerning
$K\rightarrow\pi\nu\bar{\nu}$, if one assumes that $C_{B}\ll C_{Z}$, as in the
SM, then $K\rightarrow\pi\nu\bar{\nu}$ is strongly limited by $\varepsilon
^{\prime}$:%
\begin{equation}
C_{A}=C_{B}=0\;\Rightarrow\left\{
\begin{array}
[c]{c}%
\;\;\;\;\;\;\;\;\;\;\;\;\;0<\mathcal{B}(K_{L}\rightarrow\pi^{0}\nu\bar{\nu
})<16\times10^{-11}\;,\\
7\times10^{-11}<\mathcal{B}(K^{+}\rightarrow\pi^{+}\nu\bar{\nu})<12\times
10^{-11}\;.
\end{array}
\right.
\end{equation}
However, the current $K^{+}\rightarrow\pi^{+}\nu\bar{\nu}$ experimental limit
can be saturated when $C_{B}\approx C_{Z}$, in which case $K_{L}\rightarrow
\pi^{0}\nu\bar{\nu}$ could reach the model-independent upper limit of
Eq.~(\ref{KLnn})
\begin{equation}
\mathcal{B}(K_{L}\rightarrow\pi^{0}\nu\bar{\nu})\approx4.3(\mathcal{B}%
(K^{+}\rightarrow\pi^{+}\nu\bar{\nu})-\mathcal{B}(K^{+}\rightarrow\pi^{+}%
\nu\bar{\nu})^{\text{SM}})<1.2\times10^{-9}\;. \label{LargeCB}%
\end{equation}

With $\varepsilon^{\prime}$ so constraining, even a slight cancellation among
the electroweak penguins could have a significant outcome for
$\operatorname{Im}C_{\gamma}^{+}$. This could occur in most models since the
$\mathcal{H}_{\text{PB}}$ operators are usually not independent but arise
simultaneously. Indeed, the intermediate loop particles are in general coupled
to both the $\gamma$ and $Z$ bosons. Let us stress, as said before, that we do
not expect a fine-tuning among these electroweak penguins, at most some
cancellations, because their $SU(2)_{L}$-breaking properties are significantly
different. Still, it is worth to investigate this possibility, so let us relax
the one-operator-at-a-time procedure.\begin{figure}[t]
\centering
\begin{tabular}
[b]{ll}%
$a.\text{\raisebox{-4.5cm}{\includegraphics[width=7.7cm]{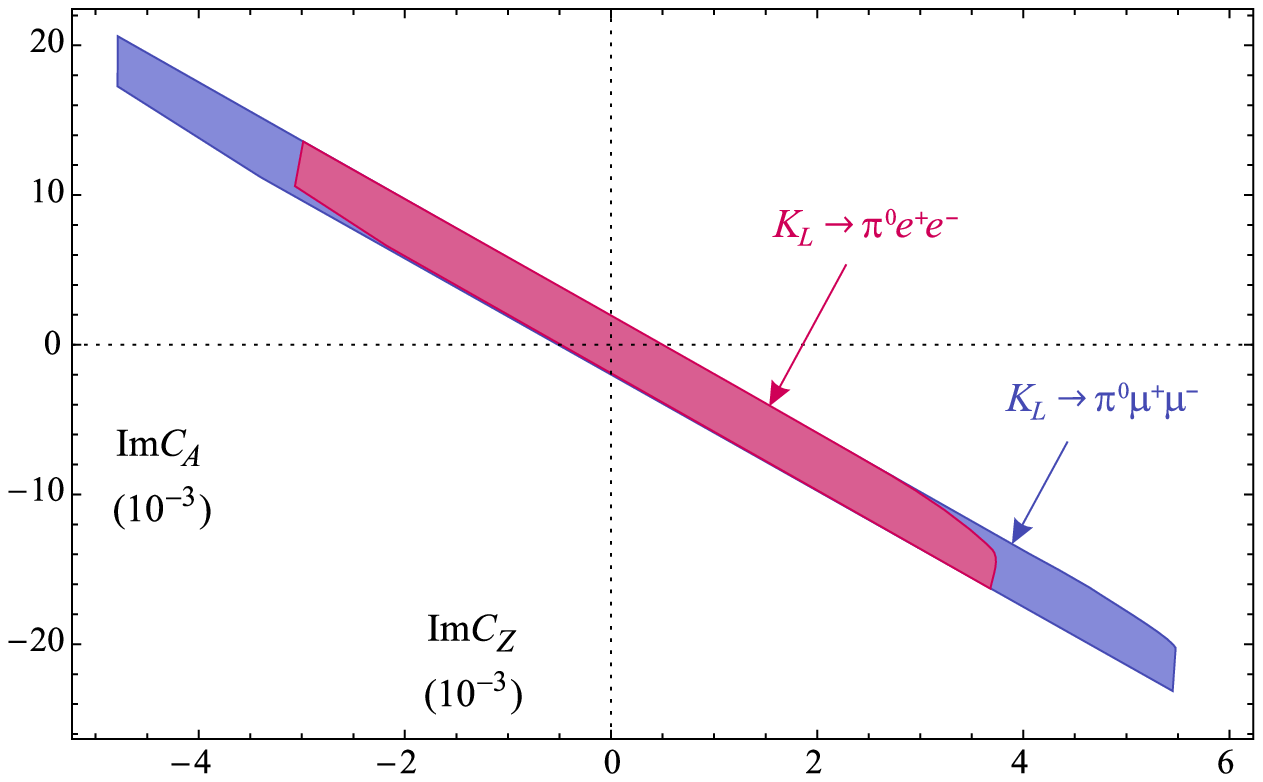}}  }$ &
$b.\text{\raisebox{-4.5cm}{\includegraphics[width=7.7cm]{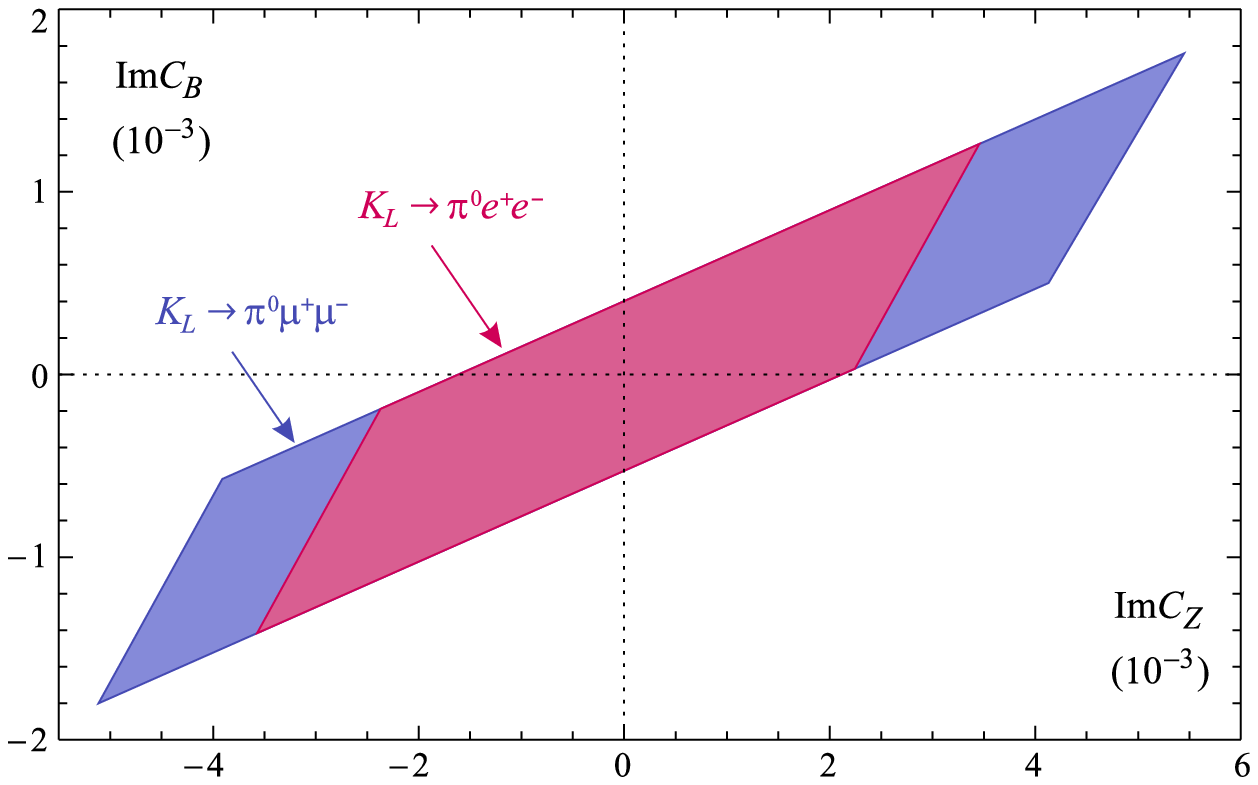}}  }$\\
$c.\text{\raisebox{-4.5cm}{\includegraphics[width=7.7cm]{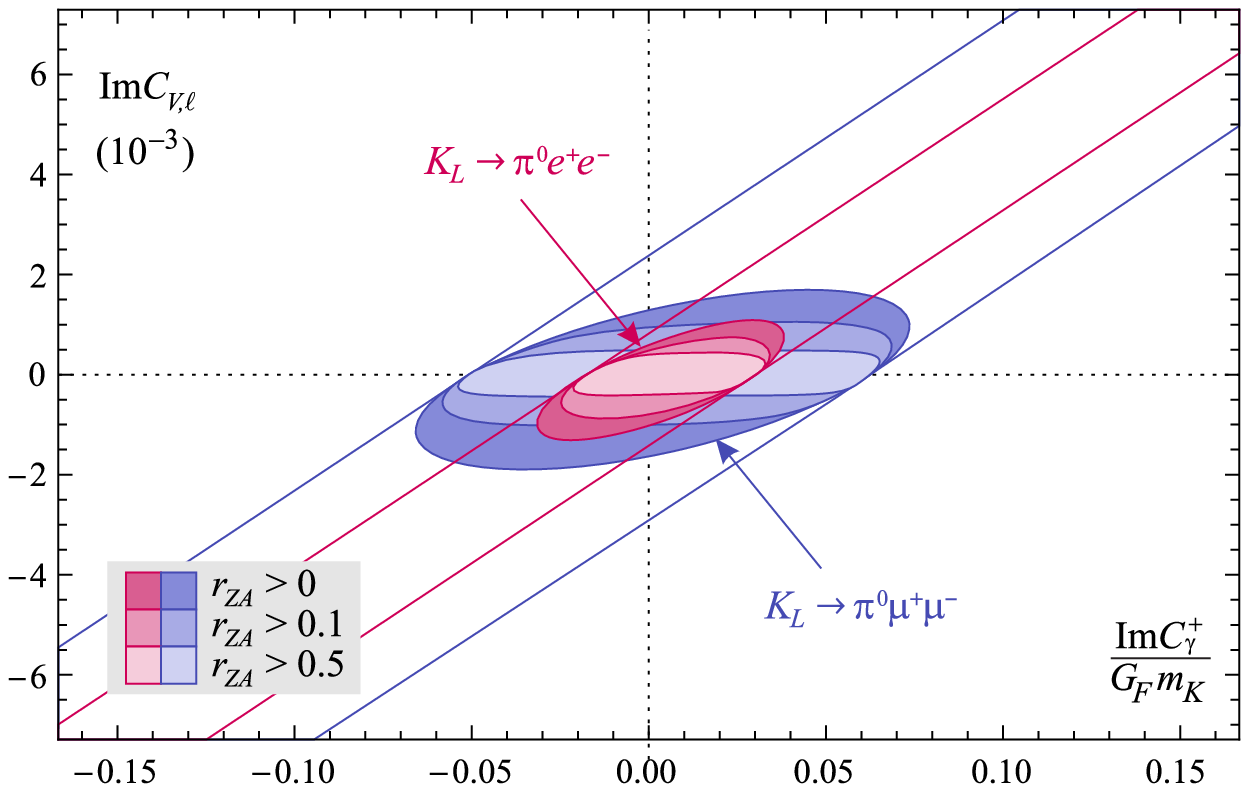}}  }$ &
$d.\text{\raisebox{-4.5cm}{\includegraphics[width=7.7cm]{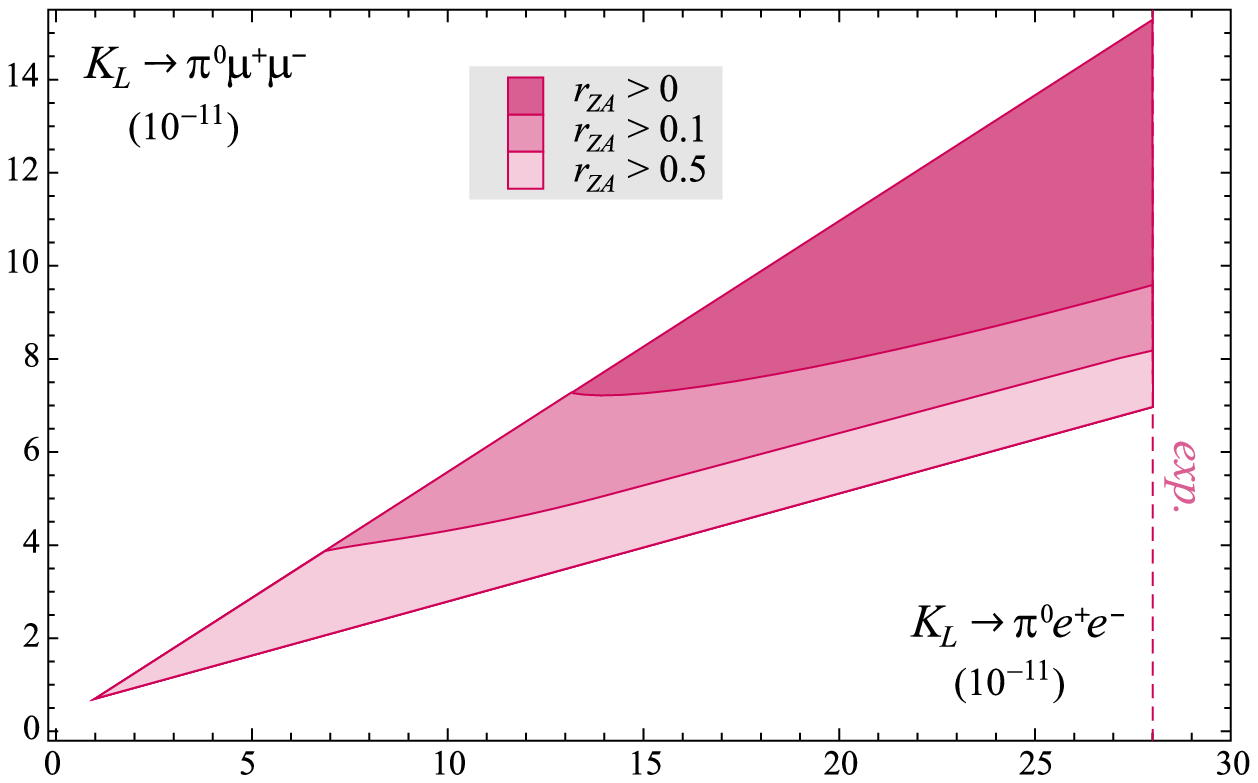}}  }$%
\end{tabular}
\caption{Loop-level FCNC scenario, with all the electroweak operators as well
as $Q_{\gamma}^{\pm}$ simultaneously turned on. ($a-b$) Correlations between
$\operatorname{Im}C_{A}$, $\operatorname{Im}C_{B}$, and $\operatorname{Im}%
C_{Z}$, as implied by the experimental bounds on $K^{+}\rightarrow\pi^{+}%
\nu\bar{\nu}$, $K_{L}\rightarrow\pi^{0}\ell^{+}\ell^{-}$, and $\varepsilon
^{\prime}$. ($c$) Contours in the $\operatorname{Im}C_{V,\ell}%
-\operatorname{Im}C_{\gamma}^{+}$ plane, with the color lightness indicating
the level of fine-tuning between $C_{A}$ and $C_{Z}$, see Eq.~(\ref{rZA}).
($d$) The correlation between $K_{L}\rightarrow\pi^{0}e^{+}e^{-}$ and
$K_{L}\rightarrow\pi^{0}\mu^{+}\mu^{-}$, again as a function of the
fine-tuning between $C_{A}$ and $C_{Z}$. Compared to Fig.~\ref{Fig12},~a
larger range is attainable. Note that here, the theoretical errors in
$K_{L}\rightarrow\pi^{0}\ell^{+}\ell^{-}$ are discarded for clarity.}%
\label{Fig14}%
\end{figure}

Once Eq.~(\ref{epeNP}) is added to $K\rightarrow\pi\nu\bar{\nu}$ and
$K_{L}\rightarrow\pi^{0}\ell^{+}\ell^{-}$, the system is sufficiently
constrained and the bounds can be resolved even when all the semileptonic
operators are turned on simultaneously (all the bounds are in units of
$10^{-4}$)%
\begin{equation}%
\begin{array}
[c]{rl}%
\operatorname{Re}(\varepsilon^{\prime}/\varepsilon)\;\Rightarrow &
\;\;\;\;|\operatorname{Im}C_{A}+3.9\operatorname{Im}C_{Z}|<19\;\smallskip\\
K^{+}\rightarrow\pi^{+}\nu\bar{\nu}\;\;\Rightarrow & \;\;\wedge
\;-15<\operatorname{Im}C_{Z}-4\operatorname{Im}C_{B}<21\smallskip\\
K_{L}\rightarrow\pi^{0}e^{+}e^{-}\;\Rightarrow & \;\;\wedge
\;[\;-32<\operatorname{Im}C_{Z}<35\;\wedge\;-14<\rho\operatorname{Im}%
C_{\gamma}^{+}<18\;]\;\smallskip\\
K_{L}\rightarrow\pi^{0}\mu^{+}\mu^{-}\;\Rightarrow & \;\;\wedge
\;[\;-49<\operatorname{Im}C_{Z}<53\;\wedge\;-30<\rho\operatorname{Im}%
C_{\gamma}^{+}<35\;]\;\;.\;\smallskip
\end{array}
\end{equation}
We indicate the main source driving each bound, but it should be clear that
all the experimental constraints are entangled, and all are necessary to get a
finite-size area in parameter space.

Interestingly, these bounds are not very different from those derived on the
$SU(2)_{L}\otimes U(1)_{Y}$ operators of Eq.~(\ref{LLRbasis}). The reason is
that $\operatorname{Re}(\varepsilon^{\prime}/\varepsilon)$ in Eq.~(\ref{epeNP}%
) imposes the tight correlation $C_{A}\approx-4C_{Z}$, upon which $C_{Z}$,
$C_{A}$, and $C_{B}$ are all ultimately bounded by the rare decays through
$C_{\nu,\ell}$ and $C_{A,\ell}$, exactly like $C_{L}$, $C_{L}^{\prime}$, and
$C_{R}$ were (see Eq.~(\ref{LLR})). Still, the origin of the observed
correlations among $C_{\nu,\ell}$, $C_{A,\ell}$ and $C_{V,\ell}$ in these two
scenarios is obviously very different. It directly comes from the assumed NP
dynamics when using the $\mathcal{H}_{\text{Gauge}}$ basis, but is entirely
driven by the sensitivity of $\operatorname{Re}(\varepsilon^{\prime
}/\varepsilon)$ to electroweak penguins when using the $\mathcal{H}%
_{\text{PB}}$ basis.

If the electroweak operators are induced by SM-like $Z$ and $\gamma^{\ast}$
penguins, such a tight $C_{A}\approx-4C_{Z}$ correlation is rather unlikely
given the intrinsic differences between those FCNC (dim-4 versus dim-6). So,
when
\begin{equation}
r_{AZ}\equiv\frac{C_{A}+4C_{Z}}{C_{A}-4C_{Z}}\ll1\;, \label{rZA}%
\end{equation}
one would rather conclude that a non-standard FCNC, not aligned with the SM
penguins, is present. Since $C_{A}+4C_{Z}$ is the gauge-invariant combination
driving the vector coupling (which is known to dominate in $\varepsilon
^{\prime}$~\cite{PB}, as is obvious in Eq.~(\ref{epeNP})), one would need a
new enhanced penguin not coupled to the vector current, or not coupled to quarks.

The experimental signature for this scenario requires disentangling $C_{A}$
and $C_{Z}$. Since the experimental $K^{+}\rightarrow\pi^{+}\nu\bar{\nu}$
bound can be saturated with the help of $C_{B}$ only, it has no discriminating
power in $r_{AZ}$. The maximal attainable value for $\operatorname{Im}%
C_{\gamma}^{+}$, and thus for the CP-asymmetries, is not very sensitive to
$r_{AZ}$ either, see Fig.~\ref{Fig14}. On the other hand, the correlation
between $K_{L}\rightarrow\pi^{0}e^{+}e^{-}$ and $K_{L}\rightarrow\pi^{0}%
\mu^{+}\mu^{-}$ shown in Fig.~\ref{Fig14} could signal such a scenario.
Indeed, without fine-tuning, one is back to the situation shown in
Fig.~\ref{Fig13}, i.e. both rates saturated by a large $Q_{\gamma}^{+}$
contribution in their vector current when they deviate from their SM
predictions. On the other hand, as $r_{AZ}$ decreases, more and more of the
model-independent region in the $K_{L}\rightarrow\pi^{0}e^{+}e^{-}$%
--$K_{L}\rightarrow\pi^{0}\mu^{+}\mu^{-}$ plane gets covered.

\subsubsection{QCD penguins\label{QCDLoop}}

If $SU(3)_{C}\otimes U(1)_{em}$ stays unbroken at the low scale, the FCNC
loops must involve intermediate charged and colored particle(s). The photonic
penguin is thus necessarily accompanied by the gluonic one. Further, if NP
enhances significantly the chromomagnetic operators $Q_{g}^{\pm}$ (defined in
Eq.~(\ref{ChromoMag})), the magnetic operators $Q_{\gamma}^{\pm}$ are then
directly affected through the RGE~(\ref{OPEgg}),
\begin{equation}
C_{\gamma}^{\pm}(\mu_{c})=\eta^{2}\left[  C_{\gamma}^{\pm}(\mu_{NP}%
)+8(1-\eta^{-1})C_{g}^{\pm}(\mu_{NP})\right]  \;,\;\;C_{g}^{\pm}(\mu_{c})=\eta
C_{g}^{\pm}(\mu_{NP})\;.\label{RGE2}%
\end{equation}
So, $C_{g}^{\pm}(\mu_{NP})$ act as lower bounds for $C_{\gamma}^{\pm}(\mu
_{c})$. The opposite cannot be asserted from Eq.~(\ref{RGE2}) since the
$\mathcal{O}(\alpha)$ mixings $Q_{\gamma}^{\pm}\rightarrow Q_{g}^{\pm}$ are
missing. However, those mixings are presumably long-distance dominated, hence
have to be dealt with at the matrix-element level. For instance, in the case
of $\varepsilon^{\prime}$, the $Q_{\gamma}^{-}$ contribution is subleading
even when $\operatorname{Im}C_{\gamma}^{-}$ saturates the experimental limit
on the $K^{+}\rightarrow\pi^{+}\pi^{0}\gamma$ CP-asymmetry, see
Eq.~(\ref{Bound}). So, the mixing effects do not forbid a large splitting
$C_{\gamma}^{\pm}(\mu_{c})\gg C_{g}^{\pm}(\mu_{c})$.

Still, owing to their similar dynamics, $C_{\gamma}^{\pm}(\mu_{NP})$ and
$C_{g}^{\pm}(\mu_{NP})$ may have similar sizes. Then, since $Q_{g}^{+}$
contributes to $\varepsilon^{\prime}$, both magnetic operators are tightly
bounded%
\begin{equation}
\frac{|\operatorname{Im}C_{\gamma}^{-}|}{G_{F}m_{K}}\approx
\frac{|\operatorname{Im}C_{g}^{-}|}{G_{F}m_{K}}\lesssim5\times10^{-4}\;,
\end{equation}
if we require $|\operatorname{Re}(\varepsilon^{\prime}/\varepsilon
)_{g}|<\operatorname{Re}(\varepsilon^{\prime}/\varepsilon)^{\exp}$ and set
$B_{G}=1$. This is extremely constraining, and would rule out any effect of
the magnetic operators in rare decays or in CP-asymmetries.

The presence of the other FCNC could significantly alter this bound. So, let
us again turn on all the penguin operators but freeze the relation among the
magnetic ones, $|\operatorname{Im}C_{\gamma}^{+}|=1.5|\operatorname{Im}%
C_{g}^{-}|$. Also, we neglect the chromoelectric operators (the usual QCD
penguins), as their impact is less important~\cite{BurasJ04}. Then, using
Eq.~(\ref{epeNP}) together with~(\ref{QgE}), the bounds can be resolved except
when $\varepsilon^{\prime}$ and $K_{L}\rightarrow\pi^{0}\ell^{+}\ell^{-}$ just
happen to depend on the same combination of $\operatorname{Im}C_{A}$ and
$\operatorname{Im}C_{\gamma,g}^{+}$, which occurs for $\operatorname{Im}%
C_{\gamma}^{+}\approx-3\operatorname{Im}C_{g}^{-}$ (with $B_{G}=+1$).

\begin{figure}[t]
\centering
\begin{tabular}
[b]{cc}%
$a.\text{\raisebox{-4.4cm}{\includegraphics[width=7.7cm]{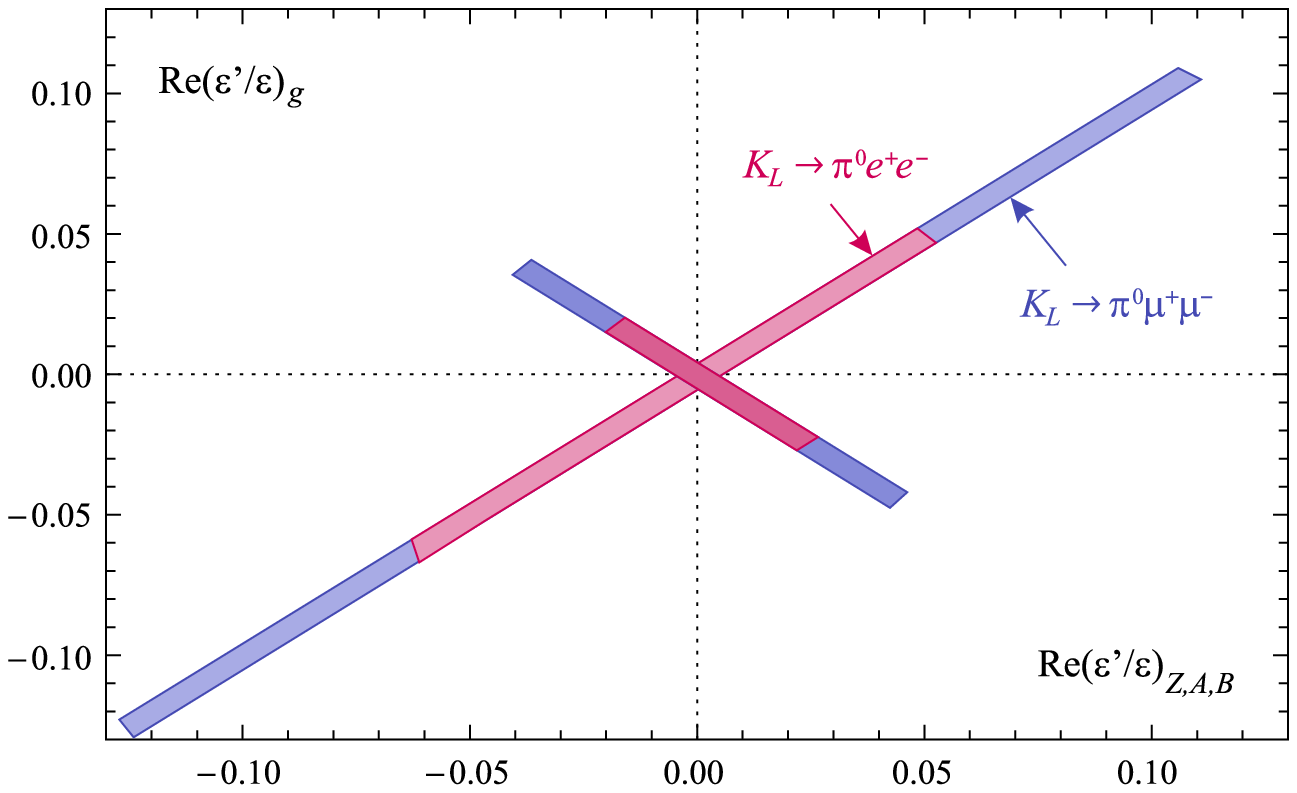}}  }$ &
$b.\text{\raisebox{-4.4cm}{\includegraphics[width=7.7cm]{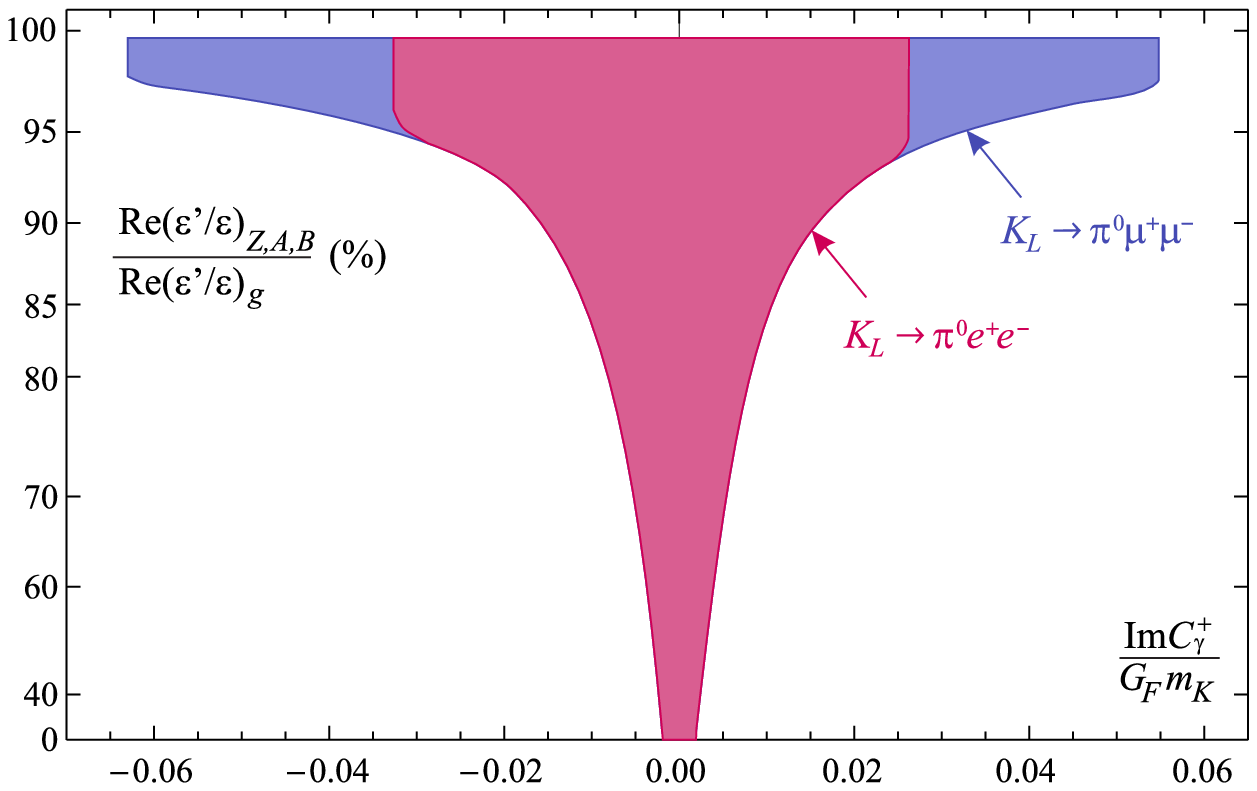}}  }$\\
\multicolumn{2}{c}{$c.\text{\raisebox{-4.4cm}%
{\includegraphics
[width=7.7cm]{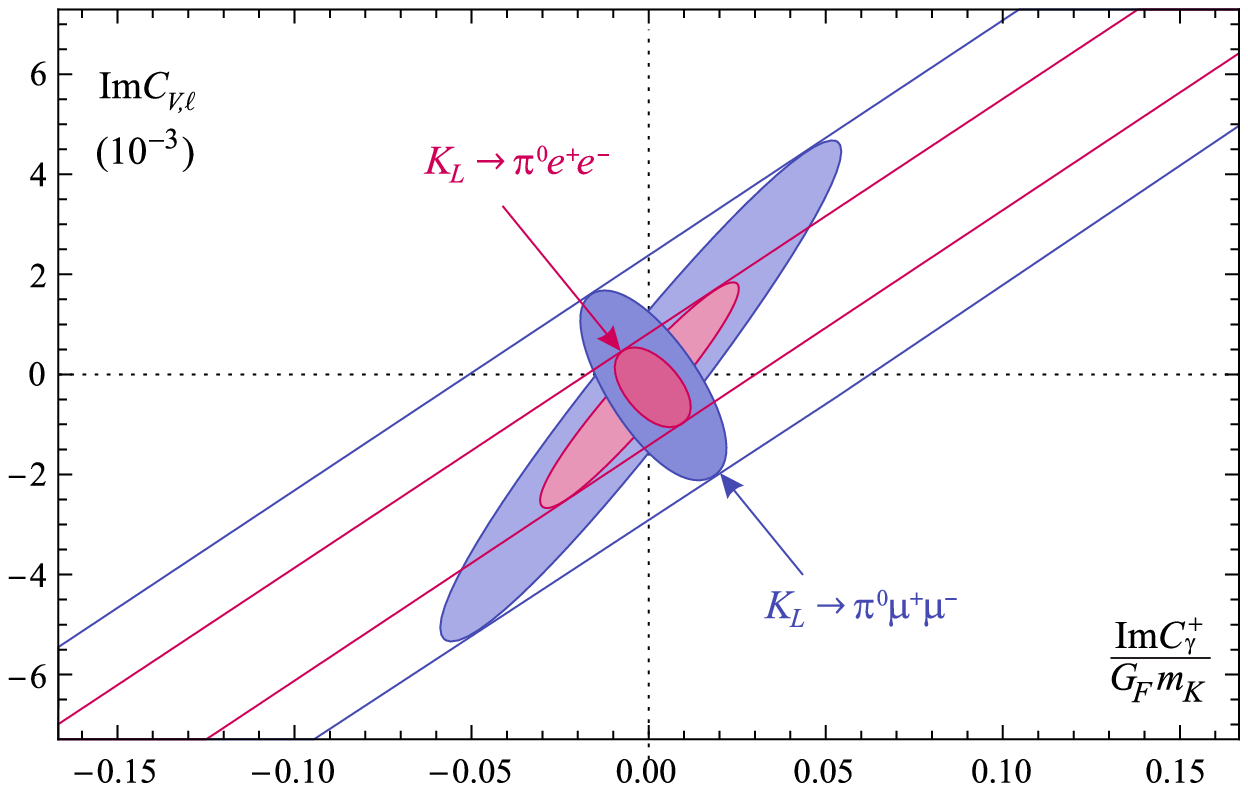}}  }$}%
\end{tabular}
\caption{Loop-level FCNC scenario, with all the electroweak operators as well
as $Q_{\gamma,g}^{\pm}$ simultaneously turned on, but imposing
$\operatorname{Im}C_{\gamma}^{+}=\pm1.5\operatorname{Im}C_{g}^{-}$. ($a$)
Correlation between the electroweak and gluonic contributions to
$\varepsilon^{\prime}$, imposing $|\operatorname{Re}(\varepsilon^{\prime
}/\varepsilon)^{\text{NP}}|<2\operatorname{Re}(\varepsilon^{\prime
}/\varepsilon)^{\exp}$. ($b$) The $\operatorname{Im}C_{\gamma}^{+}$ range as a
function of the fine-tuning between $\operatorname{Re}(\varepsilon^{\prime
}/\varepsilon)_{EW}$ and $\operatorname{Re}(\varepsilon^{\prime}%
/\varepsilon)_{g}$. ($c$) The corresponding contours in the $\operatorname{Im}%
C_{V,\ell}-\operatorname{Im}C_{\gamma}^{+}$ plane. In ($a$) and ($c$), the
lighter (darker) colors denote destructive (constructive) interference between
$Q_{A}$ and $Q_{\gamma}^{+}$ in $K_{L}\rightarrow\pi^{0}\ell^{+}\ell^{-}$.}%
\label{Fig15}%
\end{figure}

In this scenario, the driving force is the cancellation between the two
largest contributions to $\varepsilon^{\prime}$, i.e. between
$\operatorname{Im}C_{g}^{-}$ and $\operatorname{Im}(4C_{Z}+C_{A})$. The
electroweak operators are not fine-tuned except for the $\operatorname{Im}%
C_{Z}-\operatorname{Im}C_{B}$ correlation imposed by the rare decays, which
stays as in Fig.~\ref{Fig14}. So, in this scenario, large effects are possible
in $K\rightarrow\pi\nu\bar{\nu}$ thanks to $Q_{B}$ and $Q_{Z}$, while
$K_{L}\rightarrow\pi^{0}\ell^{+}\ell^{-}$ receive sizeable contributions in
both their vector and axial-vector currents. Contrary to the situation without
$Q_{g}^{\pm}$, these latter decays can no longer be used to probe the
cancellations in $\varepsilon^{\prime}$ since they do not directly depend on
the chromomagnetic operators.

Actual numbers for the bounds on the Wilson coefficients would not make much
sense here, because the fine-tuning in $\operatorname{Re}(\varepsilon^{\prime
}/\varepsilon)$ reaches horrendous values before the rare decay constraints
can kick in. As shown in Fig.~\ref{Fig15}, individual contributions to
$\operatorname{Re}(\varepsilon^{\prime}/\varepsilon)$ can be as large as
$10\%$. Instead, let us freeze the situation and set the $Q_{g}^{-}$
contribution to $\operatorname{Re}(\varepsilon^{\prime}/\varepsilon)$ at
$2\times10^{-2}$. As shown in Fig.~\ref{Fig15}, this requires a large but not
impossible 90\% cancellation between the electroweak and the gluonic penguins.

To uniquely identify this cancellation, the best strategy relies on the direct
CP-asymmetries (see Fig.~\ref{Fig15}). The first step is to exploit the RGE
constraint $C_{\gamma}^{\pm}(\mu_{c})\gtrsim C_{g}^{\pm}(\mu_{c})$, which
implies that the asymmetries in Eq.~(\ref{CmAsym}) are all at the percent
level%
\begin{equation}
\frac{\operatorname{Im}C_{\gamma}^{-}}{G_{F}m_{K}}\gtrsim
\frac{\operatorname{Im}C_{g}^{-}}{G_{F}m_{K}}\approx\frac{\operatorname{Re}%
(\varepsilon^{\prime}/\varepsilon)_{g}}{3B_{G}}\approx10^{-2}\;.
\label{ScenarioLoop}%
\end{equation}
Since $\varepsilon_{+0\gamma}^{\prime}$, $\varepsilon_{+-\gamma}^{\prime}$,
and $\varepsilon_{||}^{\prime}$ are mostly insensitive to the hadronic penguin
fraction in $\varepsilon^{\prime}$, they would cleanly signal the presence of
NP in $Q_{\gamma}^{-}$. The second step derives from the pure $\Delta I=1/2$
nature of the chromomagnetic operator. Since it enters only in $K\rightarrow
(\pi\pi)_{0}$, its presence would be felt in $\varepsilon_{\perp}^{\prime}$
(see Eq.~(\ref{CpAsym})), in addition to that of $Q_{\gamma}^{+}$. So, using
Eq.~(\ref{QgE}) and enforcing $|\operatorname{Im}C_{\gamma}^{+}%
|=1.5|\operatorname{Im}C_{g}^{-}|$, we can write%
\begin{equation}
|\varepsilon_{\perp}^{\prime}/\varepsilon|_{g}=\frac{\sqrt{2}}{\omega
}\operatorname{Re}(\varepsilon^{\prime}/\varepsilon)_{g}\approx
0.65\;,\;\;|\varepsilon_{\perp}^{\prime}/\varepsilon|_{\gamma}=\frac{1}%
{4|\varepsilon|}\operatorname{Re}(\varepsilon^{\prime}/\varepsilon)_{g}%
\approx2.2\;,
\end{equation}
with $\omega^{-1}=\operatorname{Re}A_{0}/\operatorname{Re}A_{2}\approx22.4$
the $\Delta I=1/2$ enhancement factor, and $B_{G}=+1$. By contrast,
electroweak penguins contribute mostly to the $K\rightarrow(\pi\pi)_{2}$
amplitude, and have thus a negligible impact on $\varepsilon_{\perp}^{\prime}$
compared to $Q_{g}^{-}$. So, in principle, by combining $\varepsilon_{\perp
}^{\prime}$ with $\varepsilon_{+0\gamma}^{\prime}$, $\varepsilon_{+-\gamma
}^{\prime}$, or $\varepsilon_{||}^{\prime}$, it is possible to evidence NP in
both $Q_{\gamma}^{\pm}$ and $Q_{g}^{-}$. Of course, this whole program is very
challenging experimentally, but completing the first step may be feasible,
since $Q_{\gamma}^{-}$ could push $\varepsilon_{+0\gamma}^{\prime}$ and
$\varepsilon_{+-\gamma}^{\prime}$ up to less than an order of magnitude away
from their current limits.

\subsubsection{Minimal Supersymmetric Standard Model\label{MSSM}}

The MSSM with R-parity is a particular implementation of the loop-level FCNC
scenario discussed in the previous section. All the bounds derived there are
thus not only valid, but could become tighter. Indeed, the various FCNC could
be more directly correlated once the NP dynamics is specified. In addition,
the MSSM introduces only a finite number of new sources of flavor-breaking
through its soft-breaking squark mass terms and trilinear couplings.

The most important correlation is that between the gluonic and photonic
penguins, as analyzed in details in Ref.~\cite{CDI99,BurasCIRS99}. Both can be
generated by gluino-down squark loops, so that~\cite{GabbianiGMS96}%
\begin{subequations}
\begin{align}
C_{\gamma}^{\pm}(m_{\tilde{g}})  &  =\frac{\pi\alpha_{S}(m_{\tilde{g}}%
)}{m_{\tilde{g}}}\left[  (\delta_{LR}^{D})_{21}\pm(\delta_{RL}^{D}%
)_{21}\right]  F(x_{qg}),\;\;F(x_{qg})\approx F(1)=\frac{2}{9}\;,\\
C_{g}^{\pm}(m_{\tilde{g}})  &  =\frac{\pi\alpha_{S}(m_{\tilde{g}})}%
{m_{\tilde{g}}}\left[  (\delta_{LR}^{D})_{21}\pm(\delta_{RL}^{D})_{21}\right]
G(x_{qg}),\;\;G(x_{qg})\approx G(1)=-\frac{5}{18}\;,
\end{align}
\end{subequations}
where $x_{qg}=m_{\tilde{q}}^{2}/m_{\tilde{g}}^{2}$, $m_{\tilde{q}(\tilde{g})}$
the squark (gluino) mass, and $F(x_{qg})$, $G(x_{qg})$ the loop functions. The
chirality flips are induced by the $SU(2)_{L}$ breaking trilinear term
$\mathbf{A}^{D}$, parametrized through the mass insertions $(\delta_{RL}%
^{D})_{21}=(\delta_{LR}^{D})_{12}^{\ast}$. At the low-scale, the Wilson
coefficients obey%
\begin{equation}
C_{\gamma}^{\pm}(\mu_{c})=\left(  \eta\frac{F(x_{qg})}{G(x_{qg})}%
+8(\eta-1)\right)  C_{g}^{\pm}(\mu_{c})\approx-1.6C_{g}^{\pm}(\mu_{c})\;.
\label{MSSMRGE}%
\end{equation}
In the absence of any other supersymmetric contributions to $\varepsilon
^{\prime}$, this leads to the tight constraint~\cite{MIdS2}
\begin{equation}
\operatorname{Re}(\varepsilon^{\prime}/\varepsilon)\;\Rightarrow
\;\frac{|\operatorname{Im}C_{g}^{-}(\mu_{c})|}{G_{F}m_{K}}\lesssim
5\times10^{-4}\;\rightarrow\;|\operatorname{Im}(\delta_{RL}^{D})_{21,12}%
|\lesssim2\times10^{-5}\;. \label{CgNP}%
\end{equation}

Before discussing how this bound could get relaxed by NP effects in the other
FCNC, let us consider the MFV prediction for $\delta_{RL}^{D}$, to get a
handle on the ``minimal'' size of $C_{\gamma,g}^{\pm}$. The $U(3)^{5}$
flavor symmetry-breaking of $\mathbf{A}^{D}$ imposes an expansion at least linear in
the Yukawa couplings \cite{MFV}%
\begin{equation}
\mathbf{A}^{D}\sim A_{0}\mathbf{Y}_{d}(a_{0}\mathbf{1}+a_{1}\mathbf{Y}%
_{u}^{\dagger}\mathbf{Y}_{u}+...)\;,
\end{equation}
with $v_{d}\mathbf{Y}_{d}=\mathbf{m}_{d}$, $v_{u}\mathbf{Y}_{u}=\mathbf{m}%
_{u}V$, $v_{u,d}$ the vacuum expectation values of the $H_{u,d}^{0}$ Higgs
boson, $A_{0}$ setting the SUSY breaking scale, and $a_{i}$ some free
$\mathcal{O}(1)$ parameters (which can be complex~\cite{ComplexMFV}). In that
case, $(\delta_{LR}^{D})_{IJ}\sim m_{d^{J}}/m_{\tilde{d}}\sim10^{-4}$, and no
visible deviations could arise in $\varepsilon^{\prime}$ or in the other
CP-violation parameters~(\ref{CmAsym}). Turned around, this means that these
observables are particularly sensitive to deviation with respect to MFV. Since
this framework is only one particular realization of the flavor sector of the
MSSM, motivated in part by the tight constraints in the $b\rightarrow s,d$ or
$\ell\rightarrow\ell^{\prime}$ sectors, and in part by its rather natural
occurrence starting from universal soft-breaking terms at the high scale, it
has to be confirmed experimentally also in the $s\rightarrow d$ sector.

Before exploiting the analysis of Sec.~\ref{QCDLoop}, there is another
important correlation arising in the MSSM. The $\Delta S=2$ observables can be
induced by the same source of flavor-breaking as the magnetic operators. One
derives for $m_{\tilde{g}}=500$ GeV~\cite{MIdS2}:%
\begin{subequations}
\begin{align}
\Delta M_{K}  &  \Rightarrow\sqrt{\operatorname{Re}(\delta_{RL}^{D})_{21}^{2}%
}<3\times10^{-3}\;\;\rightarrow\;\;\frac{|\operatorname{Re}C_{\gamma}^{\pm}%
|}{G_{F}m_{K}}\lesssim0.1\;,\\
\varepsilon_{K}  &  \Rightarrow\sqrt{\operatorname{Im}(\delta_{RL}^{D}%
)_{21}^{2}}<4\times10^{-4}\;\;\rightarrow\;\;\frac{|\operatorname{Im}%
C_{\gamma}^{\pm}|}{G_{F}m_{K}}\lesssim0.01\;. \label{eKMSSM}%
\end{align}
\end{subequations}
The absence of a large cancellation among the supersymmetric contributions is
explicitly assumed, for example with the processes where the flavor-breaking
originates from the $SU(2)_{L}$ conserving squark masses (most notably
$\delta_{LL}^{D}$). At this stage, we want to point out that the bounds on
$\operatorname{Re}C_{\gamma}^{\pm}$ obtained from radiative decays are
competitive with that from $\Delta M_{K}$:
\begin{subequations}
\label{radMIA}%
\begin{align}
K^{+}\overset{}{\rightarrow}\pi^{+}\pi^{0}\gamma &  \Rightarrow
\frac{|\operatorname{Re}C_{\gamma}^{-}|}{G_{F}m_{K}}\lesssim0.1\;\;\rightarrow
\;\;|\operatorname{Re}(\delta_{RL}^{D})_{21}|<3\times10^{-3}\;,\\
K^{0}\overset{}{\rightarrow}\gamma\gamma &  \Rightarrow
\frac{|\operatorname{Re}C_{\gamma}^{+}|}{G_{F}m_{K}}\lesssim0.3\;\;\rightarrow
\;\;|\operatorname{Re}(\delta_{RL}^{D})_{21}|<10^{-2}\;,
\end{align}
\end{subequations}
assuming $C_{\gamma}^{+}\approx\pm C_{\gamma}^{-}$. Compared to the bound from
$\Delta M_{K}$, radiative decays directly constrain $\operatorname{Re}%
(\delta_{RL}^{D})_{21}$, and there can be no weakening through interferences
among SUSY contributions since only $Q_{\gamma}^{\pm}$ enter.

Let us consider the bound from $\varepsilon_{K}$ as the maximal allowed value
for $\operatorname{Im}C_{\gamma}^{\pm}$. We can now directly connect the
present MSSM scenario to that discussed in Sec.~\ref{QCDLoop} since the bound
(\ref{eKMSSM}) matches that in Eq.~(\ref{ScenarioLoop}). Given the
constraint~(\ref{MSSMRGE}), which also matches that of Sec.~\ref{QCDLoop},
such values for $\operatorname{Im}C_{\gamma,g}^{\pm}$ are only possible
provided there is a large electroweak-gluonic penguin cancellation in
$\varepsilon^{\prime}$, of about 90\% of their respective contributions, see
Fig.~\ref{Fig15}.

This cannot be excluded a priori, even though the electroweak penguins are not
directly correlated with gluonic penguins in the MSSM. With the $SU(2)_{L}$
conserving mass insertions $\delta_{LL}^{D}$ limited by the $\Delta S=2$
observables, electroweak penguins arise essentially from the flavor-breaking
in the up-squark sector. Indeed, when $\mathbf{A}^{U}=A_{0}\mathbf{Y}_{u}%
+...$, the quadratic combination of mass-insertion $(\delta_{LR}^{U}%
)_{13}(\delta_{LR}^{U})_{23}^{\ast}$ gets significantly enhanced by the large
top mass~\cite{ColangeloI98}. This scenario was analyzed in details e.g. in
Refs.~\cite{BurasCIRS99,IsidoriMPST06}, where significant deviations with
respect to the SM where found to be possible for $K\rightarrow\pi\nu\bar{\nu}%
$. In particular, the box diagram was found to be sizeable in
Ref.~\cite{BurasEJR05}. Though these scenarios concentrated on the low to
moderate $\tan\beta\equiv v_{u}/v_{d}$ regime, the situation is similar at
large $\tan\beta$. Indeed, on one hand, $C_{\gamma,g}^{\pm}$ and thus
$\operatorname{Re}(\varepsilon^{\prime}/\varepsilon)_{g}$ could reach larger
values even under MFV since $\mathbf{Y}_{d}=\mathbf{m}_{d}/v_{d}$ gets
enhanced, but on the other, the charged Higgs contribution to the electroweak
penguins can kick in, making them sensitive to the flavor-breakings in the
$\delta_{RR}^{D}$ sector\footnote{At large $\tan\beta$, Higgs mediated
penguins could also appear. Those are embedded in helicity-suppressed scalar
and pseudoscalar semileptonic operators. We refer to Ref.~\cite{MST06} for an
analysis of their possible impact.}.

Altogether, there can be two different situations in the MSSM:

\begin{itemize}
\item If there is a large cancellation between gluonic and electroweak penguins in $\varepsilon^{\prime}$, large enhancements are possible in the rare decays. This is the scenario of Sec.~\ref{QCDLoop}. The $K^{+}\rightarrow\pi^{+}\nu\bar{\nu}$ mode can saturate its current limit, and $K_{L}\rightarrow\pi^{0}\nu\bar{\nu}$ can reach the model-independent bound~(\ref{LargeCB}). The $K_{L}\rightarrow\pi^{0}e^{+}e^{-}$ can also saturate its experimental bound, while leptonic universality then limits $K_{L}\rightarrow\pi^{0}\mu^{+}\mu^{-}$ to about $40\%$ of its current (looser) bound. As in Sec.~\ref{QCDLoop}, the direct CP-violating parameters in radiative $K$ decays could reach the percent level, see Fig.~\ref{Fig15}, and would be the cleanest signatures for this scenario.

\item On the contrary, if there is no large cancellation in $\varepsilon^{\prime}$, say not beyond about $10\%$, then $C_{\gamma}^{\pm}$ are indirectly limited by the tight correlation~(\ref{MSSMRGE}), and all the direct CP-violating parameters would be small, presumably beyond the experimental reach. Further, a fine-tuning between the $Z$ and virtual $\gamma$ penguins able to push $r_{AZ}$ in Eq.~(\ref{rZA}) to small values is not possible. Both are driven by the same mass insertions, with the generic result $C_{Z}>C_{A}$ (see e.g. Ref.~\cite{IsidoriMPST06}). So, this corresponds to the first scenario of Sec.~\ref{EWloop}, characterized by the bounds~(\ref{epBound}). The $K^{+}\rightarrow\pi^{+}\nu\bar{\nu}$ and $K_{L}\rightarrow\pi^{0}\nu\bar{\nu}$ could still be very large if the boxes are sizeable ($C_{Z}\approx
C_{B}$), but $K_{L}\rightarrow\pi^{0}e^{+}e^{-}$ and $K_{L}\rightarrow\pi^{0}\mu^{+}\mu^{-}$ cannot because $C_{\gamma}^{+}\approx-1.6C_{g}^{\pm}$ is too small to enhance them (see the red areas in Fig.~\ref{Fig13}$d$).
\end{itemize}

In summary, to probe for a possible large electroweak and QCD penguin cancellations in $\varepsilon^{\prime}$, the $K\rightarrow\pi\nu\bar{\nu}$ are useful only if the scaling between box and penguins is known. However, telltale signatures would be large enhancements of $K_{L}\rightarrow\pi^{0}e^{+}e^{-}$ and $K_{L}\rightarrow\pi^{0}\mu^{+}\mu^{-}$ as well as large CP-violating parameters in radiative $K$ decays.\vfill   \pagebreak 

\section{Conclusions}

In this paper, the $s\rightarrow d\gamma$ process has been thoroughly studied. The best phenomenological windows are the direct CP-violating parameters in radiative $K$ decays for real photon emissions, and the rare $K_{L}\rightarrow\pi^{0}e^{+}e^{-}$ and $K_{L}\rightarrow\pi^{0}\mu^{+}\mu^{-}$ decays for the $s\rightarrow d\gamma^{\ast}$ transition. For all these observables, a sufficiently good control over the purely long-distance SM contributions has to be achieved to access to the short-distance physics, where NP effects could be competitive. So, in the first part of this paper, the SM predictions were systematically reviewed, with the results:

\begin{enumerate}
\item \boldmath   $K^{+}\rightarrow\pi^{+}\pi^{0}\gamma$\unboldmath  : We included the $\Delta I=3/2$ contributions, which were missing in the literature, and found that they enhance the loop amplitude by
about $50\%$. As a result, the recent NA48 measurement~\cite{ExpKppg} of the direct-emission electric amplitude can be well-reproduced without the inclusion of significant counterterm contributions. Concerning direct CP-violation, we identified an observable, Eq.~(\ref{a0CP}), which is not phase-space suppressed, and could thus help increase the experimental sensitivity to $\varepsilon_{+0\gamma}^{\prime}$. Thanks to the improved experimental and theoretical analyses, the prediction for $\varepsilon_{+0\gamma}^{\prime}$ in the SM is under good control, though a large cancellation between the $Q_{3,..,10}$ (four-quark operators, see Eq.~(\ref{OPE})) and $Q_{\gamma}^{-}$ (magnetic operator, see Eq.~(\ref{HeffG})) contributions limits its overall precision, $\varepsilon_{+0\gamma}^{\prime}=5(5)\times10^{-5}$.

\item \boldmath    $K^{0}\rightarrow\pi^{+}\pi^{-}\gamma$\unboldmath : The inclusion of the $\Delta I=3/2$ contributions, together with the experimental extraction of the counterterms from $K^{+}\rightarrow\pi^{+}\pi^{0}\gamma$, permits to reach a good accuracy. Contrary to previous analyses, we found that the $Q_{3,..,10}$ contribution to the direct CP-violating parameter $\varepsilon_{+-\gamma}^{\prime}$ is suppressed by the $\Delta I=1/2$ rule and negligible against that of
$Q_{\gamma}^{-}$. Altogether, the very small value $\varepsilon_{+0\gamma}^{\prime}=0.8(3)\times10^{-5}$ is obtained in the SM.

\item \boldmath  $K^{0}\rightarrow\gamma\gamma$\unboldmath : For the direct CP-violating parameter $\varepsilon_{||}^{\prime}$, we confirmed the computation of Ref.~\cite{BuccellaDM91} for the $Q_{3,...,10}$ contribution. However, that of $Q_{\gamma}^{-}$ was missing, and lead to a factor five enhancement to $\varepsilon_{||}^{\prime}\approx1.4\times10^{-5}$ in the SM. For the parameter $\varepsilon_{\perp}^{\prime}$, the situation changes completely compared to Ref.~\cite{BuccellaDM91}. Indeed, the anatomy of $K_{L}\rightarrow\gamma\gamma$ has been clarified in Ref.~\cite{GerardST}, where the absence of QCD penguin contributions at leading order was proven. As a result, we got the striking prediction that $\varepsilon_{\perp}^{\prime}$ is a direct measure of these QCD penguins, $\varepsilon_{\perp}^{\prime}(Q_{3,...,10})=-i\operatorname{Im}A_{0}/\operatorname{Re}A_{0}$, while the $Q_{\gamma}^{+}$ contribution is much smaller in the SM. So, this $\Delta I=1/2$-enhanced observable could resolve the QCD versus electroweak penguin fraction in $\varepsilon^{\prime}$ (to which $\varepsilon_{+0\gamma}^{\prime}$, $\varepsilon_{+-\gamma}^{\prime}$, and $\varepsilon_{||}^{\prime}$ have essentially no sensitivity), and could improve the theoretical prediction of $\varepsilon_{K}$.

\item \boldmath   $K_{L}\rightarrow\pi^{0}\ell^{+}\ell^{-}$\unboldmath  : We have updated the branching ratio formulas of Refs.~\cite{BDI03,ISU04,MST06}, which now reflect the better experimental situation for $K_{L}\rightarrow\pi^{0}\gamma\gamma$, the extraction of the matrix elements from $K_{\ell3}$ performed in Ref.~\cite{MesciaS07}, and the reanalysis of the error treatment (along the lines of Refs.~\cite{DEIP98,BDI03}) for the indirect CP-violating contribution detailed in Appendix~\ref{AppKpll}.

\item \boldmath $\operatorname{Re}(\varepsilon^{\prime}/\varepsilon)$\unboldmath : We have computed the long-distance part of the magnetic operator contribution to $\varepsilon^{\prime}$, as well as to $\Delta M_{K}$ and $\varepsilon_{K}$. While it is (as expected) negligible for the last two, it could a priori be sizeable for $\varepsilon^{\prime}$ if $Q_{\gamma}^{-}$ is enhanced by NP. Even though this contribution cannot be predicted accurately, and the short-distance part is lacking, we proved that the recent NA48 bound~\cite{ExpKppg} on $\varepsilon_{+0\gamma}^{\prime}$ ensures that it does not exceed about $30\%$ of $\operatorname{Re}(\varepsilon^{\prime}/\varepsilon)^{\exp}$, and thus, for the time being, can be neglected.
\end{enumerate}

In the second part of the paper, the possible NP impacts on the $s\rightarrow d\gamma$ process were analyzed. The direct CP-violating parameters in radiative decays offer the cleanest accesses to $s\rightarrow d\gamma$ since they are free from any competing NP effect (except $\varepsilon_{\perp
}^{\prime}$) once the $Q_{3,...,10}$ contributions are fixed in terms of $\operatorname{Re}(\varepsilon^{\prime}/\varepsilon)^{\exp}$. However, these parameters are not yet tightly bounded experimentally. By contrast, the $K_{L}\rightarrow\pi^{0}\ell^{+}\ell^{-}$ decays are sensitive to both $s\rightarrow d\gamma$ and $s\rightarrow d\gamma^{\ast}$ processes, as well as to many other possible FCNC, but are already tightly bounded experimentally. So, to resolve the possible interferences among NP contributions, and thereby assess how large the CP-violating parameters could be, several scenarios were considered. The main discriminator was chosen as the assumed NP dynamics, which translates as a choice of basis for the effective four-fermion semi-leptonic operators. To summarize each scenario:

\begin{enumerate}
\item \textbf{Model-independent}: The basis~(\ref{BasisPheno}) is constructed so as to minimize the interferences between the NP contributions in physical observables~\cite{Basis}. Its main characteristics is the entanglement of the magnetic operator $Q_{\gamma}^{+}$ with the semileptonic operator $Q_{V,\ell
}=\bar{s}\gamma_{\mu}d\otimes\bar{\ell}\gamma^{\mu}\ell$, since they both produce the $\ell^{+}\ell^{-}$ pair in the same $1^{--}$ state. So, if these two interfere destructively, the CP-violating parameters in radiative decays could be large. For example, if there is a $80$\% cancellation between $Q_{\gamma}^{+}$ and $Q_{V,e}$ in $K_{L}\rightarrow\pi^{0}e^{+}e^{-}$, $\varepsilon_{+0\gamma}^{\prime}$ could saturate its current experimental limit $-22(36)\%$~\cite{ExpKppg}, see Fig.~\ref{Fig11}. By comparison, a strict enforcement of the MFV hypothesis would suppress all these CP-violating parameters down to the $10^{-4}$ range. This shows the power of these parameters in exhibiting deviations with respect to MFV.

\item \textbf{Tree-level FCNC}: The basis~(\ref{LLRbasis}) assumes that the NP is invariant under $SU(2)_{L}\otimes U(1)_{Y}$, and generates the semileptonic operators through tree-level processes. The main characteristics is the strong correlation between $K\rightarrow\pi\nu\bar{\nu}$, $K_{L}\rightarrow\pi
^{0}(\ell^{+}\ell^{-})_{1^{--}}$, and $K_{L}\rightarrow\pi^{0}(\ell^{+}\ell^{-})_{1^{++},0^{-+}}$ for a given lepton flavor, but the absence of leptonic universality. This is sufficient to resolve the entanglement between $Q_{\gamma}^{+}$ and $Q_{V,\ell}$. The CP-violating parameters are then
bounded by $K_{L}\rightarrow\pi^{0}e^{+}e^{-}$, see Fig.~\ref{Fig12}, with e.g. $|\varepsilon_{+0\gamma}^{\prime}|\lesssim11\%$. Also, each rare decay can saturate its experimental bound, though all cannot be large simultaneously, but for $K_{L}\rightarrow\pi^{0}\nu\bar{\nu}$ which must satisfy its model-independent bound~(\ref{KLnn}).

\item \textbf{Loop-level FCNC / electroweak penguins only}: The basis~(\ref{HPB}) provided by the SM electroweak penguin and box operators is adequate when the FCNC originates entirely from loop processes. The main characteristics of this scenario is the entanglement of the $s\rightarrow d\gamma$ and $s\rightarrow d\gamma^{\ast}$ photon penguins in $K_{L}\rightarrow\pi^{0}(\ell^{+}\ell^{-})_{1^{--}}$. However, once in this basis, it is natural to allow the photon and $Z$ to couple also to quarks, bringing
$\varepsilon^{\prime}$ in the picture. Then, the only way to have sizeable effects in rare decays is to allow for a large box operator, to fine-tune the electroweak penguins so as to avoid the large vector current contribution in $\varepsilon^{\prime}$, or to allow for $Q_{\gamma}^{\pm}$ to be large. The
main issue is thus to resolve the fine-tuning in $\varepsilon^{\prime}$. Indeed, if it is extreme, one would conclude that the chosen basis is inadequate, and NP is not aligned with the $Z$ or $\gamma$ penguins. While the direct CP-violating parameters are rather insensitive, and could reach at most a few percents, the correlation between the $K_{L}\rightarrow\pi^{0}e^{+}e^{-}$ and $K_{L}\rightarrow\pi^{0}\mu^{+}\mu^{-}$ modes can be used to signal such a fine-tuning in $\varepsilon^{\prime}$, see Fig.~\ref{Fig14}.

\item \textbf{Loop-level FCNC / electroweak and chromomagnetic penguins}. When generated at loop level, the magnetic operators are always accompanied by the chromomagnetic operators since the $SU(3)_{C}\otimes U(1)_{em}$ quantum numbers must flow through the loop. Their relative strength, however, cannot
be assessed model-independently. If one forces the two to be of similar strengths, the main characteristic of this scenario is then the tight fine-tuning required by $\varepsilon^{\prime}$ between the gluonic and the electroweak penguins, see Fig.~\ref{Fig15}. To resolve this, rare decays are
rather ineffective, but the direct CP-violating parameters are perfectly suited since they directly measure $Q_{\gamma}^{\pm}$. The parameter $\varepsilon_{\perp}^{\prime}$ is particularly interesting, since it is also directly sensitive to the $\Delta I=1/2$ chromomagnetic operator $Q_{g}^{-}$
through its dependence on $\operatorname{Im}A_{0}/\operatorname{Re}A_{0}$.

\item \textbf{Loop-level FCNC / MSSM}. The main characteristics of the MSSM is the strict correlation between the magnetic and chromomagnetic penguins, Eq.~(\ref{MSSMRGE}). Depending on the level of fine-tuning between gluonic and electroweak penguins in $\varepsilon^{\prime}$, this scenario collapses either
to scenario 3 or 4. In the former case, both magnetic penguins have to be small since they are correlated, and the MSSM further forbids the specific fine-tuning between the electroweak penguins required by $\varepsilon^{\prime}$. As a result, the rare decays are tightly constrained, see Fig.~\ref{Fig13}, with the possible exception of $K\rightarrow\pi\nu\bar{\nu}$ if the box amplitudes are exceptionally large. It should be stressed though that the cancellation between the gluonic and electroweak penguins required in $\varepsilon^{\prime}$ need not be extreme to leave room for sizeable
supersymmetric contributions to both $K_{L}\rightarrow\pi^{0}\ell^{+}\ell^{-}$ and direct CP-violating parameters, see Fig.~\ref{Fig15}. Finally, radiative decays were found to provide a competitive bound on $\operatorname{Re}\delta_{12}^{D}$, see Eq.~(\ref{radMIA}).
\end{enumerate}

In conclusion, the stage is now set theoretically to fully exploit the $s\rightarrow d\gamma$ transition. The SM predictions are under good control, the sensitivity to NP is excellent, and signals in rare and radiative $K$ decays not far from the current experimental sensitivity are possible. Thus, with the advent of the next generation of $K$ physics experiments, the complete set of flavor changing electromagnetic processes, $s\rightarrow d\gamma$, $b\rightarrow(s,d)\gamma$, and $\ell\rightarrow\ell^{\prime}\gamma$, could become one of our main windows into the flavor sector of the NP which
will hopefully show up at the LHC.

\subsection*{Acknowledgements}

We would like to thank Jean-Marc G\'{e}rard for the interesting discussions and his suggestions. P.~M. thanks the Karlsruhe Institute of Technology, where part of this work was completed, for
its hospitality.

\vfill                 

\appendix     \pagebreak 

\addtocontents{toc}{\protect\setcounter{tocdepth}{1}}

\section{The $K\rightarrow\pi\pi\gamma$ decays in Chiral Perturbation
Theory\label{AppA}}

At $\mathcal{O}(p^{2})$, the direct emission vanishes while $E_{IB}$ is fully
predicted in terms of the $\mathcal{O}(p^{2})$ $K\rightarrow\pi\pi$
amplitudes. Including $\mathcal{O}(p^{4})$ corrections, the IB amplitudes
become
\begin{equation}
E_{IB}^{++0}=-\frac{em_{K}^{3}A\left(  K^{+}\rightarrow\pi^{+}\pi^{0}\right)
^{phys}}{K_{1}\cdot qP\cdot q}\;,\;E_{IB}^{1+-}=-\frac{em_{K}^{3}A\left(
K_{1}\rightarrow\pi^{+}\pi^{-}\right)  ^{phys}}{K_{1}\cdot qK_{2}\cdot q}\;,
\end{equation}
while $E_{IB}^{2+-}=E_{IB}^{200}=E_{IB}^{100}=0$ in the limit of
CP-conservation ($\sqrt{2}|K_{2,1}\rangle\equiv|K^{0}\rangle\pm|\bar{K}%
^{0}\rangle$ in the usual ChPT conventions~\cite{DAmbrosioI96}). The subscript
''$phys$'' means the full $\mathcal{O}(p^{4})$ on-shell decay amplitudes, i.e.
with physical (renormalized) weak couplings, masses, decay constants, and
including the strong phases arising from the $\pi\pi$ loops \cite{Bijnens98}.

Once the IB amplitudes are correctly renormalized, the left-over
$\mathcal{O}(p^{4})$ contributions are purely of the direct-emission type,
i.e. vanish in the limit $q\rightarrow0$ (which translates as $E_{DE}%
\rightarrow c^{st}$, given the factored out projector in Eq.~(\ref{GenAmpli}%
)). The loop contributions, still in the limit of CP-conservation, are
\begin{subequations}
\label{elec}%
\begin{align}
E_{loop}^{++0}  &  =-\dfrac{e(m_{K}^{2}-m_{\pi}^{2})m_{K}}{8\pi^{2}F_{\pi}%
}\left[  h(z_{1})+g(z_{2})-4A^{+}h_{\pi\pi}\left(  -z_{3}\right)
+2A^{K}h_{KK}\left(  -z_{3}\right)  \right]  \;,\\
E_{loop}^{1+-}  &  =-\dfrac{e(m_{K}^{2}-m_{\pi}^{2})m_{K}}{8\pi^{2}F_{\pi}%
}\left[  h(z_{1})+h(z_{2})-8A^{0}h_{\pi\pi}\left(  -z_{3}\right)
-4A^{K}h_{KK}\left(  -z_{3}\right)  \right]  \;,\\
E_{loop}^{2+-}  &  =-\dfrac{e(m_{K}^{2}-m_{\pi}^{2})m_{K}}{8\pi^{2}F_{\pi}%
}\left[  h(z_{1})-h(z_{2})\right]  \;,\\
E_{loop}^{200}  &  =-\dfrac{e(m_{K}^{2}-m_{\pi}^{2})m_{K}}{8\pi^{2}F_{\pi}%
}\left[  g(z_{1})-g(z_{2})\right]  \;,\\
E_{loop}^{100}  &  =0\;,
\end{align}
\end{subequations}
where $h(z)=A^{8}h_{K\eta}(z)+A^{0}h_{\pi K}(z)-A^{+}h_{K\pi}(z)$ and
$g(z)=2A^{+}(h_{\pi K}(z)+h_{K\pi}(z))$. The loop functions $h_{ij}(z)$ are
given in Ref.~\cite{DAmbrosioI94} in terms of the subtracted three-point
Passarino-Veltman function $C_{20}$, and the $A^{i}$ are expressed in terms of
the $\mathcal{O}(p^{2})$ on-shell (but not necessarily physical) $K\rightarrow
PP$ amplitudes:
\begin{subequations}
\begin{align}
A^{+}  &  =\dfrac{A\left(  K^{+}\rightarrow\pi^{+}\pi^{0}\right)  }{2F_{\pi
}(m_{K}^{2}-m_{\pi}^{2})}=\frac{5}{6}G_{27}^{3/2}-\dfrac{1}{2}A^{ew}\;,\\
A^{0}  &  =\dfrac{A\left(  K_{1}\rightarrow\pi^{+}\pi^{-}\right)  }{2F_{\pi
}(m_{K}^{2}-m_{\pi}^{2})}=G_{8}+\frac{1}{9}G_{27}^{1/2}+\frac{5}{9}%
G_{27}^{3/2}-A^{ew}\;,\\
A^{8}  &  =\dfrac{-\sqrt{3}A\left(  K^{+}\rightarrow\pi^{+}\eta_{8}\right)
}{2F_{\pi}(m_{K}^{2}-m_{\pi}^{2})}=G_{8}-\frac{4}{9}G_{27}^{1/2}+\frac{5}%
{18}G_{27}^{3/2}-\frac{3}{2}A^{ew}\;,\\
A^{ew}  &  =\dfrac{A\left(  K^{+}\rightarrow K^{+}K_{S}\right)  }{2F_{\pi
}(m_{K}^{2}-m_{\pi}^{2})}=\frac{2e^{2}F_{\pi}^{3}G_{ew}}{2F_{\pi}(m_{K}%
^{2}-m_{\pi}^{2})}\;,
\end{align}
\end{subequations}
with $\left|  G_{8}\right|  =9.1\times10^{-12}\,$MeV$^{-2}$, $\left|
G_{27}\right|  =|G_{27}^{1/2}|=|G_{27}^{3/2}|=5.3\times10^{-13}\,$MeV$^{-2}$,
and $\operatorname*{sign}(G_{8}/G_{27})=+1$. The vanishing of $E_{loop}^{100}$
is a consequence of the CP symmetry combined with Bose symmetry. All the loop
amplitudes are finite, but some separately finite counterterms contribute
($N_{i}\equiv N_{14}-N_{15}-N_{16}-N_{17}$)%
\begin{equation}
(E_{CT}^{++0},\;E_{CT}^{1+-},\;E_{CT}^{2+-})=-\dfrac{2eG_{8}m_{K}^{3}}{F_{\pi
}}(-N_{i},\;2\operatorname{Re}N_{i},\;2i\operatorname{Im}N_{i})\;,\;\;E_{CT}%
^{2+-}=E_{CT}^{200}=E_{CT}^{100}=0\;.
\end{equation}
Finally, the $Q_{\gamma}^{-}$ operator enters as%
\begin{equation}
(E_{\gamma}^{++0},\;E_{\gamma}^{1+-},\;E_{\gamma}^{2+-})=\frac{eB_{T}m_{K}%
^{2}}{3(2\pi)^{2}F_{\pi}}(-C_{\gamma}^{-},\;\operatorname{Re}C_{\gamma}%
^{-},\;i\operatorname{Im}C_{\gamma}^{-})\;,\;\;E_{\gamma}^{2+-}=E_{\gamma
}^{200}=E_{\gamma}^{100}=0\;\;.
\end{equation}
Note that these $Q_{\gamma}^{-}$ contributions cannot be absorbed into the
$N_{i}$.

For $K\rightarrow\pi^{+}\pi^{0}\gamma$, the function $E^{loop}(W^{2}%
,T_{c}^{\ast})$ occurring in Eq.~(\ref{Eloop}) is
\begin{equation}
G_{8}E^{loop}(z_{1},z_{2})=\operatorname{Re}\left[h(z_{1})+g(z_{2})-4A^{+}h_{\pi
\pi}\left(  -z_{3}\right)\right] \;,
\end{equation}
as obtained from Eq.~(\ref{elec}) by neglecting $\operatorname{Re}A^{ew}%
\ll\operatorname{Re}G_{8,27}$ (since $G_{ew}$ is entirely generated by the
electroweak penguins). The real part refers to the weak phases only.
Performing the multipole expansion and expressing the $K\rightarrow PP$
amplitudes parametrically in terms of the $K\rightarrow\pi\pi$ isospin
amplitudes%
\begin{equation}
A_{0}=\sqrt{2}F_{\pi}(m_{K}^{2}-m_{\pi}^{2})\left[  G_{8}+\frac{1}{9}%
G_{27}^{1/2}-\frac{2}{3}A^{ew}\right]  \;,\;A_{2}=2F_{\pi}(m_{K}^{2}-m_{\pi
}^{2})\left[  \frac{5}{9}G_{27}^{3/2}-\frac{1}{3}A^{ew}\right]  \;,
\end{equation}
we find%
\begin{subequations}
\begin{align}
G_{8}E_{1}^{loop}(z_{3}\overset{}{=}2z) &  =\dfrac{-em_{K}}{(4\pi F_{\pi})^{2}%
}\left[  A_{0}h_{0}(z)+A_{2}h_{2}(z)+A_{\delta2}\delta h_{2}(z)\right]  \;,\\
h_{0}(z) &  =\sqrt{2}(h_{K\eta}(z)+h_{\pi K}(z))\;,\\
h_{2}(z) &  =4h_{\pi K}(z)+\frac{3}{2}h_{K\pi}(z)-6|h_{\pi\pi}\left(
-2z\right)  |-\frac{1}{2}h_{K\eta}(z)\;,\\
\delta h_{2}(z) &  =3h_{K\eta}(z)-6h_{KK}\left(  -2z\right)  \;,
\end{align}
\end{subequations}
where $A_{\delta2}=-(2/3)F_{\pi}(m_{K}^{2}-m_{\pi}^{2})A^{ew}$. For the small
$\delta h_{2}(z)$ term, we can further set $\operatorname{Im}A_{\delta
2}\approx\operatorname{Im}A_{2}$ since CP-violation from $Q_{8}$ dominates in
the $\Delta I=3/2$ channel. Eq.~(\ref{ep0g}) is then found by defining
$(\delta)h_{20}(z_{3})=(\delta)h_{2}(z)/h_{0}(z)$. Let us stress that $A_{0}$,
$A_{2}$ are just convenient parameters to keep track of the weak phases of
$G_{8}$, $G_{27}$, and $G_{ew}$. As such, they do not include any strong
phase. Further, the strong phase originating from $h_{\pi\pi}$ is discarded
since already taken care of through the multipole expansion (the absolute
value is adequate since $\operatorname{Re}h_{\pi\pi}\left(  -z_{3}\right)  >0$
over the phase-space).

Similarly, the $K^{0}\rightarrow\pi^{+}\pi^{-}\gamma$ direct emission
amplitude occurring in Eq.~(\ref{hpm}) is the dipole part of the amplitude in
Eq.~(\ref{elec}),%
\begin{subequations}
\begin{align}
E_{+-}(z_{3}\overset{}{=}2z) &  =-\dfrac{2em_{K}}{(4\pi F_{\pi})^{2}}\left[
A_{0}h_{0}^{\prime}(z)+A_{2}h_{2}^{\prime}(z)+A_{\delta2}\delta h_{2}^{\prime
}(z)\right]  -\frac{4eG_{8}m_{K}^{3}}{F_{\pi}}N_{i}\;,\\
h_{0}^{\prime}(z) &  =\sqrt{2}(h_{K\eta}(z)+h_{\pi K}(z)-4|h_{\pi\pi}\left(
-2z\right)  |)\;,\\
h_{2}^{\prime}(z) &  =-\frac{1}{2}h_{K\eta}(z)+h_{\pi K}(z)-\dfrac{3}%
{2}h_{K\pi}(z)-4|h_{\pi\pi}\left(  -2z\right)  |\;,\\
\delta h_{2}^{\prime}(z) &  =3h_{K\eta}(z)+6h_{KK}\left(  -2z\right)  \;.
\end{align}
\end{subequations}
Again, defining $(\delta)h_{20}^{\prime}(z_{3})=(\delta)h_{2}^{\prime
}(z)/h_{0}^{\prime}(z)$ immediately leads to Eq.~(\ref{epmg}).

It is worth noting that contrary to what is generally stated, the amplitude
for $K_{L}\rightarrow\pi^{0}\pi^{0}\gamma$ does not vanish at $\mathcal{O}%
(p^{4})$, but is suppressed by the $\Delta I=1/2$ rule. Being in addition a
pure quadrupole emission, the rate is tiny%
\begin{equation}
\mathcal{B}(K_{L}\rightarrow\pi^{0}\pi^{0}\gamma)_{G_{27}}=7.3\times
10^{-13}\;.
\end{equation}
For comparison, Ref.~\cite{EPRKppg} found using dimensional arguments that the
$G_{8}$ contribution at $\mathcal{O}(p^{6})$ is of the order of $10^{-10}$,
much larger but still far below the experimental bound $2.43\times10^{-7}$.

\subsection{$\varepsilon_{+0\gamma}^{\prime}$ beyond $\mathcal{O}(p^{4})$}

To get an estimate of the possible impact of higher order corrections, let us
include the counterterms $\bar{N}$ in Eq.~(\ref{ep0g}), so that%
\begin{equation}
\varepsilon_{+0\gamma}^{\prime}(z)=\frac{\sqrt{2}|\varepsilon^{\prime}%
|}{\omega}f(z,\Omega,\delta_{N})\;,\;\;f(z,\Omega,\delta_{N})=\frac{1+\omega
\Omega(h_{20}(z)+\delta h_{20}(z))-\operatorname{Im}\delta_{N}}{(\Omega
-1)(1+\omega h_{20}(z)-\operatorname{Re}\delta_{N})}-\frac{1}{\Omega-1}-1\;,
\end{equation}
with%
\begin{equation}
\operatorname{Re}\delta_{N}=\frac{1}{h_{0}(z)}\frac{\sqrt{2}m_{K}^{2}}%
{m_{K}^{2}-m_{\pi}^{2}}\operatorname{Re}\bar{N}\;,\;\;\operatorname{Im}%
\delta_{N}=\frac{\sqrt{2}}{h_{0}(z)}\frac{m_{K}^{2}}{m_{K}^{2}-m_{\pi}^{2}%
}\operatorname{Im}\bar{N}\frac{\operatorname{Re}A_{0}}{\operatorname{Im}A_{0}%
}\;.
\end{equation}
Parametrically, $\bar{N}$ accounts for all the $\mathcal{O}(p^{4})$
counterterms, as well as for the momentum-independent parts of higher order
effects. To proceed, some assumptions have to be made on its weak phase. From
the experimental data, we know that $\operatorname{Re}\bar{N}$ is of the
typical size expected for $\mathcal{O}(p^{6})$ corrections instead of
$\mathcal{O}(p^{4})$. Since both $Q_{6}$ and $Q_{8}$ contribute at
$\mathcal{O}(p^{6})$ through two-loop graphs, $\bar{N}$ a priori receives
contributions from all the penguin operators, besides the current-current
operators. On the other hand, the electromagnetic operators are too small to
affect $\operatorname{Re}\bar{N}$, allowing their impact to be pulled out and
treated separately (see main text).

So, inspired by the $\mathcal{O}(p^{4})$ loop result, we parametrically write:%
\begin{equation}
\bar{N}=b\left(  \left(  1-a\right)  A_{0}+aA_{2}+i\delta a\operatorname{Im}%
A_{2}\right)  \;, \label{model0}%
\end{equation}
with $b\sim\mathcal{O}(p^{6})/\mathcal{O}(p^{4})$. Assuming the corrections
parametrized in terms of $A_{0}$ and $A_{2}$ are of the same sign as at
$\mathcal{O}(p^{4})$, we take $a\in\lbrack0,1]$ to span from the pure QCD
penguin to the pure electroweak penguin scenario, and $a\approx(1+\omega
)^{-1}\approx0.95$ if the $\mathcal{O}(p^{4})$ scaling between the $G_{8}$ and
$G_{27}$ contributions survives at $\mathcal{O}(p^{6})$. In a way similar to
what happens at $\mathcal{O}(p^{4})$, the parameter $\delta a$ allows for
additional $Q_{8}$ contributions in the imaginary parts. Since at
$\mathcal{O}(p^{4})$, it comes entirely from $K\rightarrow\pi\eta$ and
$K\rightarrow KK$ vertices and misses the $K\rightarrow\pi\pi$ vertex and its
associated loop, we expect $\delta a\ll1$. With this,%
\begin{equation}
\frac{\operatorname{Im}\delta_{N}}{\operatorname{Re}\delta_{N}}%
=\frac{(1-a)+(a+\delta a)\omega\Omega}{(1-a)+a\omega}\;. \label{model}%
\end{equation}
By varying $\Omega\in\lbrack-1,$ $+0.8]$, $a\in\lbrack0,1]$, $|\delta
a|\leq0.1$, and $\operatorname{Re}\bar{N}$ within $1\sigma$ of the
range~(\ref{CT}), we get the final prediction~(\ref{Finalep0g}).

\section{Updated error analysis for $\mathcal{B}(K_{L}\rightarrow\pi^{0}%
\ell^{+}\ell^{-})$\label{AppKpll}}

Besides minor changes in the conventions, essentially to pull out an outdated
value of $\operatorname{Im}\lambda_{t}$ from the coefficients in
Ref.~\cite{MST06}, we have updated most of the numbers in Eq.~(\ref{MasterKL})
to reflect a better treatment of the errors. For $C_{dir}^{\ell}$, the smaller
errors are taken from Ref.~\cite{MS}, relying on precise extraction from
$K_{\ell3}$ decays.

The new value of $C_{\gamma\gamma}^{\mu}$ reflects the improved experimental
situation on $K_{L}\rightarrow\pi^{0}\gamma\gamma$, whose rate went down and
is now in perfect agreement between KTeV~\cite{KTeVgg} and NA48~\cite{NA48gg}.
We note that this agreement, together with that on the contribution of the
resonances (assuming vector meson dominance (VMD)), renders the error on
$C_{\gamma\gamma}^{\mu}$ extremely conservative~\cite{ISU04}.

For the coefficients $C_{mix}^{\ell}$ and $C_{int}^{\ell}$, the changes are
deeper. These coefficients are sensitive to the $K_{S}\rightarrow\pi\ell
^{+}\ell^{-}$ amplitude, which is entirely dominated by the virtual photon
penguin:%
\begin{equation}
A(K_{1}(P)\rightarrow\pi^{0}\gamma^{\ast}(q))=\frac{eG_{F}}{8\pi^{2}}%
W_{S}\left(  z\right)  \left(  q^{2}P^{\mu}-q^{\mu}P\cdot q\right)
\;,\;\;W_{S}\left(  z\right)  =a_{S}+b_{S}z+W_{S}^{\pi\pi}\left(  z\right)
\;, \label{KSAS}%
\end{equation}
where $z=q^{2}/M_{K^{0}}^{2}$ and $\alpha_{em}\approx1/137$. As detailed in
Ref.~\cite{DEIP98}, the only assumption behind the parametrization of the
$W_{S}(z)$ form-factor is that all the intermediate states other than $\pi\pi$
are well described by a linear polynomial in $z$, and thus can be absorbed in
the unknown substraction constants $a_{S}$ and $b_{S}$. The $\pi\pi$ loop
function $W_{S}^{\pi\pi}\left(  z\right)  $, the only one to develop an
imaginary part, was estimated including both the phenomenological
$K_{S}\rightarrow\pi^{+}\pi^{-}\pi^{0}$ vertex (i.e., including slopes), and
the physical $\pi^{+}\pi^{-}\rightarrow\gamma^{\ast}$ vertex (i.e., with its
VMD behavior). Because $K_{S}\rightarrow\pi^{+}\pi^{-}\pi^{0}$ is dominantly
CP-violating, and $b_{S}$ is higher order in the chiral expansion, the leading
term $a_{S}$ dominates.

Given the current error on the $K_{S}\rightarrow\pi^{0}\ell^{+}\ell^{-}$
rates, setting $b_{S}/a_{S}=0.4$ and keeping only quadratic terms in
$a_{S}^{2}$ give reasonable predictions for the $K_{L}$ rates. However, in
preparation for better measurements, we prefer to systematically account for
the momentum dependence of the form-factor in extracting the coefficients of
the master formula~(\ref{MasterKL}). To this end, and contrary to previous
parametrizations, we find that it is not convenient to use $a_{S}$ as the
parameter entering Eq.~(\ref{MasterKL}), because this necessarily overlooks
the other terms of $W_{S}(z)$.

To construct the alternative parameter $\bar{a}_{S}$ occurring in
Eq.~(\ref{MasterKL}), we start by defining for the muon and electron modes:%
\begin{equation}
a_{\ell(,\Lambda)}^{2}=\frac{\int_{(\Lambda)}d\Phi_{\ell}|W_{S}\left(
z\right)  |^{2}}{\int_{(\Lambda)}d\Phi_{\ell}}\;,\;d\Phi_{\ell}=\beta_{\ell
}\left(  z\right)  \beta_{\pi}^{3}\left(  z\right)  (1+2r_{\ell}^{2}/z)dz\;,
\end{equation}
with $\beta_{\ell}\left(  z\right)  =\sqrt{1-4r_{\pi}^{2}/z}$, $\beta_{\pi
}\left(  z\right)  =\lambda^{1/2}(1,r_{\pi}^{2},z)$, $\lambda(a,b,c)=a^{2}%
+b^{2}+c^{2}-2(ab+ac+bc)$, and $r_{i}=m_{i}/m_{K}$. The expansions of
$a_{\ell(,\Lambda)}^{2}$ in terms of $a_{S}$ and $b_{S}$ read:
\begin{subequations}
\begin{align}
a_{e}^{2}  &  =a_{S}^{2}+0.278a_{S}b_{S}-0.015a_{S}+0.031b_{S}^{2}%
-0.005b_{S}+0.0003\;,\\
a_{e,\Lambda}^{2}  &  =a_{S}^{2}+0.443a_{S}b_{S}-0.029a_{S}+0.057b_{S}%
^{2}-0.009b_{S}+0.0005\;,\\
a_{\mu}^{2}  &  =a_{S}^{2}+0.585a_{S}b_{S}-0.052a_{S}+0.091b_{S}%
^{2}-0.018b_{S}+0.0011\;.
\end{align}
\end{subequations}
The subscript $\Lambda$, if present, indicates a cut for $z>\Lambda
^{2}/M_{K^{0}}^{2}$. Experimentally, it is set at $\Lambda=165$ MeV for the
electron mode to deal with $K_{S}\rightarrow\pi^{0}\pi^{0}$ backgrounds. In
terms of these, the $K_{S}$ rates are,
\begin{subequations}
\begin{align}
\mathcal{B}(K_{S}  &  \rightarrow\pi^{0}e^{+}e^{-})_{\Lambda}=2.41\cdot
10^{-9}\;a_{e,\Lambda}^{2}\overset{\exp}{=}(3.0_{-1.2}^{+1.5}\pm
0.2)\cdot10^{-9}\;\text{\cite{NA48asE}},\\
\mathcal{B}(K_{S}  &  \rightarrow\pi^{0}\mu^{+}\mu^{-})=0.990\cdot
10^{-9}\;a_{\mu}^{2}\,\,\overset{\exp}{=}(2.9_{-1.2}^{+1.4}\pm0.2)\cdot
10^{-9}\;\text{\cite{NA48asM}}.
\end{align}
\end{subequations}
The numerical coefficients have no significant errors since they are functions
of the masses, $G_{F}$, $\alpha_{em}$, and $\tau_{S}$ only. To optimize the
theoretical and experimental information, we want to average these two
measurements. This makes sense because, as $0.1<b_{S}/a_{S}<0.7$ and
$0.8<|a_{S}|<1.6$, the following ratio is very stable, even though depends on
the sign of $a_{S}$:%
\begin{equation}
r_{e/\mu}=a_{\mu}^{2}/a_{e,\Lambda}^{2}=1.035(24)\;[1.071(25)]\;,
\label{ratio1}%
\end{equation}
with $a_{S}<0$ indicated inside brackets. The error is mostly driven by the
range on $b_{S}$, but given that VMD would fix $b_{S}/a_{S}\approx m_{K}%
^{2}/m_{\rho}^{2}\approx0.4$, we think $0.1<b_{S}/a_{S}<0.7$ is very
conservative. Note that with the cut $\Lambda>2m_{\mu}$, this ratio would be
closer to one and even more stable as the $a_{e,\Lambda}^{2}$ and $a_{\mu}%
^{2}$ expansions in $a_{S}$ and $b_{S}$ tend to coincide. We therefore define
the average of $a_{\mu}^{2}$ and $a_{e,\Lambda}^{2}\times r_{e/\mu}$ with
$\Lambda=165$ MeV as $\bar{a}_{S}=1.25(22)$. The difference between $a_{S}<0$
and $a_{S}>0$ is negligible compared to the experimental errors. The error on
$r_{e/\mu}$ is \textit{not} included in $\bar{a}_{S}$, but instead in the
coefficients of Eq.~(\ref{MasterKL}).

The pure indirect CP-violating contribution is found from $\Gamma
(K_{L}\rightarrow\pi^{0}\ell^{+}\ell^{-})_{ICPV}=|\varepsilon|^{2}\Gamma
(K_{S}\rightarrow\pi^{0}\ell^{+}\ell^{-})$ with $|\varepsilon|=(2.228\pm
0.011)\times10^{-3}$. This immediately gives the coefficients $C_{mix}^{\mu}$
in Eq.~(\ref{MasterKL}) for the muon mode, to which we assign an error of
$2.3\%$ due to Eq.~(\ref{ratio1}). For the electron mode, there is an
additional source of error due to the extrapolation from $\Lambda=165$ MeV
down to $\Lambda=2m_{e}$. To control that, we use%
\begin{equation}
a_{e,\Lambda}^{2}/a_{e}^{2}=1.053(29)\;[1.076(30)]\;, \label{ratio2}%
\end{equation}
as $0.1<b_{S}/a_{S}<0.7$ and $0.8<|a_{S}|<1.6$. This means that the
phase-space increase as $\Lambda\rightarrow2m_{e}$ is dampened by the
form-factor. We add the error from Eq.~(\ref{ratio1}) and~(\ref{ratio2}) in
quadrature to assign a $3.6\%$ error on $C_{mix}^{e}$ in Eq.~(\ref{MasterKL}).
Note that this extrapolation error may be dropped if the $\Lambda$ cut is also
needed for $K_{L}\rightarrow\pi^{0}e^{+}e^{-}$, which may be the case to deal
with the (CP-violating) backgrounds from $K_{L}\rightarrow\pi^{0}\pi^{0}$ decays.

We proceed similarly for the interference term:%
\begin{equation}
C_{int}^{\ell}\times\bar{a}_{S}=53.37w_{7V}\times\int d\Phi_{\ell}%
\;f_{+}\left(  z\right)  \frac{\operatorname{Im}\left(  \varepsilon
W_{S}\left(  z\right)  \right)  }{\operatorname{Im}\varepsilon}\overset
{\phi_{\varepsilon}\approx45^\circ}{=}53.37w_{7V}\times\int d\Phi_{\ell}\;f_{+}\left(  z\right)  W_{S}\left(
z\right)  \;,
\end{equation}
with $f_{+}\left(  z\right)  $ the form-factor of the FCNC matrix element
$\langle\pi^{0}|\bar{s}\gamma^{\mu}d|K^{0}\rangle$. The error on the numerical
prefactor is negligible. Let us rewrite $C_{int}^{\ell}$ in terms of $a_{\ell
}$:%
\begin{equation}
\left\{
\begin{array}
[c]{l}%
C_{int}^{e}\times\bar{a}_{S}=7.793w_{7V}\times a_{e,\Lambda}\times r_{im}%
^{e}\;,\\
C_{int}^{\mu}\times\bar{a}_{S}=1.650w_{7V}\times a_{\mu}\times r_{im}^{\mu}\;,
\end{array}
\right.  \;\;\;r_{im}^{\ell}\equiv\frac{\int d\Phi_{\ell}\;f_{+}\left(
z\right)  W_{S}\left(  z\right)  }{\int d\Phi_{\ell}\times\sqrt{\int_{\Lambda
}d\Phi_{\ell}|W_{S}\left(  z\right)  |^{2}/\int_{\Lambda}d\Phi_{\ell}}}\;.
\end{equation}
The ratios $r_{im}^{\ell}$ can be studied as $0.1<b_{S}/a_{S}<0.7$ and
$0.8<|a_{S}|<1.6$, and are found very stable:%
\begin{equation}
r_{im}^{e}=0.965(13)\;[-0.957(14)]\;,\;\;\;r_{im}^{\mu}%
=1.0455(8)\;[-1.0530(6)]\;. \label{raio3}%
\end{equation}
The error on $r_{im}^{e}$ is larger than that on $r_{im}^{\mu}$ because of the
extrapolation from $\Lambda=165$ MeV down to $\Lambda=2m_{e}$. So, in terms of
the average $\bar{a}_{S}$, and including the $\sim2\%$ error due to
Eq.~(\ref{ratio1}) gives the coefficients in Eq.~(\ref{MasterKL}).

Finally, it should be stressed that the intrinsic errors on the coefficients
$C_{mix}^{\ell}$ and $C_{int}^{\ell}$ are already below $5\%$ thanks to the
ratios~(\ref{ratio1},~\ref{ratio2},~\ref{raio3}), but could in principle be
improved in the future by better constraining $b_{S}/a_{S}$ using the
experimental $m_{\ell\ell}$ spectra for both $K_{S}\rightarrow\pi^{0}\ell
^{+}\ell^{-}$ decay modes.

\end{document}